\documentclass[twocolumn]{aastex701}

\DeclareFixedFont{\ttb}{T1}{txtt}{bx}{n}{8} % for bold
\DeclareFixedFont{\ttm}{T1}{txtt}{m}{n}{8}  % for normal

\usepackage{color}
\definecolor{deepblue}{rgb}{0,0,0.5}
\definecolor{deepred}{rgb}{0.6,0,0}
\definecolor{deepgreen}{rgb}{0,0.5,0}

\usepackage{listings}

\newcommand\pythonstyle{\lstset{
language=Python,
basicstyle=\ttm,
morekeywords={self},              % Add keywords here
keywordstyle=\ttb\color{deepblue},
emph={MyClass,__init__},          % Custom highlighting
emphstyle=\ttb\color{deepred},    % Custom highlighting style
stringstyle=\color{deepgreen},
frame=tb,                         % Any extra options here
showstringspaces=false
}}

\lstnewenvironment{python}[1][]
{
\pythonstyle
\lstset{#1}
}
{}

\newcommand\pythoninline[1]{{\pythonstyle\lstinline!#1!}}

\usepackage{graphicx}	% Including figure files
\usepackage{amsmath}	% Advanced maths commands

\begin{document}

\title{\texttt{diffhydro}: Inverse Multiphysics Modeling and Embedded Machine Learning in Astrophysical Flows}

\author[orcid=0000-0001-7832-5372,sname='Horowitz']{Benjamin Horowitz}
\affiliation{Kavli IPMU (WPI), UTIAS, The University of Tokyo, Kashiwa, Chiba 277-8583, Japan}
\affiliation{Center for Data-Driven Discovery, Kavli IPMU (WPI), UTIAS, The University of Tokyo, Kashiwa, Chiba 277-8583, Japan}
\affiliation{Lawrence Berkeley National Lab, 1 Cyclotron Road, Berkeley, CA 94720, USA}
\email[show]{ben.horowitz@ipmu.jp}

\author[0009-0007-5730-7838,gname=Zarija,sname=Luki\'c]{Zarija Luki\'c}
\affiliation{Lawrence Berkeley National Lab, 1 Cyclotron Road, Berkeley, CA 94720, USA}
\email{}  

\author[0000-0001-7457-8487,gname=Kentaro,sname=Nagamine]{Kentaro Nagamine}
\affiliation{Theoretical Astrophysics, Department of Earth \& Space Science, Graduate School of Science, The University of Osaka, 1-1 Machikaneyama, Toyonaka, Osaka 560-0043, Japan}
\affiliation{Theoretical Joint Research, Forefront Research Center, Graduate School of Science, The University of Osaka, 1-1 Machikaneyama, Toyonaka, Osaka 560-0043, Japan}
\affiliation{Kavli IPMU (WPI), UTIAS, The University of Tokyo, Kashiwa, Chiba 277-8583, Japan}
\affiliation{Department of Physics \& Astronomy, University of Nevada, Las Vegas, 4505 S. Maryland Pkwy, Las Vegas, NV 89154-4002, USA}
\affiliation{Nevada Center for Astrophysics, University of Nevada, Las Vegas, 4505 S. Maryland Pkwy, Las Vegas, NV 89154-4002, USA}
\email{kn@astro-osaka.jp}

\author[0000-0002-5712-6865,gname=Yuri,sname=Oku]{Yuri Oku}
\affiliation{Center for Cosmology and Computational Astrophysics (C3A),
Institute for Advanced Study in Physics, Zhejiang University, Hangzhou 310027, People’s Republic of China}
\email{oku@zju.edu.cn}

\begin{abstract}
%We introduce a substantially expanded differentiable astrophysical hydrodynamics framework, \texttt{diffhydro}, advancing our previous proof-of-concept demonstrations to physically richer ISM and gravity–turbulence regimes at scale. The new release adds radiative heating/cooling, OU–driven turbulence, self-gravity via multi-grid Poisson, and custom adjoints that enable efficient end-to-end gradients and multi-device scaling. Built in JAX, the solver’s modular finite-volume components are compiled by XLA into fused accelerator kernels, delivering high-throughput forward runs and tractable differentiation through long integrations. Standard tests (Sedov–Taylor, Kelvin–Helmholtz, driven/decaying turbulence) agree well with Athena++. We demonstrate distributed simulations up to $1024^{3}$ and gradient-based reconstructions of complex initial conditions in turbulent, self-gravitating, radiatively cooling flows. We also introduce a solver-in-the-loop neural “corrector” that reduces coarse-grid errors during time integration while preserving stability. Overall, \texttt{diffhydro} enables scalable PDE-constrained inference and integrated hybrid physics–ML modeling for astrophysical applications.

We present the extension of the differentiable hydrodynamics code, \texttt{diffhydro}, enabling scalable PDE-constrained inference and integrated hybrid physics–ML models for a wide range of astrophysical applications.  New physics additions include radiative heating/cooling, OU–driven turbulence, and self-gravity via multigrid Poisson.  We demonstrate good agreement with Athena++ code on standard validation tests like Sedov–Taylor, Kelvin–Helmholtz, and driven/decaying turbulence. We further introduce a solver-in-the-loop neural corrector that reduces coarse-grid errors during time integration while preserving stability. The addition of custom adjoints facilitates efficient end-to-end gradients and multi-device scaling.  We present simulations up to $1024^{3}$ elements, run on distributed GPU systems, and we show gradient-based reconstructions of complex initial conditions in turbulent, and self-gravitating, radiatively cooling flows.  The code is written in JAX, and the solver’s modular finite-volume components are compiled by XLA into fused accelerator kernels, delivering high-throughput forward runs and tractable differentiation through long integrations.

\end{abstract}


\section{Introduction}
Hydrodynamical simulations are central to modelling astrophysical gas across a wide range of scales, from the turbulent interstellar medium to the assembly and evolution of galaxies. With recent and forthcoming facilities such as \textit{JWST} \citep{2006SSRv..123..485G}, \textit{Rubin} \citep{2019ApJ...873..111I}, and \textit{XRISM} \citep{2020arXiv200304962X}, observations are now placing increasingly detailed constraints on the morphology, thermodynamics, and kinematics of multiphase gas. These data motivate simulation frameworks that can be coupled more directly to observations, both for predictive forward modelling and for inferring the physical conditions that give rise to observed structures in their dynamical environment. 

Established hydrodynamics codes—including \textsc{Athena++} \citep{2020ApJS..249....4S}, \textsc{Arepo} \citep{Arepo_public}, \textsc{Gizmo} \citep{Gizmo}, \textsc{Ramses} \citep{Ramses}, \textsc{nyx} \citep{Lukic2015}, and others—have achieved widespread success as accurate and efficient forward solvers. Yet their design priorities have historically centred on throughput and scalability for forward modelling, and they are usually implemented in low-level languages with static, MPI-dominated execution patterns. While ideal for large parameter sweeps, this structure makes it cumbersome to embed the solvers inside modern optimization or inference loops that require repeated, tightly coupled interaction with the numerical update. As a result, tasks such as initial-condition reconstruction, parameter estimation, and subgrid-model calibration typically fall back on outer-loop heuristics, surrogate models, or finite-difference sensitivities, which can be expensive and are prone to compounding numerical error \citep[e.g.][]{2021MNRAS.506.4011E,2022ApJ...941...42H,2023MNRAS.526.6103K,2025MNRAS.536..254M,2025arXiv250904067C}.

\begin{figure*}
    \centering

\begin{flushleft}
\emph{(a) Traditional Forward Simulation (Sec. \ref{sec:hydro}, \ref{sec:cooling}, \ref{sec:sn}, \ref{sec:selfgrav})}
\end{flushleft}
    \includegraphics[width=0.85\linewidth]{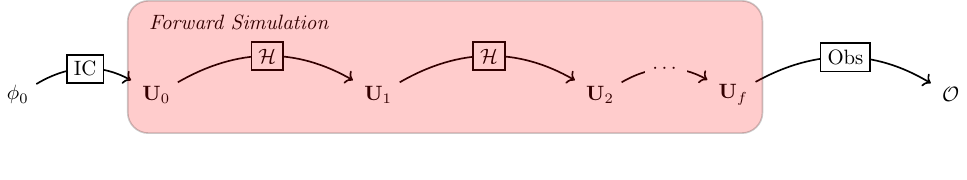}
    \vspace{-25pt}
    \begin{flushleft}
\emph{(b) Initial Condition/Parameter Reconstruction (Sec. \ref{sec:opt_ic})}
\end{flushleft}
    \includegraphics[width=0.99\linewidth]{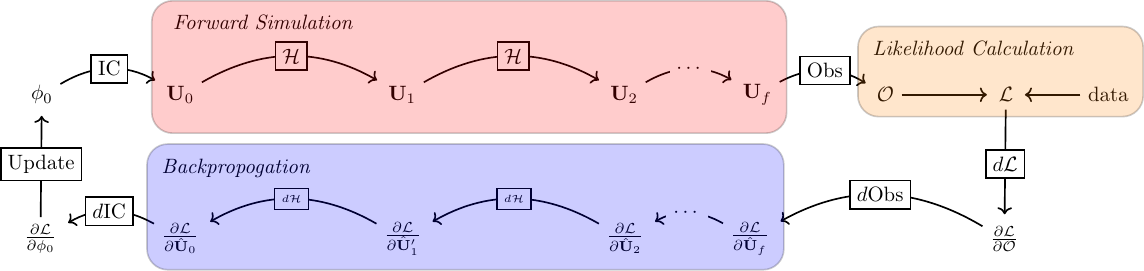}
        \begin{flushleft}
\emph{(c) Solver-in-the-Loop Optimization (Sec. \ref{sec:cil})}
\end{flushleft}
        \includegraphics[width=0.95\linewidth]{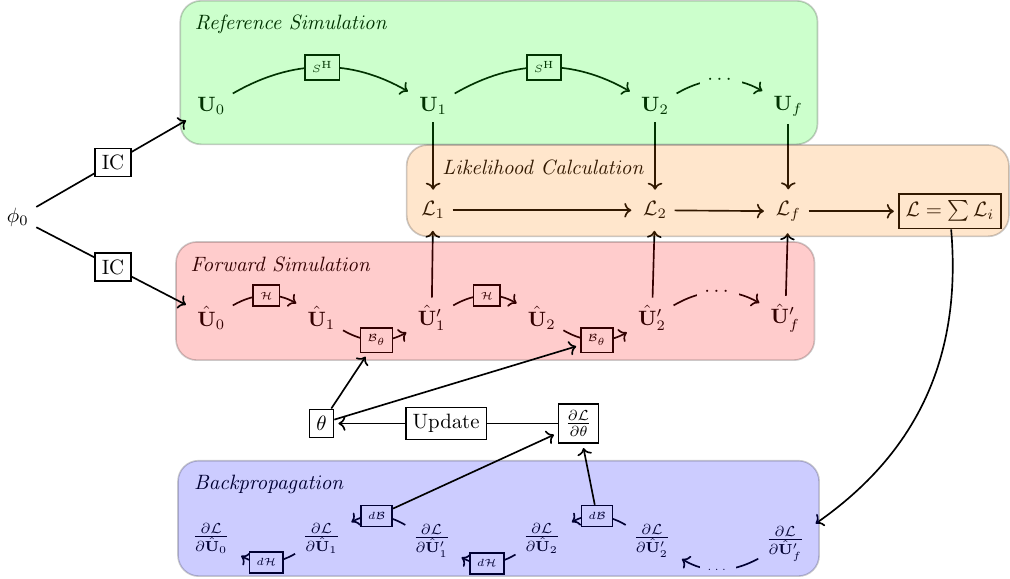}

    \caption{
Schematic overview of forward and inverse workflows enabled by \texttt{diffhydro}.
\textit{(a) Traditional simulation:} an initial condition (IC) specified by parameters $\boldsymbol{\phi}_0$ is mapped to the conservative state $\mathbf{U}_0$ and advanced forward through repeated applications of the hydrodynamic update operator $\mathcal{H}$, yielding a final state $\mathbf{U}_f$ from which observables $\mathcal{O}$ are computed.
\textit{(b) Initial--condition/parameter reconstruction:} the same forward model is paired with a likelihood $\mathcal{L}(\mathcal{O};\mathrm{data})$; automatic differentiation backpropagates gradients through the observable postprocessing map and each timestep ($d\mathrm{Obs}$ and $d\mathcal{H}$) to obtain $\partial \mathcal{L}/\partial \boldsymbol{\phi}_0$, enabling gradient-based updates of ICs or physical parameters.
\textit{(c) Solver--in--the--loop optimization:} a reference simulation provides target states or observables at multiple times, defining a summed loss $\mathcal{L}=\sum_i \mathcal{L}_i$. A differentiable forward run $\hat{\mathbf{U}}$ is controlled by trainable parameters $\boldsymbol{\theta}$ (e.g., closure/source-term models $\mathcal{B}_\theta$); gradients $\partial \mathcal{L}/\partial \boldsymbol{\theta}$ are obtained by backpropagating through both $\mathcal{H}$ and $\mathcal{B}_\theta$, allowing optimization of embedded models while preserving the underlying finite-volume numerics.
}
    \label{fig:diffsim}
\end{figure*}

\emph{Differentiable simulation} offers an alternative route, in which the hydrodynamical update itself participates in gradient-based optimization (see \citet{2021arXiv210905237T} for an introduction). Reverse-mode automatic differentiation makes it possible to obtain gradients with respect to initial conditions, physical parameters, or closure terms, enabling tighter integration between simulations and data. This setup allows machine-learning components to be trained ``in the loop’’ of the governing PDE \citep{2020arXiv200700016U} to accelerate underlying numerics or learn un-modelled physics. We show these different workflows schematically in Fig \ref{fig:diffsim}. Recent advances in differentiable programming and accelerator-native numerical computing now make such approaches technically viable at astrophysically relevant scales. 

These developments motivate the construction of solvers that support end-to-end differentiability while remaining accurate, stable, and computationally efficient. In \citet{2025arXiv250202294H} (hereafter HL25), we introduced a proof-of-concept differentiable hydrodynamics solver coupled to a particle-mesh dark-matter module and demonstrated reconstruction applications. The present work refocuses and extends that effort toward astrophysical gas dynamics, with design choices loosely inspired by the \textsc{Athena++} suite, and with substantially expanded numerical and physical capabilities. In particular, relative to HL25 we have:

\begin{enumerate}
\item Overhauled the interface to simplify integration of new physics and ML components through a systematic, modular flux/force task list.
\item Added an extensive library of numerical recipes for reconstruction, Riemann solvers, and related methods for convective/conductive fluxes.
\item Implemented a differentiable multigrid Poisson solver for gravity and related elliptic fluxes/forces.
\item Introduced new physics modules, including (non-equilibrium) heating/cooling, turbulence, conduction, and multigrid Poisson gravity.
\item Reworked the code for scalable multi-device parallelism, with demonstrations up to $1024^{3}$.
\item Implemented techniques to reduce memory and accelerator-time costs during backpropagation, including adjoint methods and frozen gradients.
\item Demonstrated forward model performance and initial-field reconstruction in strongly nonlinear settings (turbulence, instabilities, and self-gravitating flows).
\item Demonstrated solver-in-the-loop approaches using embedded neural networks to accelerate simulations and explore data-driven closures.
\end{enumerate}

Contemporary with this effort, there has been significant activity in the machine-learning community aimed at building differentiable fluid solvers, largely driven by industrial and engineering applications \citep{2023CoPhC.28208527B,2023arXiv230713533K,2025arXiv250523940F}. A particularly mature example is \texttt{JAXFLUIDS} \citep{Bezgin2025}, which provides a performant \texttt{JAX} implementation of standard hydrodynamical components. However, its design centers on compact CFD-style workflows and does not readily accommodate the multiphysics extensions that are routine in astrophysical gas dynamics—e.g., differentiable self-gravity, radiative heating/cooling, stochastic turbulence driving, and additional stiff or elliptic operators—nor the task-based modularity needed to insert learned closures or correctors directly into the update without significant modification. These requirements motivate a solver architecture that treats fluxes and forces as composable, differentiable operators and remains scalable on modern accelerators. 

Although its design philosophy differs from traditional astrophysical production codes, \texttt{diffhydro} retains the core features of modern finite-volume solvers: modular reconstruction and flux operators, multiple integrators, and support for multi-device parallelism via XLA’s \texttt{shardmap}. The code is intended to function as a fully capable forward-modeling tool while also enabling automatic differentiation, thereby expanding the range of supported workflows. It can be used as a conventional hydrodynamics solver with numerical flexibility and performance comparable to established GPU-based frameworks, while additionally providing end-to-end gradients for optimization, inference, and solver-in-the-loop machine learning. Because its discretizations follow standard numerical practices, results can be validated directly against trusted production codes on matched setups, enabling straightforward verification and diagnosis of solver-sensitive differences. In this sense, \texttt{diffhydro} is best viewed as a complementary tool in the hydrodynamics ecosystem, preserving the practical utility of conventional solvers while adding differentiability as an orthogonal capability.

The paper proceeds from general design principles to specific physical modules and then to differentiable use cases. In Sec. \ref{sec:codearch} (\emph{Code Architecture and Design Considerations}) we lay out the core software and numerical choices that enable \texttt{diffhydro} to behave like a modern forward finite-volume code while remaining end-to-end differentiable: a functional, modular layout; accelerator-native execution and multi-device parallelism; and custom adjoints for the non-smooth operators that arise in shock-capturing schemes. %This architectural overview is intended both as a guide for users who want to run the code in a conventional way and as a foundation for understanding how gradients are made stable and practical.

Sections~3–6 then introduce the physical solvers built on top of this design. Sec. \ref{sec:hydro} (\emph{Hydrodynamics Solver}) briefly describes the base finite-volume method and introduces our turbulent forcing terms. Sec. \ref{sec:cooling} (\emph{Radiative Heating and Cooling}) adds a differentiable source-term module, discussing both physical assumptions and numerical integration in a way consistent with the hydro core. Sec. \ref{sec:sn} (\emph{Example: Forward Model Supernova Remnant}) demonstrates how these ingredients combine in a realistic forward modeling problem, and Sec. \ref{sec:selfgrav} (\emph{Self Gravity}) presents the complementary gravity solvers used in later applications. With these pieces established, we turn in Sec. \ref{sec:opt_ic} and \ref{sec:cil} to differentiable workflows that are effectively impossible to realize with standard codes: Sec. \ref{sec:opt_ic} (\emph{Example: Optimizing Initial Conditions}) uses gradients to solve an inverse problem against a target distribution, while Sec. \ref{sec:cil} (\emph{Example: Solver-in-the-loop}) illustrates solver-in-the-loop learning via a learned correction model. We conclude in Sec. \ref{sec:conc} by summarizing the role of \texttt{diffhydro} as a forward-compatible, differentiable companion to existing hydrodynamics codes, and by outlining directions for extending the framework. We also provide additional information on our numerical methods in App. \ref{ap:hydro} and additional examples/comparisons in App. \ref{ap:examples}.

%Differentiability is enabled through custom adjoint implementations of key operators together with rematerialisation and checkpointing strategies to control memory usage during reverse-mode differentiation.

\section{Code Architecture and Design Considerations}
\label{sec:codearch}
\texttt{diffhydro} is a fully differentiable, accelerator-native hydrodynamics framework
built on top of the \textsc{jax} ecosystem.  In contrast to traditional astrophysical simulation codes---which often rely on MPI-parallel C++ or FORTRAN kernels, specialized memory layouts, and intrusive control flow---\texttt{diffhydro} is structured around a collection of functional, composable operators acting on global array fields.  These operators are automatically transformed by the XLA compiler \citep{frostig2019compiling} through \textsc{jit}, \textsc{vmap}, \textsc{pmap}, and \textsc{shard\_map}, producing highly optimized fused GPU/TPU kernels \citep{ashari2015optimizing,2023arXiv230113062S} while preserving full compatibility with reverse-mode automatic differentiation.  The result is a flexible and research-oriented tool enabling differentiable fluid dynamics, simulation-based inference, and machine-learning-based model discovery.

\subsection{Functional and modular design}

Every numerical component in \texttt{diffhydro}---including reconstruction stencils,
Riemann solvers, flux operators, gravitational solvers, turbulence forcing, etc. ---is expressed as a pure function,
\begin{equation}
(\mathbf{U}^{n+1}, \theta^{n+1}) = \mathcal{H}(\mathbf{U}^n,\theta^n),
\end{equation}
where $\mathbf{U}$ denotes the hydrodynamical state vector, $\theta$ denotes a set of auxiliary parameters of interest (additional fields, neural network parameters, etc).
This functional programming style eliminates global side effects and enables the XLA compiler to apply aggressive graph-level optimizations, fuse kernels, and remove Python overhead.  It also ensures that every operator participates naturally in reverse-mode automatic differentiation, greatly simplifying the implementation of scientific workflows that require gradients through the full time evolution.

\texttt{diffhydro}'s modules are organised into lightweight, interchangeable components: reconstruction methods (MUSCL, PLM, PPM, WENO, TENO), Riemann solvers (Rusanov, HLL, HLLC, HLLD), a library of explicit Runge--Kutta integrators, and a suite of physics modules. We have ported many implementations from \texttt{JAXFLUIDS} into this workflow. This modularity allows users to build custom solvers/fluxes/forces by selecting individual components without modifying the core hydrodynamics code.

Below we show a simple example of a forcing term in this setup, in this case for constant gravitational force in the x-direction. Both the state vector, $\textbf{u}$, and the parameters in \texttt{params} allow automatic differentiation.

\vspace{10pt}

\begin{python}
class constant_gravity:
    def timestep(self, u):
        #forcing doesn't effect timestep
        return jnp.inf

    def force(self, i, u, params, dt):
        #i indicates timestep for periodic forcings
        g = params["grav"]
        rho = u[0]
        momx = u[1]

        s_g = jnp.zeros_like(u)
        # momentum source
        s_g = s_g.at[1].set(rho * g)
        # energy source
        s_g = s_g.at[-1].set(momx * g)

        return u + s_g * dt, params
\end{python}

\subsection{Accelerator-native execution and parallelism}

All major operations in \texttt{diffhydro} are expressed as vectorized array functions acting on full three-dimensional fields.  When \textsc{jit}-compiled, reconstruction, Riemann solves, flux divergences, and source-term evaluations are fused into a small number of device kernels, ensuring high arithmetic intensity and maximizing GPU/TPU utilization.  Multi-device parallelism is achieved via domain decomposition and halo exchange implemented through \texttt{shard\_map}, enabling the code to scale efficiently across many accelerators.

Because \texttt{diffhydro} operates on fixed-shape arrays within the compiled XLA graph, its computation is free of the branching overhead and dynamic memory patterns associated with adaptive geometry algorithms.  Uniform array shapes allow the compiler to exploit large-scale SIMD parallelism across the entire domain.

While not implemented in this version, additional performance optimizations could be gained via implementation of certain routines (e.g. multigrid solver) as base CUDA functions and called via JAX wrapper code. Differentiability could still be maintained via adjoint construction (see Sec. \ref{subsec:adj}). However, this would reduce the readability of the code and increase development time.

\subsection{Differentiable physics and custom adjoints}
\label{subsec:adj}
A defining aim of \texttt{diffhydro} is end--to--end differentiability, allowing the solver to be embedded in gradient--based workflows such as PDE--constrained inference, optimal control, and data--driven closure modeling, where initial conditions or physical parameters are optimized directly against an objective. The practical obstacle is that naive reverse--mode differentiation through a long time integration requires storing essentially all intermediate states, leading to memory costs that scale with the number of timesteps and become prohibitive for large 3D runs.

Adjoint methods \citep{pontryagin2018mathematical} provide a (partial) remedy: instead of retaining the full forward history, gradients are reconstructed by propagating a dual system backward in time, equivalent to applying the transpose of the linearized discrete update. In linear--algebra terms, the gradient of a scalar objective with respect to many inputs can be obtained by solving a transposed system once, making the cost largely independent of the number of control variables \citep{giles2000introduction}. For time--dependent PDEs the same idea yields a backward--in--time adjoint evolution whose sensitivity is exactly consistent with the discrete numerical scheme, a perspective widely used in CFD and control \citep{mcnamara2004fluid}. Recent work \citep{li2022pmwd} has shown how this adjoint formulation can be made practical within differentiable programming frameworks for large--scale astrophysical PDE systems, motivating our approach here.

\texttt{diffhydro} adopts this technique and supplies custom adjoints (vector--Jacobian products) for operators whose naive reverse pass would be memory-- or iteration--intensive, including the FFT--based Poisson solver, the multigrid gravitational solver, stochastic turbulence forcing fields, and additional linear/implicit operators as needed. For such steps the reverse pass avoids differentiating through internal FFT stages or multigrid iterations, and instead re--evaluates the corresponding linear operator with adjoint sources. For example, if the forward solve applies $\Phi = A^{-1}F$ (a Poisson solve), the adjoint requires only another Poisson solve with the adjoint field as the source to recover the gradient with respect to $F$, yielding the correct discrete gradient at a cost comparable to one additional solve and without storing the full internal state of the forward operator \citep{giles2000introduction}.

These custom adjoints combine naturally with checkpointing: we store a sparse set of forward states and recompute short segments during the backward sweep (handled in JAX by \texttt{jax.checkpoint}). Together, checkpointing and operator adjoints enable reverse--mode differentiation through thousands of hydrodynamical timesteps at tractable memory cost, making differentiable self--gravitating, radiatively cooling turbulence practical within \texttt{diffhydro}.

\subsection{Uniform Cartesian meshes and extendable geometries}

In the current version, \texttt{diffhydro} employs uniform Cartesian meshes rather than adaptive grids.  This greatly simplifies the implementation of high-order reconstruction stencils, vectorized finite-volume flux operators, and spectral gravity solvers, and it ensures consistent array shapes for XLA compilation. In addition, many implementations of machine learning-based closure models rely on convolutional operators, which naturally live on static Cartesian meshes.

While not currently supported in the current production version, we are working to extend the framework to adaptive mesh refinement and moving mesh methods. However, due to the array-based structure of \texttt{jax}, these approaches will still maintain Cartesian-like topology. We note that the choice to use Cartesian meshes and not support cylindrical/spherical geometry is becoming a common choice for GPU-supported HPC hydrodynamics codes \citep[i.e.,][]{2024arXiv240916053S}.

\subsection{Interactive and Python-first workflow}

\texttt{diffhydro} is designed around an interactive, Python-first workflow in which simulation setups, solver choices, and analysis pipelines are constructed directly in ordinary Python scripts or Jupyter notebooks, rather than being mediated through large configuration files or domain-specific languages.  This design
philosophy emphasizes transparency and rapid experimentation: users can mix analysis, diagnostic plots, parameter sweeps, solver definitions and neural network components in a single script/notebook, leveraging the full flexibility of the Python ecosystem.

All solver components---reconstruction methods, Riemann solvers,
integrators, and physics modules---are ordinary Python objects with well-defined functional interfaces, enabling users to compose and modify simulation pipelines programmatically.  While human-readable configuration files are supported for reproducibility and batch execution, the primary mode of user interaction is therefore interactive and script-driven, reflecting the development-oriented nature of \texttt{diffhydro} and its emphasis on rapid iteration, debuggability, and integration with the larger JAX scientific computing ecosystem. 

\subsection{Extensibility and research-driven development}

The architecture is intentionally designed for rapid prototyping and scientific
experimentation.  
New numerical methods can be introduced with minimal effort:
\begin{itemize}
    \item A new reconstruction method requires defining a single vectorized kernel.
    \item A Riemann solver is added by supplying a flux stencil with a common interface.
    \item A new Runge--Kutta method plugs into the method-of-lines integrator.
    \item New physics (gravity, turbulence, conduction) is introduced by registering
          a differentiable source-term operator with a force/flux and timestep method.
\end{itemize}
This flexibility makes \texttt{diffhydro} an effective platform for simulation-based research, parameter inference, differentiable physics, and the development of machine-learning-guided hydrodynamical models.

\section{Hydrodynamics Solver}
\label{sec:hydro}
The gas dynamics are evolved according to the compressible Euler equations in conservative form,
\begin{equation}
\frac{\partial \mathbf{U}}{\partial t}
+ \nabla \cdot \mathbf{F}(\mathbf{U})
= \mathbf{S}(\mathbf{U}),
\end{equation}
where
\[
\mathbf{U} = (\rho,\; \rho \mathbf{v},\; E), \qquad
P = (\gamma-1)\left(E - \tfrac12 \rho |\mathbf{v}|^2\right),
\]
and where $\mathbf{S}$ collects source terms (gravity, conduction, turbulence forcing). In this work, no magnetic fields, relativistic corrections, or geometric terms are considered but these are in active development; additional community contributions are welcome.

All hydrodynamical operations are implemented in \textsc{jax} and are therefore expressed as pure array functions acting on full 3D tensors. This enables XLA to fuse the entire hydrodynamic update into a small set of GPU/TPU kernels via \textsc{jit} compilation, while preserving full compatibility with reverse-mode automatic differentiation and parallel sharding. %Domain decomposition and halo exchange are handled internally via\texttt{shard\_map} and the halo-aware rolling operation(\texttt{roll\_with\_halo}).

Despite this compiled optimization, we found \texttt{diffhydro} exhibits approximately 10–50\% slower forward performance relative to Kokkos-optimized \texttt{AthenaK} simulations when employing comparable numerical methods. This computational overhead can be attributed to fundamental architectural trade-offs that prioritize automatic differentiation capabilities over hand-optimized throughput. The \texttt{JAX}-compiled approach delegates performance optimization to the compiler rather than relying on explicit, architecture-specific kernels. By contrast, \texttt{AthenaK} employs a Kokkos backend that utilizes carefully optimized, templated C++ kernels with memory layouts designed for efficient execution on GPU architectures. In \texttt{diffhydro}, several computational components that would conventionally be implemented as specialized CUDA/C++ routines are currently maintained in pure JAX to ensure code transparency and enable end-to-end automatic differentiation. Consequently, these components do not yet benefit from the same level of manual kernel optimization. Furthermore, \texttt{diffhydro}'s differentiable architecture necessitates additional computational overhead, including gradient checkpointing and custom adjoint implementations for computationally expensive operators. The forward evaluation therefore encompasses bookkeeping operations that are not required in AthenaK's implementation. 

We refer readers to HL25 for an overview of the basic hydrodynamical solver scheme. We provide additional information on changes from HL25, specific design philosophy, and additional tests in Appendix \ref{ap:hydro}.

%\subsection{Rayliegh Taylor Instability}
%The Rayleigh--Taylor (RT) instability arises when a dense fluid overlies a lighter one in a gravitational field, or more generally when an effective acceleration acts from the lighter toward the denser medium.  
%Linear perturbations grow at a rate 
%\begin{equation}
%    \gamma_{\mathrm{RT}} = \sqrt{A g k},
%\end{equation}
%where $g$ is the acceleration, $k$ the perturbation wavenumber, and $A=(\rho_{\mathrm{h}}-\rho_{\mathrm{l}})/(\rho_{\mathrm{h}}+\rho_{\mathrm{l}})$ is the Atwood number. In hydrodynamical simulations, the RT instability is highly sensitive to numerical diffusion: an artificially broadened density interface may inhibit the early growth of perturbations or modify the dominant wavelength.  Capturing the emergence of characteristic RT ``fingers'' and ``bubbles'' therefore requires high-order spatial reconstruction, carefully chosen limiters, and adequate resolution of the interface.  RT-driven mixing plays a central role in a wide range of astrophysical settings, including the breakup of dense shells in SNRs and superbubbles, the buoyant rise of hot gas in galactic haloes, and the fragmentation of stratified media undergoing acceleration.  When combined with radiative cooling, RT growth can accelerate the condensation of dense clumps and enhance the development of a strongly multiphase medium, making its accurate representation crucial for modelling feedback, mixing layers, and the evolution of unstable interfaces.

\begin{figure}
    \centering
    \includegraphics[width=0.99\linewidth]{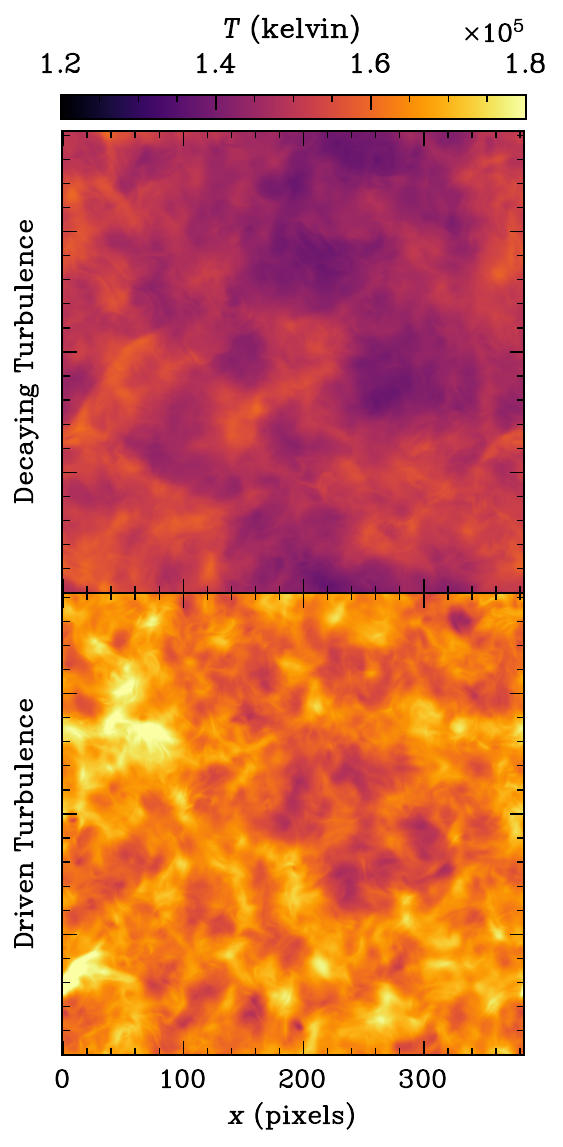}
    \caption{\emph{Forward model of Turbulence.} The same turbulent initial conditions in a $384^3$ box are evolved with and without driving with \texttt{diffhydro}. The driving force injects kinetic energy, which decays dynamically into thermal energy.}
    \label{fig:turb}
\end{figure}

\subsection{Turbulence}
\label{subsubsec:turb}
Incorporating both driven and decaying turbulence into our hydrodynamical framework is essential for modeling realistic astrophysical flows and for systematically probing the role of chaotic motions in structure formation \citep{2004ARA&A..42..211E,2004RvMP...76..125M,2013MNRAS.436.1245F}. Many key environments of interest—such as the multiphase interstellar medium, star-forming molecular clouds, galactic discs, and accretion flows—are generically turbulent, with energy continuously injected on large scales by feedback, shear, and instabilities, and then cascaded and dissipated on small scales. Explicit driven turbulence \citep{Eswaran1988,2008ApJ...688L..79F} allows us to approximate this quasi-steady cascade by imposing a controlled forcing pattern in wavenumber and solenoidal/compressive content, thereby reproducing the observed linewidths, density contrasts, and anisotropies characteristic of these systems. 

Complementarily, decaying turbulence runs, in which an initially turbulent field is allowed to relax without further driving, provide a clean way to isolate the intrinsic decay rate, dissipation mechanisms, and numerical viscosity of the scheme, and to study how transient turbulent support influences collapse, fragmentation, and mixing \citep{1998ApJ...508L..99S,2007ApJ...665..416K}. Using both modes within the same code therefore enables controlled, apples-to-apples comparisons between idealized, decaying setups and more realistic, continuously driven states, and allows us to calibrate and validate the numerical methods under conditions that closely mirror those encountered in astrophysical applications.

We initialize a three-dimensional turbulent velocity field in Fourier space
following standard procedures used in simulations of compressible turbulence
\citep[e.g.][]{1998ApJ...508L..99S, Federrath2010}.  For a periodic domain of size
$L_{\rm box}$, the discrete Fourier modes are
\begin{equation}
    \mathbf{k} = (k_x, k_y, k_z) 
    = \frac{2\pi}{L_{\rm box}} (n_x, n_y, n_z),
\end{equation}
with $(n_x,n_y,n_z)$ integer FFT mode numbers.  We restrict the initial
velocity fluctuations to a large-scale band by selecting only modes with
\begin{equation}
    k_{\min} \;\le\; 
    \frac{|\mathbf{k}|\, L_{\rm box}}{2\pi}
    \;\le\; k_{\max}.
\end{equation}

For each allowed mode we draw three independent complex Gaussian random fields
$G_i(\mathbf{k})$ with Hermitian symmetry so that the real-space velocity field
is strictly real.  We impose an isotropic velocity power spectrum
$E(k)\propto k^{p}$ by multiplying the random amplitudes by
\begin{equation}
    A(k) \;\propto\; k^{p/2},
    \label{eq:turbpl}
\end{equation}
so that the velocity Fourier amplitudes scale as 
$|\hat{\mathbf{u}}|^2 \propto k^p$.

We then decompose each mode into solenoidal and compressive components using
the projection tensors
\begin{equation}
    \mathsf{P}_{ij} = \delta_{ij} - \frac{k_i k_j}{k^2}, \qquad
    \mathsf{C}_{ij} = \frac{k_i k_j}{k^2},
\end{equation}
and construct a mixed projection
\begin{equation}
    \mathsf{M}_{ij} 
    = \zeta\,\mathsf{P}_{ij} + (1-\zeta)\,\mathsf{C}_{ij},
    \label{eq:turbproj}
\end{equation}
where $\zeta$ is the solenoidal fraction.  
The Fourier-space velocity field is then
\begin{equation}
    \hat{u}_i(\mathbf{k}) 
    = M_{ij}(\mathbf{k})\,G_j(\mathbf{k})\,A(k),
\end{equation}
which we transform to real space with an inverse FFT.

To match a desired initial RMS Mach number $\mathcal{M}_{\rm target}$, we
rescale the velocity field by a factor
\begin{equation}
    \alpha 
    = \frac{\mathcal{M}_{\rm target}\,\langle c_s\rangle}
           {\langle |{\bf u}|^2\rangle^{1/2}},
\label{eq:turbscale}
\end{equation}
where $c_s$ is the local sound speed and angle brackets denote spatial
averages.  Conservative hydrodynamic variables are then initialized as
\begin{equation}
    \rho = \rho_0, \qquad 
    \mathbf{m} = \rho\,\mathbf{u}, \qquad
    E = \frac{p_0}{\gamma-1} + \frac{1}{2}\rho |\mathbf{u}|^2 .
    \label{eq:turbic}
\end{equation}

The decaying turbulent field will be a reoccurring component of examples studied in this work as it not only is quite common in astrophysical hydrodynamical systems, but also provides a robust test of our hydrodynamical solver. We show an example of decaying turbulence in Figure \ref{fig:turb} (top). We will use it to model the ambient ISM around our supernova simulation in Sec. \ref{sec:sn} as well as a basis for generating complex initial conditions in Sec. \ref{sec:inversesolving}.

\subsubsection{Ornstein--Uhlenbeck Turbulent Forcing/Driving}
\label{subsec:OUprocess}

To maintain statistically steady turbulence, we apply an Ornstein--Uhlenbeck (OU) stochastic forcing field in Fourier space
\citep{1930PhRv...36..823U,Eswaran1988, Federrath2010}.  
At each timestep, the Fourier-space acceleration field 
$\hat{\mathbf{a}}(\mathbf{k})$ is updated according to
\begin{equation}
    \hat{\mathbf{a}}(t+\Delta t)
    = \exp(-\Delta t/\tau)\,\hat{\mathbf{a}}(t)
    + \sqrt{1-\exp(-2\Delta t/\tau)}\,\hat{\boldsymbol{\eta}},
\end{equation}
where $\tau$ is the autocorrelation time and
$\hat{\boldsymbol{\eta}}$ is a Gaussian random field.
We restrict the forced modes to 
$k_{\min}\le k \le k_{\max}$ and project the acceleration field onto a
specified solenoidal/compressive mixture using the tensors above.
The $k=0$ component is explicitly removed.

Transforming to real space gives the physical acceleration field
$\mathbf{a}(\mathbf{x})$, which we normalize to a prescribed RMS amplitude
$a_{\rm rms}$:  
\begin{equation}
    \mathbf{a} \;\rightarrow\; 
    \mathbf{a}\,
    \frac{a_{\rm rms}}{\sqrt{\langle |\mathbf{a}|^2\rangle}} .
\end{equation}

The forcing is coupled to the Euler equations through source/forcing terms in the momentum and energy equations,
\begin{equation}
    \Delta \mathbf{m} = \rho\,\mathbf{a}\,\Delta t, \qquad
    \Delta E = \rho\,\mathbf{u}\cdot\mathbf{a}\,\Delta t,
\end{equation}
ensuring that the injected energy is self-consistently accounted for in the evolution of the flow. We show an example of driven turbulence vs. decaying turbulence in Figure \ref{fig:turb}.

Setup of driven turbulence requires special care in a differentiable framework and its exact implementation will be dependent on the application. The random nature of the Ornstein-Uhlenbeck driving is not itself an issue for a differentiable model. The stochasticity can be handled via the ``reparameterization trick," where the random key is treated as an input to the model (i.e. in an analogous way as training variational autoencoders). The Markov Chain nature of the turbulent driving process requires additional care, as the previous driving has to be held-over to future time-steps. The code can either treat this in an object oriented way (i.e. as a class property) for forward evaluation or in a functional way (i.e. held as part of the auxiliary parameters) for ease of backpropagation. In practice, care should be taken to vary the input seed if one wants to infer proper posteriors in an ensemble-averaged way. We further discuss these considerations in the context of initial condition reconstruction in Sec. \ref{subsec:IC_FT}.

\section{Radiative Heating and Cooling}
\label{sec:cooling}
\begin{figure}
    \centering
        \includegraphics[width=0.99\linewidth]{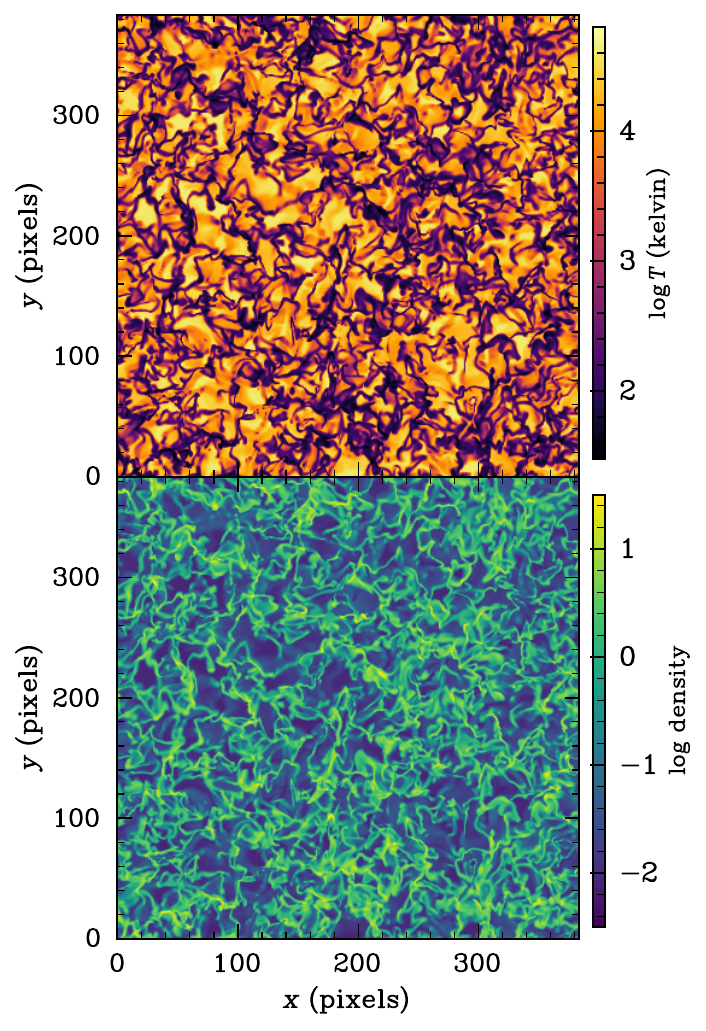}
    
    \caption{\emph{Example output from the cooling/heating test problem in \texttt{diffhydro}.} Shown are mid-plane slices through an $384^3$ ISM-like box evolved from turbulent initial conditions with Ornstein–Uhlenbeck forcing and the cooling/heating module enabled. \textbf{Top:} logarithmic gas temperature, spanning the cold to warm neutral phases, with cooling producing sharp temperature contrasts in compressed regions. \textbf{Bottom:} logarithmic mass density in the same slice, exhibiting a connected network of overdense filaments and shells generated by the driven turbulence. The close spatial correspondence between cold structures and density enhancements, and the warm/hot volume-filling background, illustrates the emergence of a multiphase medium and demonstrates that the implemented cooling/heating routines couple self-consistently to the turbulent flow.}
    \label{fig:ism}
\end{figure}

\begin{figure}
    \centering
    \includegraphics[width=0.99\linewidth]{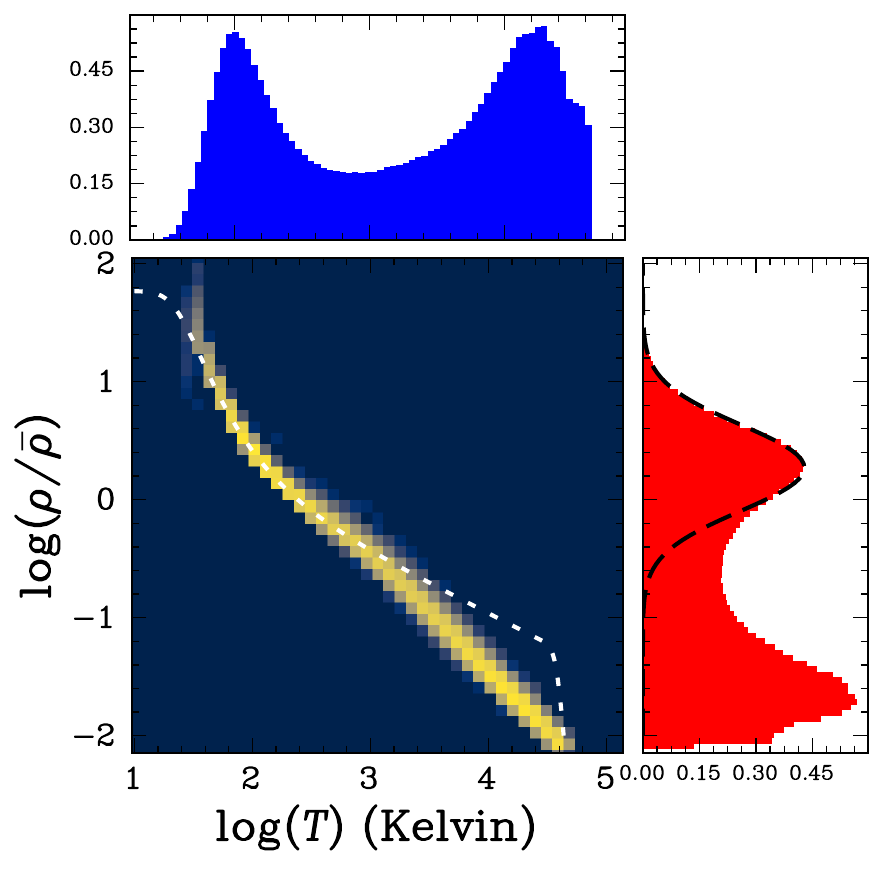}
    \caption{\emph{Thermodynamic phase structure of the turbulent two-phase ISM.}
Joint probability distribution of gas temperature and overdensity from the driven, cooling–heating hydrodynamic simulation shown in the accompanying ISM slice (Figure \ref{fig:ism}) showcasing the bi-phase structure. Colors indicate the logarithm of the mass-weighted probability density. The dashed curve shows the thermal equilibrium heating–cooling balance for the adopted microphysics. At low temperatures, the gas is roughly at thermal equilibrum and therefore follows the heating/cooling curve. At high temperature, the dense ridge follows an approximately isobaric locus, reflecting turbulent mixing and shock heating rather than local thermal equilibrium. Cold and warm gas accumulate near the stable branches of the equilibrium curve, while hot gas occupies a broad, off-equilibrium tail produced by intermittent turbulent dissipation and long cooling times. Marginalized distributions in temperature (top) and overdensity (right) highlight the bimodal phase structure of the medium.}
    \label{fig:biphase_ism}
\end{figure}
Radiative heating and cooling govern the thermal evolution of diffuse astrophysical gas across a broad range of environments. 
In the neutral interstellar medium (ISM), the competition between photoelectric heating and low--temperature atomic cooling establishes a thermally bistable structure comprising the warm and cold neutral media \citep{Koyama2002}. 
In supernova remnants (SNRs) and superbubbles, metal--line cooling at $T\sim10^{5{-}6}\,\mathrm{K}$ determines the post--shock thermodynamics and sets the momentum injection into the surrounding ISM \citep{Kim2015}. 
Similar processes regulate the condensation of hot gas in galactic winds, the cooling layers behind shocks in jets, and the multiphase structure of galactic discs and circumgalactic flows. 
In all such systems, a physically motivated prescription for the net radiative heating and cooling rate is required in order to capture the correct thermodynamic trajectories and emergent phase structure.

We write the operator--split form of the total energy equation as
\begin{equation}
  \frac{\partial E}{\partial t} + 
  \nabla\!\cdot\!\bigl[(E+P)\mathbf{v}\bigr]
  = \dots + \mathcal{H} - \mathcal{C},
\end{equation}
where the volumetric heating and cooling terms are
\begin{equation}
  \mathcal{H} = n_{\mathrm{H}}^{\chi}\Gamma, \qquad
  \mathcal{C} = n_{\mathrm{H}}^2\Lambda,
\end{equation}
where $\chi$ determines the heating - density scaling. The internal energy density is related to the thermal pressure through an ideal--gas equation of state (EOS),
\begin{equation}
  P = (\gamma-1) E_{\mathrm{th}}, \qquad 
  T = \frac{\mu m_{\mathrm{H}}}{k_{\mathrm{B}}}\frac{P}{\rho},
\end{equation}
with constant adiabatic index $\gamma$ and mean molecular weight $\mu$.

% --------------------------------------------------------------------------
\subsection{Heating/Cooling Models}
\label{sec:cooling:hierarchy}

A wide range of thermal physics models have been developed for numerical hydrodynamics. In high–resolution studies of the interstellar medium, a commonly adopted and computationally efficient choice is the simplified two–branch equilibrium prescription of \citet{Kim2015}. In this approach, low-temperature cooling ($T<10^{4}$ K) is taken from the analytic fit of \citet{Koyama2002}, which captures metal–line cooling in predominantly neutral gas, while a constant photoelectric heating rate $\Gamma_{0}$ represents the dominant heating mechanism in the atomic ISM. At higher temperatures, cooling is described by the collisional–ionization–equilibrium (CIE) curve of \citet{Sutherland1993} (SD93), so that
\begin{equation}
    \Lambda(T) =
    \begin{cases}
        \Lambda_{\mathrm{KI}}(T), & T < 10^{4}\,\mathrm{K},\\[3pt]
        \Lambda_{\mathrm{CIE}}(T), & T \ge 10^{4}\,\mathrm{K},
    \end{cases}
    \label{eq:twobranch}
\end{equation}
with $\Gamma(T)=\Gamma_{0}$ for $T<10^{4}$\,K and $\Gamma(T)=0$ otherwise. For this work we focus on fixed metallicity, though the same framework can be straightforwardly extended to spatially or temporally varying metallicity by introducing an additional lookup table. We note a similar approach was implemented for differentiable simulation spectral absorption line modeling in \citet{2024A&C....4800858D}.

Alternative, more nuanced options exist in the literature. The \textsc{grackle} library \citep{2017MNRAS.466.2217S} provides tabulated cooling and heating as functions of temperature, density, metallicity, and a redshift–dependent UV background, and supports a non-equilibrium primordial chemistry network, allowing a wider range of environments to be modeled at increased computational cost. At the high–fidelity end, \textsc{Cloudy} \citep{1998PASP..110..761F} solves detailed photoionization and collisional–excitation equilibria for hundreds of atomic and molecular species, and is typically used offline to generate multidimensional cooling grids for specific radiation fields and abundances; these grids can then be interpolated during dynamical simulations. (Indeed, the Sutherland–Dopita CIE curve is itself rooted in \textsc{Cloudy}–like calculations.)

For the purpose of this paper, however, we restrict attention to the first class of models: the two–branch equilibrium cooling/heating described in Eq. \ref{eq:twobranch}. This choice captures the dominant thermal processes in solar–metallicity galactic gas while remaining computationally tractable for our differentiable framework. Incorporating richer thermal physics—such as \textsc{grackle} or \textsc{Cloudy}–derived tables—is possible in principle, but would require careful differentiable implementation and extensive verification, so we leave such extensions to future work.
% --------------------------------------------------------------------------
\subsection{Physical Assumptions}
\label{sec:cooling:physics}

Our implementation follows the simplified model of \citet{Kim2015}. 
Below $10^{4}$\,K, the cooling function is given by the Koyama--Inutsuka fit,
\begin{equation}
    \Lambda_{\mathrm{KI}}(T) 
    = \Lambda_{1} T^{\alpha_{1}}
    + \Lambda_{2} T^{\alpha_{2}},
\end{equation}
and we adopt a spatially uniform photoelectric heating rate $\Gamma_{0}$. 
Above $10^{4}$\,K, we evaluate the CIE cooling rate through interpolation in $\log T$--$\log\Lambda$ space at fixed metalicity:
\begin{equation}
    \Lambda_{\mathrm{CIE}}(T) = 
    10^{\log_{10}\Lambda_{\mathrm{SD93}}(T)}.
\end{equation}

We assume a single ideal fluid with fixed $\gamma$ and $\mu$, neglecting explicit ionization fractions, molecular processes, and non-equilibrium ionization. 
These assumptions are well suited to solar--metallicity ISM gas, shock--heated superbubbles, and other idealised flows relevant to galactic feedback.

% --------------------------------------------------------------------------
\subsection{Numerical Integration}
\label{sec:cooling:numerics}

After each hydrodynamic update, the radiative source term modifies only the conservative energy variable. 
Given $E_{\mathrm{th}}^{n}$ and $T^{n}$ at the beginning of the cooling step, the local cooling time is
\begin{equation}
    t_{\mathrm{cool}}
    = \frac{|E_{\mathrm{th}}|}{
      \bigl| n_{\mathrm{H}}\Gamma 
      - n_{\mathrm{H}}^{2}\Lambda \bigr| }.
\end{equation}
The global timestep $\Delta t$ is restricted according to
\begin{equation}
    \Delta t \;\le\;
    \min\!\left( 
    \Delta t_{\mathrm{hydro}},\;
    c_{\mathrm{cool}} 
    \min\nolimits_{x}\, t_{\mathrm{cool}}(x),\;
    \Delta t_{\max}
    \right),
\end{equation}
where $c_{\mathrm{cool}}$ is a dimensionless safety factor.

For each cell, the thermal energy is advanced using either a second--order trapezoidal update,
\begin{equation}
    E_{\mathrm{th}}^{n+1}
    = E_{\mathrm{th}}^{n}
    + \frac{\Delta t}{2}\left[
    \dot{E}(T^{n})
    + \dot{E}(T^{\mathrm{pred}})
    \right],
\end{equation}
or an explicitly subcycled update when cooling is sufficiently rapid. 
In the subcycled regime, we divide $\Delta t$ into $N_{\mathrm{sub}}$ substeps of size $\delta t=\Delta t/N_{\mathrm{sub}}$ and apply
\begin{equation}
    E_{\mathrm{th}}^{k+1}
    = \max\!\left(
        E_{\mathrm{th}}^{\mathrm{floor}},\;
        E_{\mathrm{th}}^{k}
        + \dot{E}(T^{k})\,\delta t
      \right),
\end{equation}
with a limiter enforcing a maximum fractional change per substep. 
The temperature floor corresponds to
\begin{equation}
    E_{\mathrm{th}}^{\mathrm{floor}}(\rho)
    = \frac{\rho R T_{\mathrm{floor}}}{\gamma-1},
\end{equation}
and prevents unphysical cooling to negative pressures. 

We present a representative turbulent ISM test in Fig.\,\ref{fig:ism}, designed to validate the radiative heating/cooling module and its coupling to driven dynamics. The simulation begins from turbulent initial conditions and is continuously driven with an Ornstein–Uhlenbeck acceleration field over a finite low-$k$ band, producing a shock–shear network that seeds thermal instability. With heating and cooling active, the flow rapidly separates into a warm, volume-filling phase and a cold, dense phase that condenses in thin filaments and sheets at sites of converging motions. The resulting temperature and density slices show a strong spatial anticorrelation---cold structures coincide with overdense ridges while warm/hot gas occupies the rarefied voids---illustrating the emergence of a biphase medium maintained by the competition between turbulent compression, radiative losses, and diffuse heating. This example demonstrates that the implemented source terms reproduce the expected multiphase morphology (as shown in Figure \ref{fig:biphase_ism}) and remain stable under sustained stochastic forcing.

% --------------------------------------------------------------------------
\subsection{Differentiability}
\label{sec:cooling:differentiability}

%The radiative module is implemented using \textsc{jax}, enabling end--to--end differentiability. 
Branching between the slow and subcycled paths is handled through masked array operations, ensuring the computational graph remains static. The cooling functions, interpolation operators, floors, and limiters are all expressed using differentiable primitives such as \texttt{maximum}, \texttt{minimum}, and linear interpolation in log--space. Gradients may therefore be propagated through the entire thermal update with respect to the hydrodynamic state, EOS parameters, and even the cooling tables themselves. 
%This enables inverse--modelling and optimization tasks, and positions the radiative module as a differentiable but physically grounded component within a larger hierarchy of thermal physics models.

\section{Example: Forward Model Supernova Remnant at $1024^3$ resolution}
\label{sec:sn}

\begin{figure}
    \centering
    \includegraphics[width=0.95\linewidth]{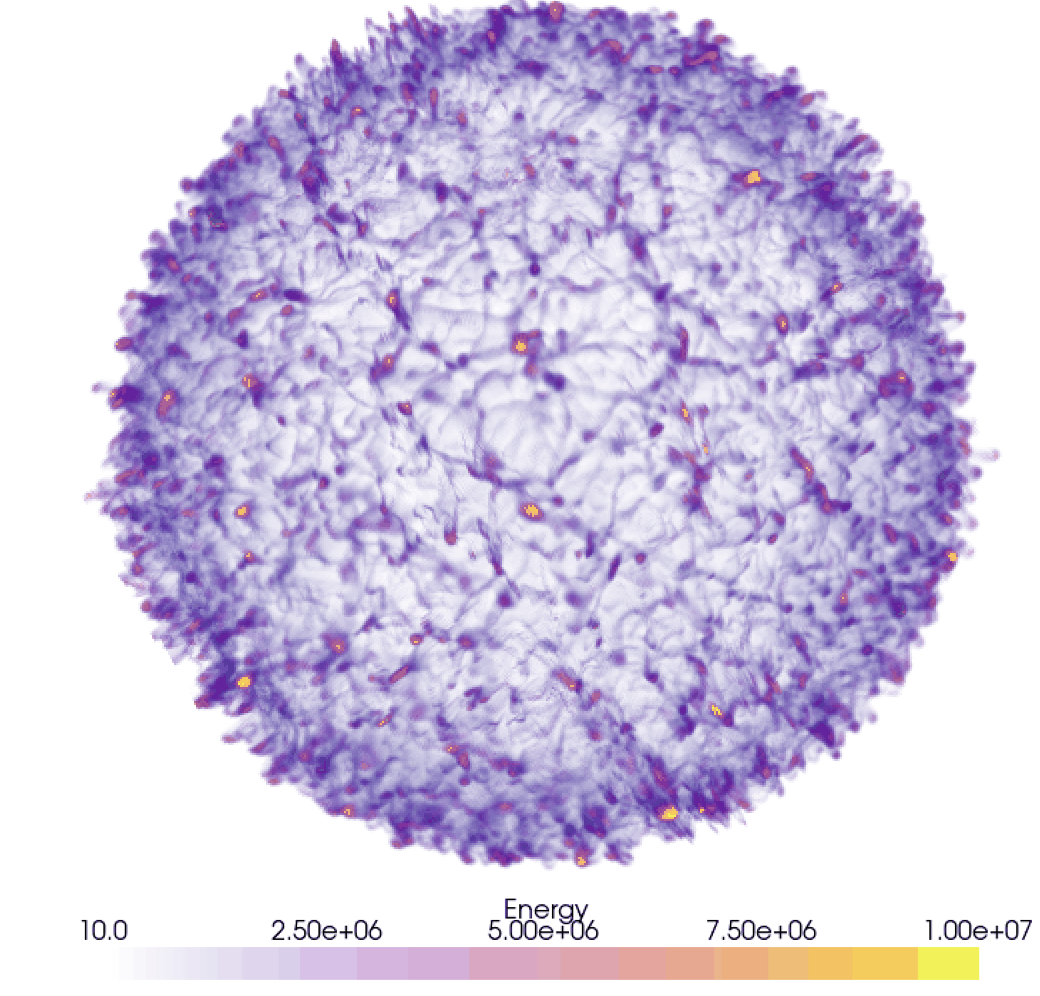}
    \caption{\emph{Volume rendering of forward model of supernova remnant.} Three-dimensional rendering of the same radiatively–cooling supernova remnant shown in Figure~\ref{fig:snr}, coloured by internal energy. The remnant forms a roughly spherical hot bubble whose surface is strongly corrugated by expansion into a turbulent, inhomogeneous ambient medium. Bright filaments and knots trace dense, rapidly–cooling material in the swept-up shell and in mixed cloud fragments, while the diffuse interior shows a web of turbulent structures generated by shock–cloud interactions and deceleration-driven instabilities. The rendering emphasizes the multi-phase character of the remnant: a hot, volume-filling cavity bounded by a rippled, cooling shell and punctuated by small-scale clumps produced by nonlinear RT/KH growth.}
    \label{fig:snr_volume}
\end{figure}

\begin{figure*}
    \centering
    \includegraphics[width=0.95\linewidth]{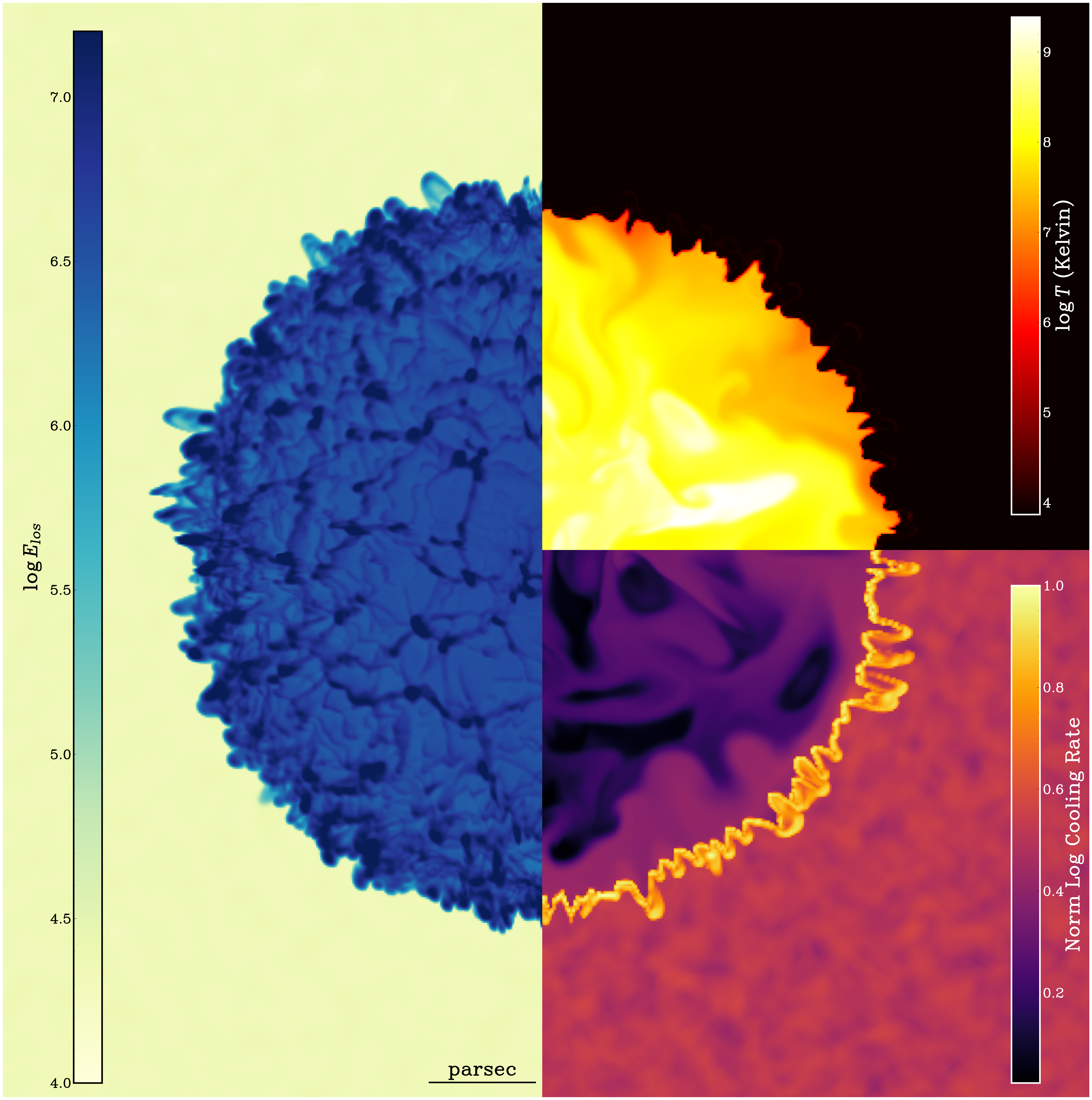}
    \caption{\emph{Supernova remnant in a turbulent, radiatively–cooling medium.} A $10^{51}$ erg explosion evolves into a multi-phase remnant at $\sim 1$ kyr whose morphology is shaped by ambient density fluctuations and cooling. The left projection panel (total energy) shows the forward solution at the same time as the diagnostic slices on the right (top: temperature, bottom: cooling rate). The shock has expanded to an approximately Sedov–Taylor radius but is visibly corrugated by interactions with overdense clumps and underdense channels. A thin, dense swept-up shell forms behind the shock; its angular variations trace the pre-existing turbulence and cooling efficiency. The hot interior is strongly vortical and mottled from repeated shock–cloud interactions, while Rayleigh–Taylor fingers protrude from the decelerating shell and are further shredded by Kelvin–Helmholtz billows along their flanks and the inner rim where tangential shear is strongest. Together, the panels highlight the coexistence of a globally self-similar expansion with rich small-scale structure produced by turbulence, strong shocks, and radiative losses.}
    \label{fig:snr}
\end{figure*}
Forward modeling of supernova remnant (SNR) formation/evolution provides a robust test of the interplay of heating, cooling, shock propagation, and fluid instabilities. The simulation of supernovas in three dimensions has been studied in the literature with varying assumptions \citep[e.g.][]{2015ApJ...814....4L,2017ApJ...834...25K,2024arXiv240916053S,2025ApJ...990...49G}, and we do not aim to present a comparison in this work.

As the SNR expands into a turbulent multiphase interstellar medium, it interacts with a highly irregular distribution of densities and velocities. As the blast wave sweeps through this structure, it encounters dense clouds, diffuse channels, and pre-existing shear flows, producing transmitted and reflected shocks that rapidly deform the initially spherical expansion. These interactions inject substantial vorticity and turbulence into the remnant interior: shock–cloud encounters generate shear layers prone to Kelvin–Helmholtz growth, while shock curvature and uneven deceleration imprint large-scale asymmetries \citep{1983ApJ...274..152V,2018A&A...617A.133M}. Even when the global evolution loosely follows Sedov–Taylor scaling, the detailed morphology is governed by the interplay between strong shocks and the ambient turbulent medium.

Radiative cooling plays a central role by converting post-shock thermal energy into a dense, compressed shell that forms around the forward shock once cooling becomes efficient. This cooled shell is intrinsically unstable: the combination of rapid compression, curvature, and lateral pressure gradients drives nonlinear thin-shell instabilities that ripple, corrugate, and locally fragment the shock front. Because cooling depends strongly on local density, regions where the shock encounters overdense clumps cool faster and thicken the shell more efficiently, reinforcing asymmetry and feeding perturbations that cascade into smaller-scale structures. The resulting shell is neither smooth nor static but instead evolves as a dynamic, turbulent surface.

As the remnant decelerates—both through mass loading from cloud interactions and energy loss via cooling—the dense shell is pushed outward by a lighter, high-pressure interior, triggering Rayleigh–Taylor instabilities \citep{1974ApJ...188..501C}. RT plumes penetrate the shell while cloud fragments and mixed material are advected inward, creating a tangled network of filaments and bubbles. Altogether, the combination of shock deformation, cooling-driven shell formation, and the growth of KH, RT, and thin-shell instabilities leads to a highly structured remnant whose morphology reflects both the initial turbulent environment and the nonlinear hydrodynamic response of the post-shock gas \citep{2024ApJ...965..168R}.

For our specific problem setup, we follow \citet{2022ApJS..262....9O}, although we are only looking at a single supernova event rather than super-bubble formation. To generate the turbulent background, we initialise a uniform-density medium with 
a pressure 
\begin{equation}
    P = 2.0 \times 10^{3} \left(\frac{n_{\mathrm{H}}}{1~\mathrm{cm}^{-3}}\right) 
k_{\mathrm{B}}~\mathrm{K\,cm^{-3}},
\end{equation}

and impose an initially driven, subsequently decaying turbulent velocity field. The forcing is normalised to a Kolmogorov power spectrum, \(E(k) \propto k^{-5/3}\), with initial driving restricted to 
\(2 \leq kL/2\pi \leq 20\). The resulting velocity dispersion is set to \(\sigma = 5~\mathrm{km\,s^{-1}}\), consistent with Larson’s scaling relation. Because this dispersion is below that expected on the spatial scale of the simulation domain, and because the system is evolved for only \(1~\mathrm{Myr}\) prior to feedback injection, the medium does not fully develop a multiphase structure via thermal instability. As a consequence, the background inhomogeneity remains weak, yielding a more nearly spherical bubble expansion and reducing the shear that would otherwise promote Kelvin--Helmholtz instabilities at phase interfaces.

The supernova event is modeled by depositing \(10^{51}~\mathrm{erg}\) of thermal energy within a sphere of radius \(r_{\rm init}\), defined such that the enclosed mass is \(1\,\mathrm{M_\odot}\). A corresponding ejecta mass of \(1\,\mathrm{M_\odot}\) is injected simultaneously. We assume a high metallicity of $Z= 1.0 Z_\odot$ to maximize the effects of cooling for demonstration.

The simulation was parallelized across 32 GPUs (V100 GPU) on 8 nodes, requiring a total of $\sim150$ GPU-hours. We show our supernova remnant in Figure~\ref{fig:snr_volume} and \ref{fig:snr}. The forward shock expands into a mildly inhomogeneous, turbulent medium, so the blast wave is only approximately spherical: as it encounters overdense clumps and underdense channels, it becomes corrugated, with local protrusions where the shock runs ahead through low-density regions and flattened segments where it is slowed by denser gas. These repeated shock–cloud interactions drive reflected shocks and strong shear inside the cavity, injecting vorticity and leaving the interior hot gas visibly turbulent and mottled, even though the overall radius evolution remains broadly Sedov–Taylor–like. Radiative cooling makes this structure strongly two-phase. A dense swept-up shell forms behind the shock and appears as a thin, high-contrast rim whose thickness and brightness vary with angle, reflecting the ambient density fluctuations: denser sightlines cool faster and build a thicker, colder shell, while more diffuse directions remain hotter and more tenuous. As the remnant decelerates, the light, high-pressure interior pushes on this heavy shell, and Rayleigh–Taylor growth produces filamentary fingers and knots that protrude inward and outward from the rim. Along these protrusions and at shear interfaces, secondary Kelvin–Helmholtz–like billows further shred material, so that cold inflows and cloud fragments continuously mix back into the hot bubble.
\begin{figure*}
    \centering
    \includegraphics[width=1.0\linewidth]{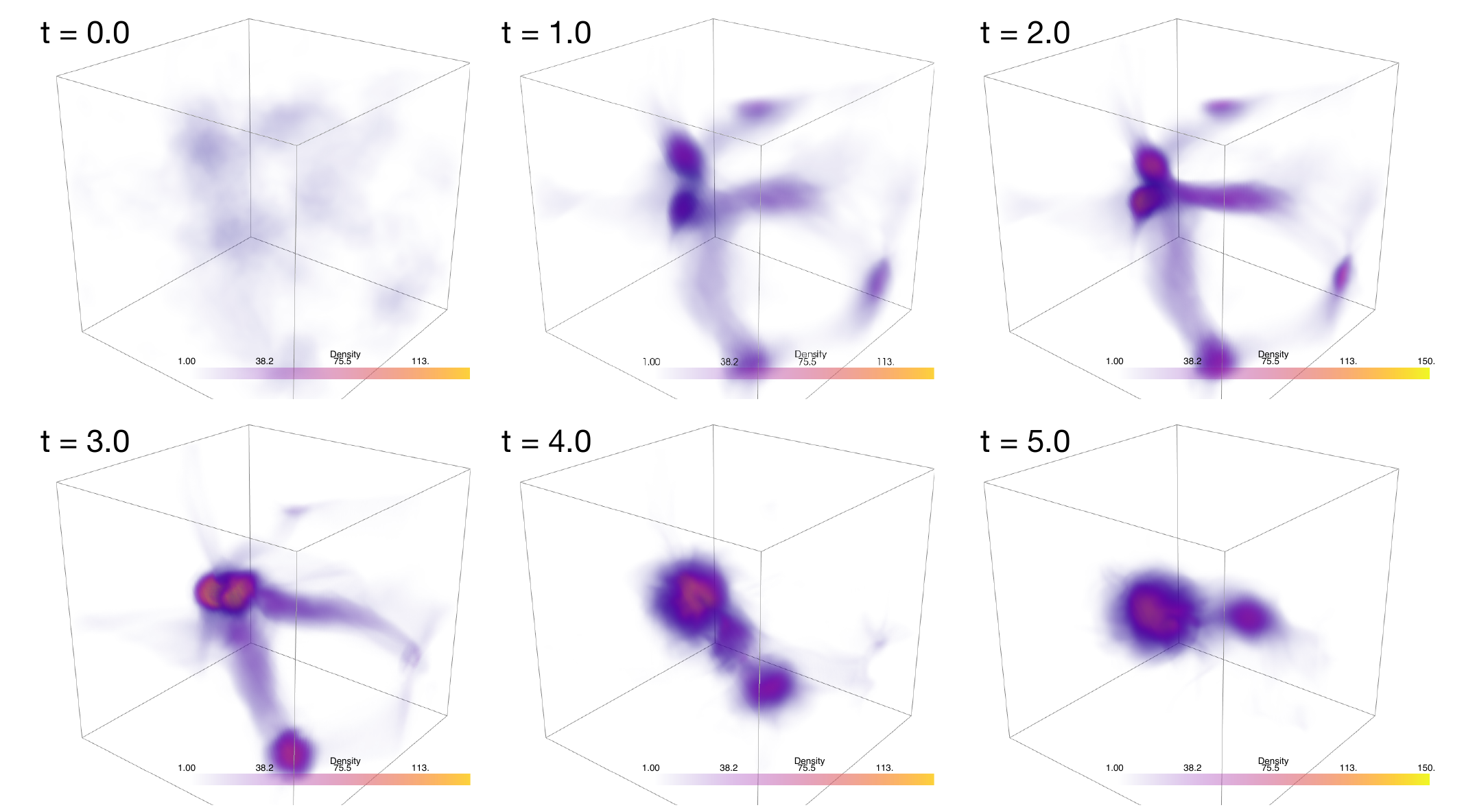}
    \caption{\textbf{Forward time sequence of volumetric density renderings from a $512^3$ periodic-box simulation with self-gravity in \textsc{diffhydro}.}
Beginning from a turbulent initial density field ($t=0$), self-gravity rapidly amplifies fluctuations, forming transient knots and filamentary structure ($t\simeq 1\text{--}3$).
Subsequent nonlinear mergers and relaxation concentrate mass into a small number of compact overdensities by late times ($t\simeq 4\text{--}5$) supported under hydrostatic equilibrium.
This simulation serves as a controlled numerical demonstration of self-gravitating turbulent dynamics, not a model of a specific astrophysical system.}
    \label{fig:turbgrav}
\end{figure*}

\section{Self Gravity}
\label{sec:selfgrav}
With radiative heating/cooling and turbulent driving in place, \texttt{diffhydro} already captures a broad class of shock-dominated and multiphase ISM dynamics, as exemplified by the supernova-remnant forward model in the preceding section. A number of the applications we ultimately target, however, require the inclusion of gas self-gravity: once density contrasts become significant, long-range gravitational forces couple back to the flow, enhancing overdensities, promoting collapse, and altering the late-time morphology. Modeling this regime within a differentiable, accelerator-native framework therefore demands an efficient Poisson solve together with an adjoint whose evaluation cost remains comparable to that of the forward pass. The following section introduces our self-gravity module and describes both the FFT-based and multigrid Poisson solvers, along with their custom adjoints, which provide a scalable gravitational backbone for the turbulent, cooling flows considered later in the paper. 

Self-gravity enters through the solution of the Poisson equation,
\begin{equation}
\nabla^2 \phi = 4\pi G \rho,
\label{eq:poisson}
\end{equation}
discretised here on a uniform Cartesian mesh. Writing the discrete Laplacian as a linear operator $L$ acting on cell-centred potentials, we solve
\begin{equation}
L \phi = 4\pi G\,\rho ,
\label{eq:disc_poisson}
\end{equation}
with forces obtained from the usual finite-difference gradient, $\mathbf{g}=-\nabla\phi$. On structured grids, the regularity of $L$ admits two widely used fast approaches: spectral inversion using FFTs, and iterative geometric multigrid (MG); both are standard in contemporary self-gravitating hydrodynamics and particle--mesh codes \citep[e.g][]{2020ApJS..249....4S}.

For periodic domains, FFT methods diagonalise Eq.~(\ref{eq:poisson}) in Fourier space. Defining $\hat{\rho}(\mathbf{k})$ and $\hat{\phi}(\mathbf{k})$ as Fourier transforms of $\rho$ and $\phi$, one obtains
\begin{equation}
\hat{\phi}(\mathbf{k}) = -\,\frac{4\pi G}{k^2}\,\hat{\rho}(\mathbf{k}) ,
\qquad (\mathbf{k}\neq 0) ,
\label{eq:fft_poisson}
\end{equation}
with the $\mathbf{k}=0$ mode fixed by the chosen gauge. The resulting solver cost is $\mathcal{O}(N\log N)$ on a mesh with $N$ cells, and delivers spectral accuracy on uniform grids. In parallel, however, three-dimensional FFTs require global data transposes (typically two all-to-all communications per solve), tying performance to network bandwidth and latency at high concurrency.

Multigrid methods instead address Eq.~(\ref{eq:disc_poisson}) via a hierarchy of coarser grids. A typical V-cycle applies a smoother $S$ to damp high-frequency errors on each level, restricts residuals to coarser meshes, and prolongates corrections back to the fine grid. Denoting the residual by $r = 4\pi G \rho - L\phi$, one MG step can be written schematically as
\begin{equation}
\phi \leftarrow \phi + P\,L_c^{-1}\,R\,r ,
\label{eq:mg_update}
\end{equation}
where $R$ and $P$ are restriction and prolongation operators and $L_c$ is the coarse-grid Laplacian. For elliptic operators on structured meshes this yields near $\mathcal{O}(N)$ complexity and, crucially, uses only local stencil communications, making MG attractive for strong scaling and for extensions beyond strictly uniform periodic problems.

Historically, MG’s linear-time scaling suggested an asymptotic advantage over FFT solvers, and early cosmological PM studies found MG to converge rapidly to FFT solutions on uniform meshes while parallelising efficiently on then-current architectures \citep{1995ApJS..100..269P,NumRec}. On modern multicore CPUs the situation is more nuanced: highly tuned FFT libraries, together with the memory-bandwidth-limited nature of stencil smoothers, can make FFT-based Poisson solves faster in practice despite their $\log N$ factor \citep{ibeid2020fft}. Conversely, on GPU-dense platforms---especially multi-GPU configurations where global transposes are costly---it is possible there would be a renewed performance edge for MG. In the context of differentiable simulations on GPUs (i.e. \citet{2021A&C....3700505M}) efficient parallelization of FFTs over GPU-nodes remains a major limitation. These considerations, along with additional applications of MG discussed in Sec. \ref{subsubsec:comparefft}, motivate our support for both approaches and the architecture-dependent choices.

Below we provide an overview of our implementations of both FFT and MG in the context of a differentiable self-gravitating hydrodynamical solver.

\subsection{Fast Fourier Transform}
\label{subsubsec:fft}
For periodic domains, the Poisson equation,
\begin{equation}
\nabla^2 \Phi = 4\pi G\left(\rho - \langle \rho \rangle\right),
\end{equation}
diagonalises in Fourier space, allowing the gravitational potential to be
computed in a single global solve. %The implementation follows the function\texttt{gravity\_accel\_rfft} in the code base.

\subsubsection{Spectral Poisson solve}

Let $\widehat{\rho}(\mathbf{k})$ denote the discrete Fourier transform of the
mean-subtracted density field. In Fourier space, the Poisson equation becomes
\begin{equation}
- k^2 \, \widehat{\Phi}(\mathbf{k}) = 4\pi G \, \widehat{\rho}(\mathbf{k}),
\end{equation}
so that the potential is given by
\begin{equation}
\widehat{\Phi}(\mathbf{k})
    = - \frac{4\pi G \, \widehat{\rho}(\mathbf{k})}{k^2},
\label{eq:phi_hat_fft}
\end{equation}
with the zero mode set to zero in order to impose $\langle \Phi \rangle = 0$.
The solver constructs wavenumber grids $(k_x,k_y,k_z)$ appropriate for the
half-spectrum layout of the real-to-complex FFT and forms the denominator
$k^2 = k_x^2 + k_y^2 + k_z^2$ accordingly.

The inverse FFT then yields the potential in real space:
\begin{equation}
\Phi(\mathbf{x}) = \mathcal{F}^{-1}\!\left[\widehat{\Phi}(\mathbf{k})\right].
\end{equation}

\subsubsection{Spectral computation of accelerations}

The gravitational acceleration is obtained by differentiating the potential,
which in Fourier space corresponds to multiplication by $i\mathbf{k}$:
\begin{equation}
\mathbf{a}(\mathbf{x})
    = - \nabla \Phi
    = - \mathcal{F}^{-1}\!\left[ i\mathbf{k} \, \widehat{\Phi}(\mathbf{k}) \right].
\end{equation}
These accelerations are then incorporated into the hydrodynamic evolution
through momentum and energy source terms.

\begin{figure}
    \centering
\begin{flushleft}
    \emph{(a) Unrolled Backpropagation:}
    \end{flushleft}
    \vspace{-8pt}
\includegraphics[width=1.0\linewidth]{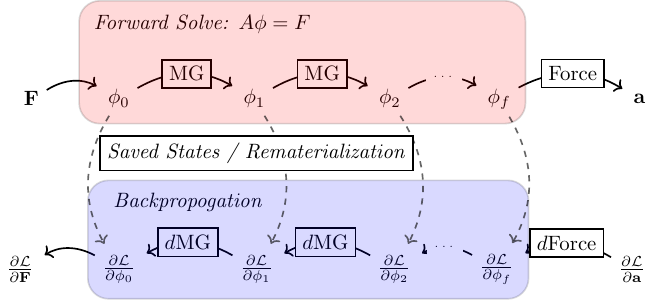}
\begin{flushleft}
\vspace{-8pt}
    \emph{(b) Adjoint Solve:}
    \vspace{-5pt}
    \end{flushleft}
    \includegraphics[width=1.0\linewidth]{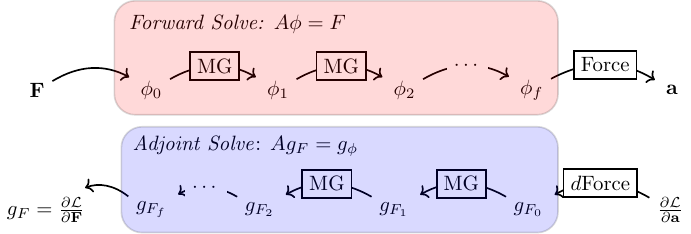}
    \caption{\textbf{Comparison of gradient computation strategies for a differentiable Poisson solver.}
(a) \emph{Unrolled backpropagation:} the Poisson solve \(A\phi = F\) is implemented as an explicit sequence of multigrid (MG) iterations, and gradients of the loss \(\mathcal{L}\) are obtained by backpropagating through the solver steps. This corresponds to differentiating the \emph{algorithm} and requires storing or recomputing intermediate iterates, leading to memory and compute costs that scale with the number of solver iterations.
(b) \emph{Adjoint method:} the solver is treated as an implicit differentiable operator, and gradients are computed by solving a single adjoint system \(A^{T}g_F = \partial \mathcal{L} / \partial \phi\). For symmetric Poisson operators \(A^{T} = A\), allowing reuse of the same MG solver. This yields gradients of the \emph{converged solution} with constant memory overhead, independent of the forward iteration count.
}
    \label{fig:adjoint}
\end{figure}
\begin{figure*}
    \centering
    \includegraphics[width=0.900\linewidth]{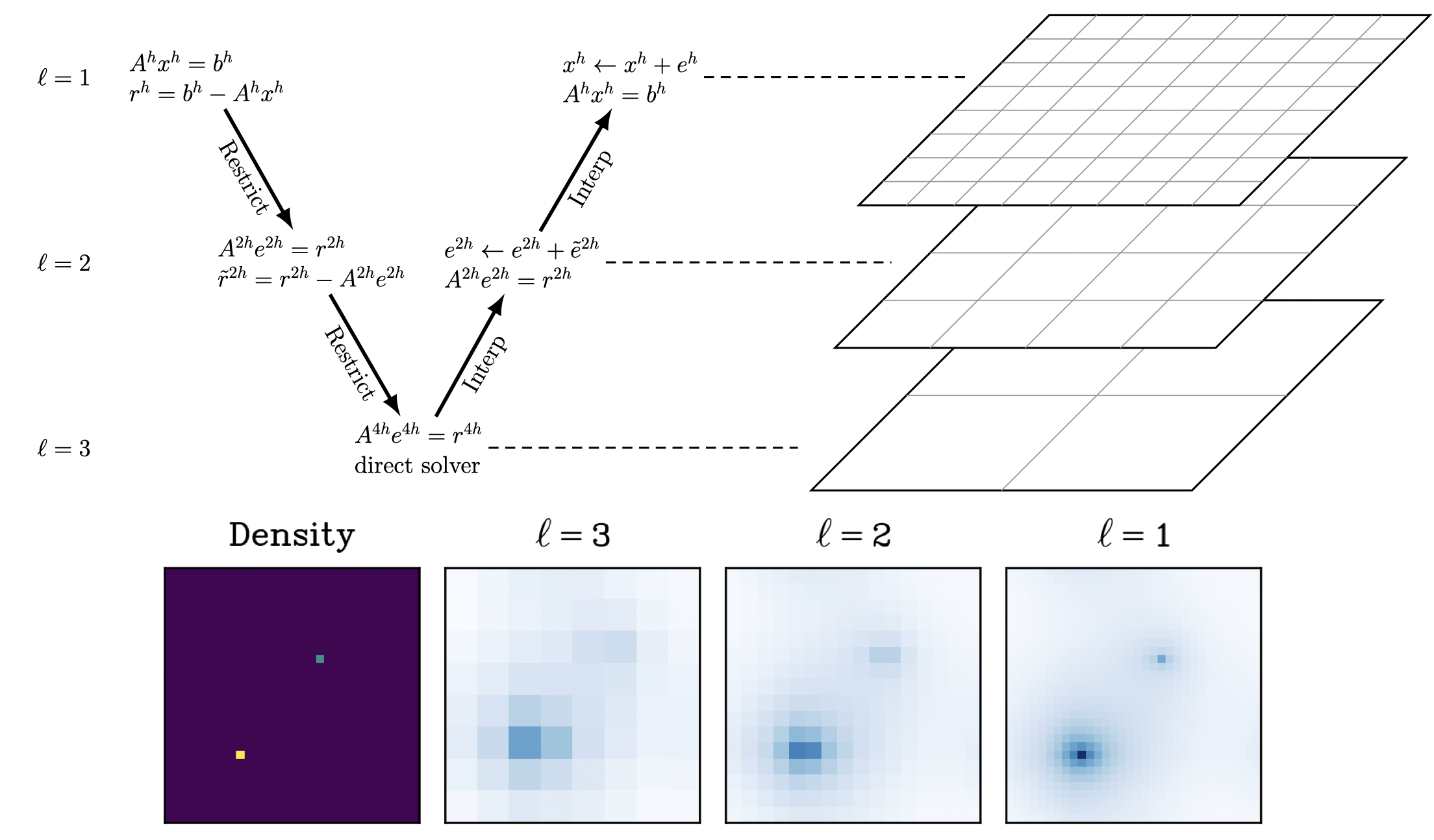}
    \caption{\textbf{Multigrid V-cycle solution of the Poisson equation for self-gravity in \textsc{diffhydro}.}
At the finest level $\ell=1$, the gravitational potential $\phi^{h}$ is computed by solving
$A^{h} x^{h} = b^{h}$ with residual $r^{h} = b^{h} - A^{h} x^{h}$.
The residual is \emph{restricted} to successively coarser meshes ($\ell=2,3$), where the error equations
$A^{2h} e^{2h} = r^{2h}$ and $A^{4h} e^{4h} = r^{4h}$ are solved, using a direct solver on the coarsest grid.
The coarse-grid error is then \emph{interpolated/prolongated} back to finer levels and used to update the solution,
$x^{h} \leftarrow x^{h} + e^{h}$, iterating until convergence. Bottom panels illustrate the density field represented on each level, highlighting how coarse grids capture long-wavelength contributions efficiently and accelerate convergence of the self-gravity force used in the hydrodynamic evolution.}

    \label{fig:vcycle}
\end{figure*}

\subsubsection{Adjoint formulation}

In practice the self-gravitation can dominate computational time and memory use due to the global nature of the force. To address these constraints, we implement a custom adjoint (vector--Jacobian product) for backpropagation (see Figure \ref{fig:adjoint}). Given incoming
gradients $(g_{a_x},g_{a_y},g_{a_z})$ with respect to the accelerations, the
adjoint of the spectral differentiation reconstructs the gradient with
respect to the potential:
\begin{equation}
\frac{\partial \mathcal{L}}{\partial \widehat{\Phi}}
    = i k_x\, \widehat{g}_{a_x}
    + i k_y\, \widehat{g}_{a_y}
    + i k_z\, \widehat{g}_{a_z},
\end{equation}
where $\widehat{g}_{a_i}$ denotes the Fourier transform of $g_{a_i}$. The
gradient with respect to the density follows from differentiating
Eq.~\eqref{eq:phi_hat_fft}:
\begin{equation}
\frac{\partial \mathcal{L}}{\partial \widehat{\rho}}
    = -4\pi G\,
      \frac{1}{k^2}\,
      \frac{\partial \mathcal{L}}{\partial \widehat{\Phi}},
\end{equation}
with the zero mode fixed to zero. Finally, an inverse FFT yields the
real-space gradient $\partial \mathcal{L}/\partial \rho$.

This custom adjoint avoids storing intermediate FFT states and keeps the
memory cost of backpropagation essentially the same as that of a forward
solve.

\subsection{Multigrid solution of the Poisson equation}
\label{subsec:multigrid}
 
Writing the right-hand side of Eq. \ref{eq:poisson} as $F = 4\pi G \rho$, the discrete problem is
\begin{equation}
A\phi = F,
\end{equation}
where $A$ is the standard seven-point second-order Laplacian,
\begin{equation}
(A\phi)_{i,j,k} = \frac{1}{h^2}
\left[
-6\phi_{i,j,k}
+ \phi_{i\pm1,j,k}
+ \phi_{i,j\pm1,k}
+ \phi_{i,j,k\pm1}
\right],
\end{equation}
with periodic wrapping ensuring the homogeneous torus topology.

To solve this equation efficiently, we employ a geometric multigrid method as shown in Figure \ref{fig:vcycle}. Multigrid exploits the fact that iterative relaxation techniques efficiently reduce high-frequency components of the error, while low-frequency components converge slowly on a fine grid but appear high-frequency when transferred to a coarser grid. By alternating between fine and coarse scales, as in a W-cycle, multigrid achieves convergence rates that are effectively independent of resolution (see Appendix \ref{ap:vwcycle}). 

Our implementation of multigrid within a differentiable framework is novel and will be explored further in future work in the context of particle-mesh simulations (Horowitz et al. In Prep), but we summarize the important implementation details below.

\subsubsection{Smoothing}
Error smoothing is accomplished via weighted Jacobi relaxation,
\begin{equation}
\phi^{(m+1)} = \phi^{(m)} + \omega\, D^{-1}(F - A\phi^{(m)}),
\end{equation}
where $D = -6h^{-2}\,I$ is the diagonal part of $A$, and $\omega=2/3$ is the usual optimal weight for the
3D periodic Laplacian. In component form,
\begin{equation}
\begin{split}
\phi^{(m+1)}_{i,j,k} &=
\frac{1-\omega}{6}\phi^{(m)}_{i,j,k} \\
&+ \frac{\omega}{6}\left[
\phi^{(m)}_{i\pm1,j,k}
+ \phi^{(m)}_{i,j\pm1,k}
+ \phi^{(m)}_{i,j,k\pm1}
- h^2 F_{i,j,k}
\right].
\end{split}
\end{equation}
A few such iterations efficiently damp high-frequency errors. We also provide routines for iterative smoothing via Red-Black Gauss-Seidel, but in practice we find Jacobi provides significantly better parallel performance and faster convergence for the systems studied.

\subsubsection{Residual and restriction}
Given an approximate potential, the residual is
\begin{equation}
r^h = F^h - A^h\phi^h.
\end{equation}
These low-frequency components are transferred to a coarser mesh via full-weighting restriction,
\begin{equation}
r^{2h}_{I,J,K}
= \frac18 \sum_{i,j,k\in\{0,1\}}
r^h_{2I+i,\;2J+j,\;2K+k},
\end{equation}
defining the coarse-grid error equation
\begin{equation}
A^{2h} e^{2h} = r^{2h}.
\end{equation}

\subsubsection{Prolongation and correction}
The coarse correction is interpolated back to the fine grid using trilinear prolongation,
\begin{equation}
e^h_{i,j,k} = \sum_{\alpha,\beta,\gamma=0}^1
w_{\alpha\beta\gamma}\;
e^{2h}_{I+\alpha,\;J+\beta,\;K+\gamma},
\end{equation}
with weights determined by the fractional offset of the fine-grid point within the coarse-grid cell.
The fine-grid potential is then updated via
\begin{equation}
\phi^h \leftarrow \phi^h + e^h,
\end{equation}
followed by several post-smoothing iterations to remove high-frequency components introduced by interpolation.

\subsubsection{Adjoint formulation}

We show a diagram of the adjoint operator in Figure \ref{fig:adjoint}. Because the Laplacian is a symmetric, negative-definite linear operator,
its inverse is also self-adjoint:
\begin{equation}
(A^{-1})^\top = A^{-1}.
\end{equation}
Given an incoming gradient $g_\phi$ from upstream in the computational graph, the adjoint equation is
\begin{equation}
A\, g_F = g_\phi.
\end{equation}
Gradients with respect to the source term $F$ are therefore computed by solving another Poisson equation,
using the same multigrid algorithm as in the forward pass:
\begin{equation}
g_F = A^{-1} g_\phi.
\end{equation}
No intermediate states of the forward multigrid cycle need to be stored; the backward pass simply performs
an additional multigrid solve. This yields a memory-efficient differentiable Poisson solver whose
backpropagation cost is dominated by a single additional cycle.

\subsubsection{Comparison to FFT}
\label{subsubsec:comparefft}
\begin{figure}
    \centering
    \includegraphics[width=1.0\linewidth]{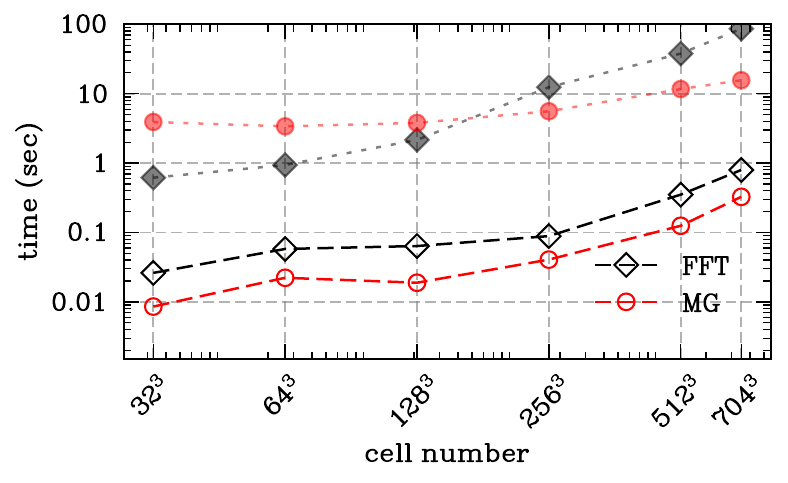}
    \caption{Forward-pass performance on 4$\times$ Tesla V100-PCIe-32GB GPUs as a function of cubic grid size. Open symbols show the wall-clock time for a single forward evaluation after compilation, while filled symbols indicate the one-time setup/compilation cost for the corresponding kernel. The FFT-based solver (black diamonds) and multigrid solver (MG; red circles) have comparable compilation overheads at small resolutions, but FFT compile time rises more steeply with grid size, consistent with larger plan construction and distributed communication setup. In evaluation mode (open symbols), MG is consistently faster than FFT over the full resolution range. Parameters for MG were chosen at each resolution to require $<0.1\%$ accuracy over all pixels for net force calculation. Absolute timings vary with system configuration and interconnect.
    }
    \label{fig:forward_timing}

    \centering
    \includegraphics[width=1.0\linewidth]{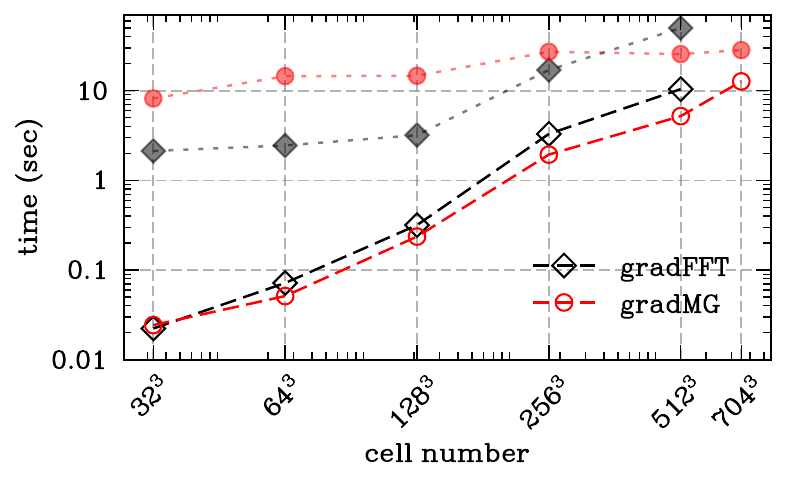}
    \caption{Backpropagation (adjoint) timing on the same 4-GPU configuration. Open symbols denote per-call backward evaluation time after compilation; filled symbols show the one-time compilation cost of the backward/adjoint kernels. The MG adjoint compiles and evaluates robustly up to the largest tested grids, with evaluation times that scale smoothly with resolution. FFT compilation and evaluation are reliable only up to $512^3$ cells in this setup; for larger grids we were unable to obtain a working distributed FFT backward pass, likely due to memory pressure and/or communication-plan constraints associated with large multi-GPU transposes. Parameters for MG were chosen at each resolution to require $<0.1\%$ accuracy over all pixels for net force calculation. Absolute timings vary with system configuration and interconnect.
    }
    \label{fig:backward_timing}
\end{figure}

While lacking the spectral accuracy of the FFT method described in Sec. \ref{subsubsec:fft}, for most applications this level of accuracy is not necessary and the multigrid solver allows tuning of accuracy to balance performance and accuracy.

We compare the timing of the FFT and MG implementation in Figures~\ref{fig:forward_timing}--\ref{fig:backward_timing}, separating the one-time compilation overhead (filled symbols) from steady-state evaluation cost per call (open symbols) for forward and backward passes on four V100 GPUs. We We use a single point mass in the center of the volume, and require the MG solver to get $<0.1\%$ accuracy; this is a most demanding case for the MG solver vs. a more spatially distributed distribution. We study up to a maximum of $704^3$, the maximum that can be backpropgated in this single node setup in our multigrid implementation.\footnote{We note that distributed backpropagation is supported by \texttt{diffhydro}, but the performance becomes heavily dependent on system architecture.} For the forward solve (Fig.~\ref{fig:forward_timing}), both methods have modest compilation costs at low resolution, but FFT compile time grows rapidly with $N$, whereas MG remains comparatively flat. After compilation, MG evaluates faster than FFT at all tested resolutions, with an increasing advantage on large meshes, indicating better scalability of the multigrid approach in the distributed GPU setting. The backward timings show the same qualitative behaviour (Fig.~\ref{fig:backward_timing}): MG adjoint compilation remains tractable and its evaluation time scales smoothly to $704^3$ cells. In contrast, we could only compile and run the FFT adjoint up through $512^3$ in this system configuration, and no stable backward evaluation was obtained at higher resolution. In addition, we found at intermediate resolutions ($256^3$ to $512^3$) we encountered frequent GPU-threading/deadlock issues in the FFT implementation which resulted in unpredictable crashes. Overall, these results suggest that even when compilation overhead is included, multigrid provides a more scalable and reliable differentiable Poisson/elliptic backend on large multi-GPU meshes, while distributed FFT approaches become increasingly costly to compile and difficult to differentiate through at high $N$.

Beyond performance there are a number of advantages/applications of the multigrid method:

\begin{itemize}
    \item \textbf{General boundary conditions and geometries:}
    Multigrid solvers accommodate arbitrary combinations of Dirichlet, Neumann, Robin, or mixed boundary conditions, and do so with essentially the same algorithmic cost as for periodic domains. 
    FFT-based Poisson solvers are optimal on uniform, translation-invariant operators with periodic boundaries; extensions to isolated/open boundaries typically require zero-padding or Green-function/multipole corrections, increasing memory and communication. 
    While sine/cosine transforms can treat simple box Dirichlet/Neumann conditions, they do not generalize to complex or spatially varying boundary conditions, nor to irregular domains.

    \item \textbf{Implicit variable-coefficient (and anisotropic) elliptic operators:}
    Many astrophysical processes lead to equations of the generic form
    $\nabla\!\cdot\!\big(\mathbf{A}(\mathbf{x},\mathbf{U})\,\nabla \mathbf{U}\big)= f$,
    where $\mathbf{A}$ is a spatially varying scalar or tensor.
    Examples include radiative/flux-limited diffusion (opacity-dependent $D(\rho,T)$), anisotropic thermal conduction along magnetic field lines, cosmic-ray diffusion/streaming, resistive/ambipolar MHD terms, viscous angular-momentum transport in disks, and species/chemical diffusion in multi-phase media.
    These operators are commonly advanced implicitly to avoid restrictive diffusive timesteps, yielding large sparse elliptic systems for which multigrid provides an efficient solver or preconditioner.
    Our library exposes these operators through user-callable forcing/flux hooks.

    \item \textbf{Non-uniform, adaptive, or moving meshes:}
    FFT methods rely on a single uniform mesh and a constant-coefficient stencil.
    Multigrid extends naturally to non-uniform spacing, block-structured AMR, nested grids, and (with appropriate restriction/prolongation) moving-mesh or mesh-refinement approaches, enabling force and transport solves across large dynamic range.
    Although the current \texttt{diffhydro} release targets fixed Cartesian meshes, the solver infrastructure is designed for these extensions.

    \item \textbf{Curvilinear coordinates and metric-weighted operators (including GR):}
    In general coordinates the Laplacian becomes a variable-coefficient operator,
    $\nabla\!\cdot\!\big(\sqrt{g}\,g^{ij}\nabla_j \Phi\big)$, 
    where metric factors vary spatially.
    This permits efficient solves in spherical/cylindrical geometries (e.g.\ stellar interiors, disks, jets) and provides a direct pathway to relativistic or pseudo-relativistic formulations by supplying a user-defined metric.

    \item \textbf{Coupled or nonlinear elliptic systems:}
    Beyond linear Poisson/diffusion, multigrid can be embedded in Newton--Krylov or FAS schemes to solve nonlinear or multi-field elliptic problems, such as flux-limited radiation closures, GR constraint equations, or stiff multi-species diffusion with coefficient dependence on the solution.
    FFT approaches generally do not extend to these cases without linearization on a uniform periodic box.
\end{itemize}

\section{Example: Optimizing Initial Conditions}
\label{sec:opt_ic}

\label{sec:inversesolving}
\begin{figure*}
    \centering
    \includegraphics[width=1.0\linewidth]{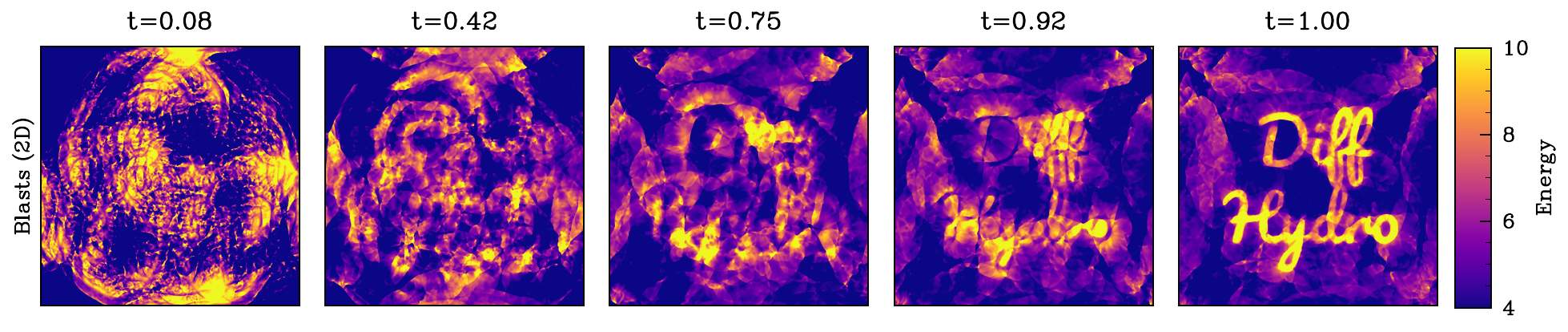}
    \includegraphics[width=1.0\linewidth]{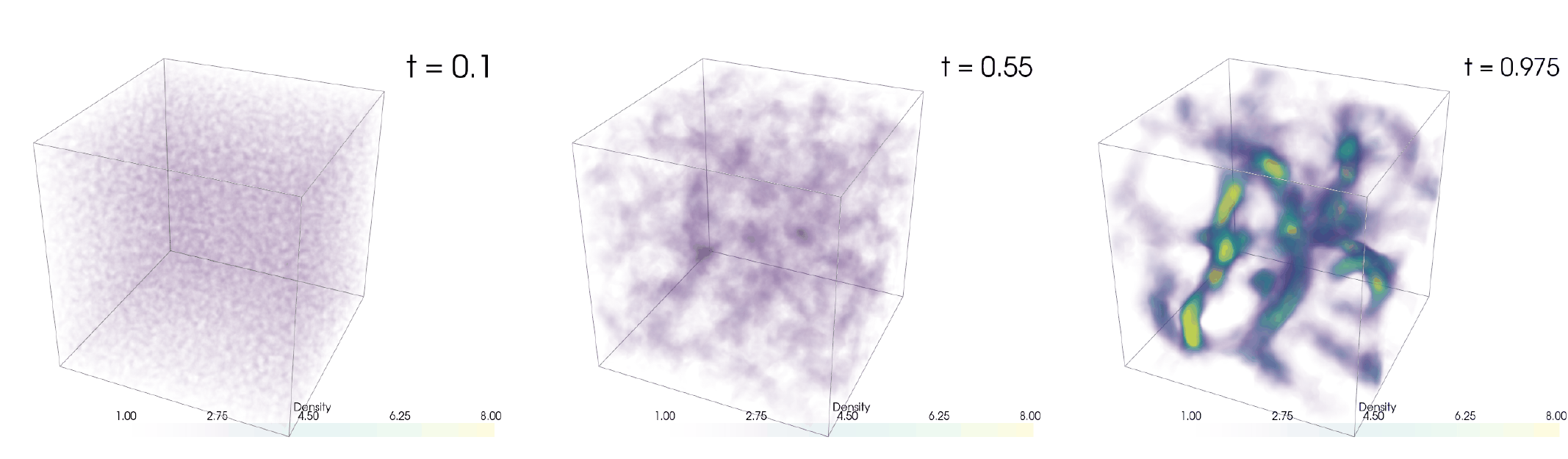}

\includegraphics[width=1.0\linewidth]{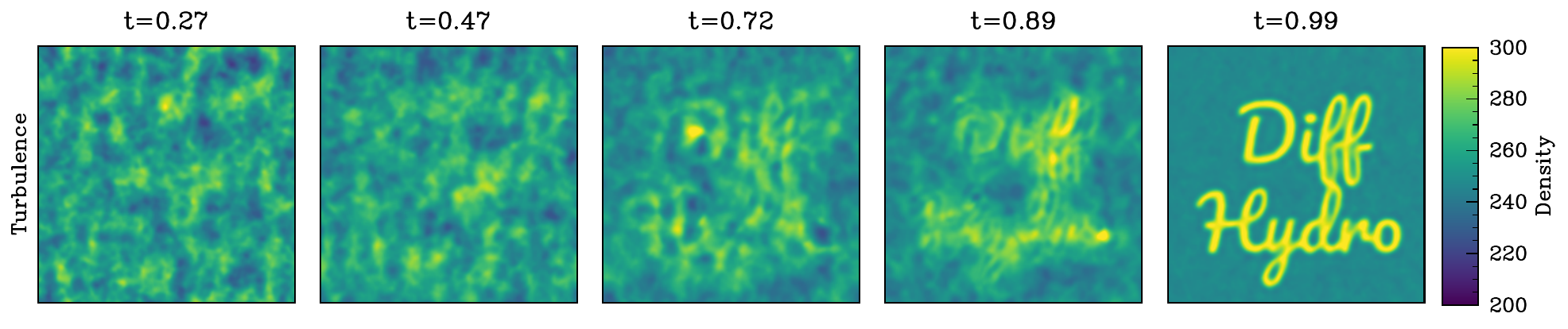}

    \caption{Inverse initial--conditions reconstructions in shock-dominated and turbulent test problems.
\textit{Top row (Blasts, 2D):} time sequence of mid-plane internal energy (energy density) in a uniform medium whose initial internal-energy field $e(\mathbf{x},0)$ has been optimized to match a prescribed late-time target in a $1024^2$ field. The optimized field produces an ensemble of Sedov--Taylor--like explosions: early evolution shows many nearly circular blast waves, which subsequently collide and merge into a curved network of shocks, contact layers, and rarefactions that reorganizes into the target large-scale pattern by $t\simeq 1$.
\textit{Middle/Bottom row (Turbulence):} Volume and projected density for a $256^{3}$ box initialized with an optimized turbulent velocity realization (Fourier phases treated as control variables) and evolved with pure hydrodynamics. Low-$k$ compressions seeded by the optimized phases sharpen through the turbulent cascade into filaments and knots, assembling into the target projected morphology at the final time.
Snapshots are shown at selected intermediate times (code units) chosen to highlight key stages of the reconstruction. Color bars indicate energy (top) and projected density (bottom). Animations are available online at \url{https://github.com/bhorowitz/DiffHydro_public/}.}
    \label{fig:icresults_turb}
\end{figure*}

Optimizing initial conditions for various observed astrophysical phenomena is one of the most natural parts of the ``differentiable" part of \texttt{diffhydro}. In this section we present a gallery of various density reconstructions performed with different physics and constraints/priors on the initial conditions. We emphasize that these particular setups are for demonstration and not meant to representative of specific astrophysical systems. We have chosen dimensional scaling, likelihoods, convergence criteria, and timescales to best show the capabilities of \texttt{diffhydro} and the underlying physical processes modeled.

For all examples in this section we use an ADAM (Adaptive Moment Estimation) optimizer as implemented in the \texttt{optax} package. ADAM is a stochastic, gradient–based optimization scheme that augments standard first–order methods with per-parameter adaptive learning rates derived from exponentially weighted estimates of the first and second moments of the gradient. Given a (stochastic) gradient \( g_t \), ADAM forms running averages
\begin{equation}
m_t = \beta_1 m_{t-1} + (1-\beta_1) g_t,\qquad
v_t = \beta_2 v_{t-1} + (1-\beta_2) g_t^{\,2},
\end{equation}
which act as estimators of the mean and un-centered variance of the gradient components. To remove the bias introduced at early iterations, these are corrected via
\begin{equation}
\hat{m}_t = \frac{m_t}{1-\beta_1^{t}},\qquad
\hat{v}_t = \frac{v_t}{1-\beta_2^{t}}.
\end{equation}
The resulting parameter update takes the form
\begin{equation}
\theta_{t+1} = \theta_t - \alpha\, \frac{\hat{m}_t}{\sqrt{\hat{v}_t} + \epsilon},
\end{equation}
which effectively normalizes each step by the local gradient variability and therefore
improves robustness in the presence of noisy or poorly scaled gradients.

In generic high-dimensional optimization problems— beyond its widespread use in neural-network training—ADAM exhibits strong performance in regimes characterized by anisotropic curvature or significant stochasticity in the objective function. Tasks such as reconstructing initial conditions in hydrodynamical simulations typically involve large dynamic ranges across parameter subspaces and loss landscapes that are highly structured, multi-modal, or elongated. ADAM’s implicit diagonal preconditioning and variance stabilization mitigate these challenges, allowing efficient exploration of rugged parameter spaces while requiring minimal manual tuning of learning rates. These properties make it a reliable baseline optimizer for large-scale inverse problems, simulation-based inference, and other scientific applications where classical second-order methods are computationally prohibitive.

For this work, we have chosen a complex target distribution which should be self-apparent in Figure \ref{fig:icresults_turb}, \ref{fig:icresults_grav}, and \ref{fig:icresults_forcing}. While it is extraordinarily unlikely any real astrophysical system will form this specific pattern, the ability of \texttt{diffhydro} to recover even extreme patterns demonstrates the power of this approach. All reconstructions were performed on a single node, with 4 Tesla V100-PCIE-32GB GPUs. Depend on the physics implemented, our examples took 1-4 minutes per gradient evaluation and were run for $\sim 100$ iterations. 

\subsection{Pure Shock (2D)}

As a simple but stringent shock-dominated inverse test, we optimize the initial internal-energy field $e(\mathbf{x},0)$ in an otherwise uniform medium at rest to match a prescribed target energy distribution at a later time. In forward evolution, localized peaks in the optimized $e(\mathbf{x},0)$ behave like an ensemble of impulsive Sedov--Taylor injections, each launching a strong, approximately self-similar blast wave into the cold background. At early times the solution is therefore a superposition of nearly circular shocks expanding from many sites. As these remnants grow, their shock fronts collide: when two Sedov waves meet, the interaction produces a compressed, over-pressured midplane, a contact layer separating material from the two explosions, and reflected shocks that propagate back into each already-shocked interior. The overlap region heats further and expands laterally, so the population of blasts rapidly reorganizes into a web of curved shocks, rarefactions, and interaction ``bridges,'' with late-time morphology set by repeated shock--shock collisions and merging. 

We perform this optimization on a $1024^2$ mesh, which cleanly resolves both the individual Sedov fronts and their subsequent interactions. The top row of Fig. \ref{fig:icresults_turb} shows that \texttt{diffhydro} can backpropagate stable gradients through multiple strong-shock encounters, allowing the optimizer to coordinate many blast sites simultaneously and recover the target large-scale structure despite the highly nonlinear and nonconvex dynamics.

\subsection{Turbulence}
\label{subsec:icturb}

In the initial--conditions optimization experiments, we parametrize turbulent velocity fields using the same Fourier--space construction described in Sec. \ref{subsubsec:turb} but with the \emph{phases treated as control variables} rather than random draws. Specifically, we define the discrete wavenumber grid for a periodic box and restrict fluctuations to a specific $k$ driving band. Within this band, each velocity component is assigned complex coefficients with fixed amplitudes following an isotropic power--law spectrum (Eq.\ref{eq:turbpl}) and a prescribed solenoidal/compressive mixture via the projection tensors $P_{ij}$, $C_{ij}$, and $M_{ij}$ (Eqs. \ref{eq:turbproj}). The only free degrees of freedom are the mode phases $\phi_i({\bf k})$, which enter through $G_i({\bf k})=\exp(i\phi_i)$; inverse FFT then yields a real--space velocity field that is rescaled to a target RMS Mach number using Eq. \ref{eq:turbscale} before assembling the conservative state (Eq. \ref{eq:turbic}). 

Optimizing $\phi_i$ therefore adjusts the spatial organization of the large--scale turbulent field while preserving the spectral slope, driving scale, and solenoidal fraction as fixed priors, enabling gradient--based reconstruction of turbulent initial conditions consistent with the turbulence initialization procedure in Sec. \ref{subsubsec:turb}. This not only enforces our turbulent field prior but also reduces the dimensionality of our field to optimize. For this work, we use $k_{max} = 10$, which corresponds to an input field of $(3, 523304)$ in a $256^3$ box, a factor of $\sim$30 reduction vs. a completely unconstrained initial velocity field. 

We optimize a mean-square-error likelihood between the final simulated projected density field and the target distribution,
\begin{equation}
\mathcal{L}_{\rm data}(\boldsymbol{\phi})
= \frac{1}{N}\sum_{\mathbf{x}}
\left[ \sum_{z}[f(\mathbf{x},z,t_f;\boldsymbol{\phi})]-T_{\rm targ}(\mathbf{x})\right]^2.
\label{eq:mse_likelihood}
\end{equation}
Figure~\ref{fig:icresults_turb} (Turbulence) visualizes the dynamics of this reconstruction in terms of the projected density field at five intermediate times. Starting from a uniform medium with the optimized turbulent velocity realization, the early-time projection ($t\simeq 0.27$) shows a fairly isotropic, mottled density pattern characteristic of compressible turbulence: low-$k$ driving rapidly generates a network of weak shocks and converging flows, producing modest contrast in filaments and sheets without any large-scale coherent morphology. By $t\simeq 0.47$ the cascade has strengthened these compressions and begun to organize them spatially; several elongated overdense structures emerge and the projection develops a clear anisotropic texture, indicating that the optimizer has selected phases that bias where strong convergences occur.

At intermediate times ($t\simeq 0.72$) the density field becomes more structured and intermittent. Shock intersections and shear layers sharpen into brighter knots and narrow filaments, while surrounding regions evacuate into broader underdense lanes. This is the stage where nonlinear mode coupling is most active: the initially band-limited velocity field has transferred power to smaller scales, and the optimizer relies on this cascade to generate high-contrast features from comparatively smooth low-$k$ control variables. %By $t\simeq 0.89$ coherent large-scale arcs and connected ridges are visible in the projection, showing that the optimized realization is not merely matching the target statistically, but is steering the turbulent flow so that specific shock complexes persist and align into the desired late-time morphology.

Finally, by $t\simeq 0.99$ the projected density closely reproduces the target distribution, assembling into a sharp, high-contrast pattern despite the strongly nonlinear dynamics and the absence of additional physics beyond pure turbulence. The time sequence highlights the essential mechanism of the inversion: optimizing the low-$k$ phases controls the placement and timing of large-scale compressions; subsequent turbulent evolution amplifies these into shock-seeded filaments and knots, which the optimizer coordinates so that their superposition at $t_f$ matches the target projection. In this sense, the figure provides an intuitive demonstration of how a compact phase parametrization can guide a fully nonlinear compressible turbulent cascade toward a specified observationally motivated endpoint. 

\begin{figure*}
    \centering
    \includegraphics[width=1.0\linewidth]{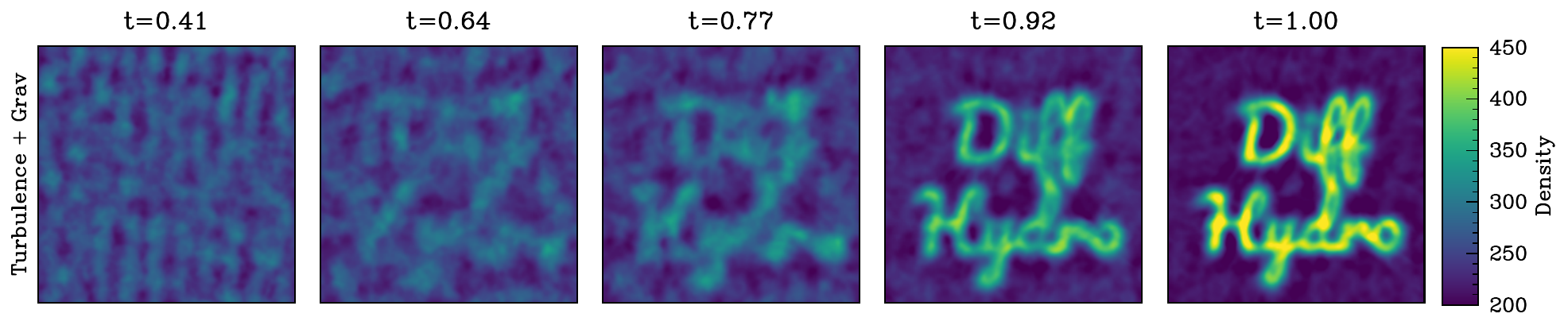}
    
    \includegraphics[width=1.0\linewidth]{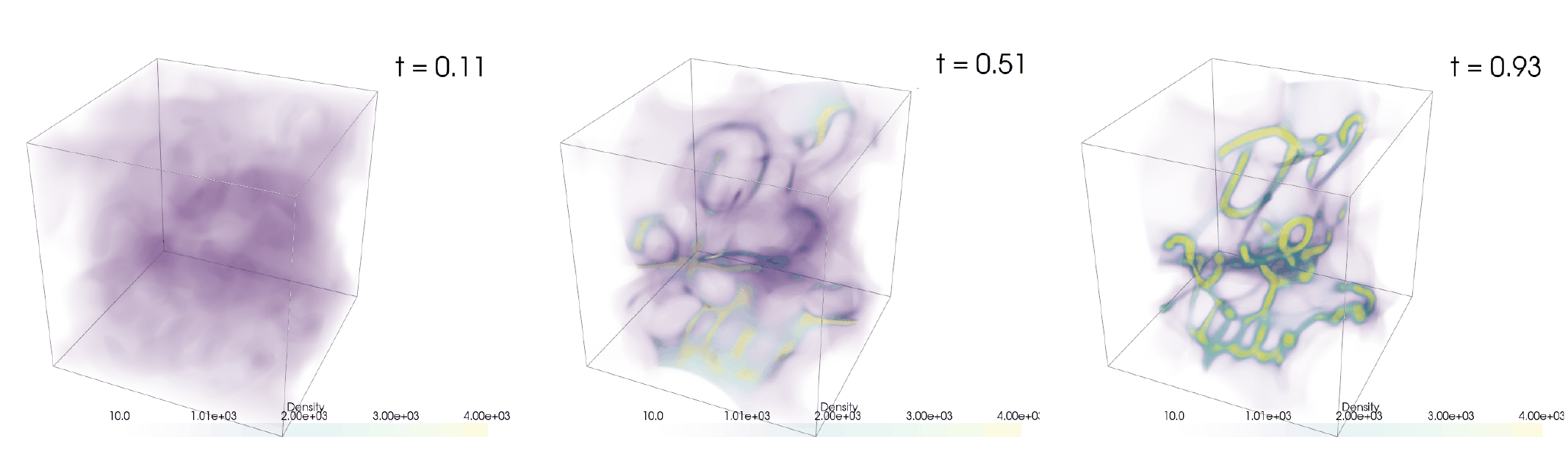}
    \includegraphics[width=1.0\linewidth]{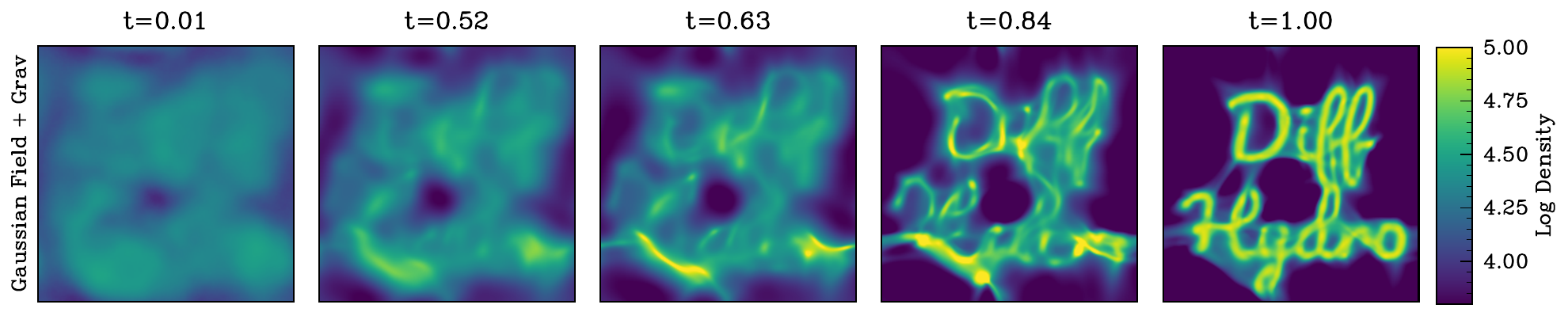}

    \caption{Evolution of density in two $256^3$ self-gravitating test boxes initialized with different priors on the initial conditions, shown as $x$-axis projections at the labeled times (in code units). \textit{Top row (Turbulence + Grav):} turbulent initial conditions plus self-gravity, where pre-existing velocity/density structure seeds a network of intermittent overdensities that sharpen into filaments during collapse. \textit{Middle/Bottom row (Gaussian Field + Grav):} Gaussian random-field initial conditions plus self-gravity, producing a smoother early evolution followed by coherent filament and clump formation. Timesteps are non-uniform and were chosen to emphazize key stages of the gravitational evolution. Color bars indicate projected density (top) and projected log density (bottom). Three dimensional volume renderings and animations are available online at \url{https://github.com/bhorowitz/DiffHydro_public/}.}
    \label{fig:icresults_grav}
\end{figure*}

\subsection{Turbulent field with Gravity}

We follow the problem and likelihood setup in Sec.~\ref{subsec:icturb}, but now also include self--gravity using the differentiable multi--grid Poisson solver of Sec.~\ref{subsec:multigrid}. The gravitational coupling is tuned so that, over the $\sim 800$ timesteps used in the optimization, gravity produces appreciable large--scale reorganization and incipient collapse without driving the system into a catastrophic singular collapse. The control variables remain the low--$k$ turbulent phases that set the initial velocity field, so the optimizer must discover an initial turbulent realization whose subsequent nonlinear evolution under the combined action of shocks, vorticity, and gravity matches the target.

The top row of Fig.~\ref{fig:icresults_grav} shows the resulting reconstruction in projected density. Early evolution resembles the purely turbulent case: the phase--optimized low--$k$ field rapidly generates compressions, shear layers, and a filamentary network in the density. As gravity becomes dynamically important, these turbulent overdensities are selectively amplified: coherent filaments thicken, knots deepen, and material streams along filaments into nascent potential wells. In other words, gravity does not merely ``add contrast,'' it couples to the turbulent morphology by turning transient shock--compressed features into longer--lived, accreting structures, while simultaneously erasing other features through long range forces. The late--time target is therefore achieved via a delicate balance between turbulent stirring (which continually seeds and distorts structure) and gravitational focusing (which nonlinearly reinforces a subset of that structure).

This reconstruction exhibits noticeably poorer convergence than the non--gravitating turbulence inversions. In practice we required three rounds of annealing (implemented as hand--tuned learning--rate reductions) to reach a visually and quantitatively satisfactory match. This behavior is expected: the target map contains subtle, high--order structure that is \emph{a priori} unlikely to arise from generic turbulent collapse, so the optimizer must search a narrow basin of initial phases that produce the correct sequence of compressions, mergers, and gravitational amplifications. Relative to the forward turbulence examples of Sec.~\ref{subsec:icturb}, self--gravity introduces additional long--range couplings and history dependence, making the likelihood surface markedly more nonconvex and multi--modal. The successful reconstruction in Fig.~\ref{fig:icresults_grav} thus provides a strong stress test of \texttt{diffhydro}’s differentiability through coupled hyperbolic--elliptic dynamics and demonstrates that gradient--based initial--condition inference remains feasible even in self--gravitating, strongly nonlinear flows.

\subsection{Gaussian Field with Gravity}

In this reconstruction we move beyond the band--limited turbulent prior of Sec.~\ref{subsec:icturb} and instead place a \emph{Gaussian random field (GRF) prior} on the initial conditions, implemented using the JAX--Cosmo \citep{2023OJAp....6E..15C} GRF machinery at fiducial Planck cosmology. The optimization is performed on a $256^3$ periodic mesh evolved with self--gravity via our differentiable multi--grid Poisson solver. The GRF prior fixes the two--point statistics of the initial field through a chosen power spectrum $P(k)$, while the \emph{phases remain the effective degrees of freedom} that control the specific realization. In contrast to the turbulence--phase experiments, this prior introduces global spatial density correlations from the outset and provides a statistically motivated baseline that is standard in cosmological inference problems.

Our objective combines a data misfit with a simple quadratic phase regularization/prior. Concretely, we minimize a mean--squared--error likelihood between the final simulated field and a target, Eq. \ref{eq:mse_likelihood}, augmented by a prior penalty
\begin{equation}
\mathcal{L}_{\rm prior}(\boldsymbol{\phi})
= \sum_{i,\mathbf{k}} \phi_i(\mathbf{k})^2 ,
\end{equation}
so that the total loss is $\mathcal{L}=\mathcal{L}_{\rm data}+\mathcal{L}_{\rm prior}$.
Operationally this is an $\ell_2$ constraint on departures of the phase parameters from their GRF baseline, which stabilizes optimization in the strongly nonconvex gravity--driven landscape and biases the solution toward realizations that remain statistically typical under the assumed Gaussian model (e.g. \citet{2019TARDIS}). Gradients are obtained by differentiating through the GRF parametrization, the hydrodynamical update, and the multi--grid gravity solve, allowing the optimizer to adjust large--scale morphology indirectly through the initial phase configuration. This approach mirrors that performed in HL25, but with added complexity of the target distribution, significantly increased resolution ($\times 8$), with a multigrid Poisson solver, but without the dark matter joint evolution.

The resulting dynamics are shown in Fig.~\ref{fig:icresults_grav} (Gaussian Field + Grav). At very early times ($t\simeq 0.01$) the density field is smooth and correlated, reflecting the GRF prior rather than shock--seeded structure. By intermediate times ($t\simeq 0.5$--$0.7$) gravitational amplification of initially mild overdensities becomes visible: broad features sharpen into coherent filaments and sheets, and contrast grows preferentially along connected ridges. As evolution proceeds ($t\simeq 0.85$) these filaments thicken and merge into a web of arcs and knots, indicative of gravity focusing correlated initial peaks rather than producing structure purely through turbulent intermittency. We see oscillatory patterns in some dense structures (i.e. lower left) as the system attempts to maintain hydrostatic equilibrium. By the final time ($t=1$) the system exhibits a pronounced network of narrow, high--contrast filaments surrounding lower--density voids, demonstrating that the optimizer can tune a statistically constrained Gaussian realization so that its nonlinear self--gravitating evolution reproduces the target’s late--time topology.

Overall, this reconstruction illustrates that \texttt{diffhydro} supports \emph{full--field inference with statistically motivated priors} at 3D resolution, even when the forward model includes long--range self--gravitating dynamics. Using a GRF prior provides a direct bridge to standard assumptions in cosmological and ISM analyses, while the phase regularization keeps the inverse problem well--posed and computationally stable. The successful match in Fig.~\ref{fig:icresults_grav} thus complements the turbulence--based reconstructions by showing that gradient--based IC recovery remains feasible when the prior is a global correlated random field rather than a narrowband turbulent driver.

\begin{figure*}
    \centering

    \includegraphics[width=1.0\linewidth]{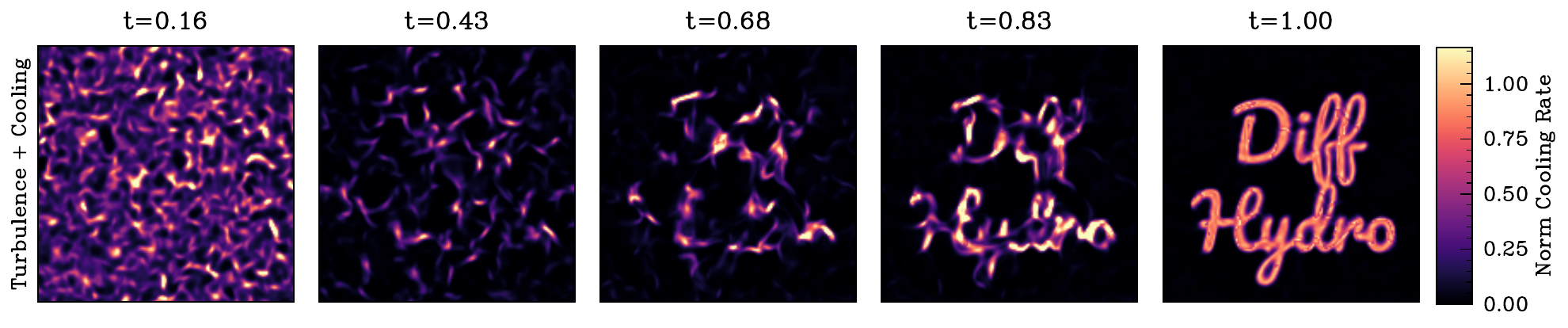}
    \includegraphics[width=1.0\linewidth]{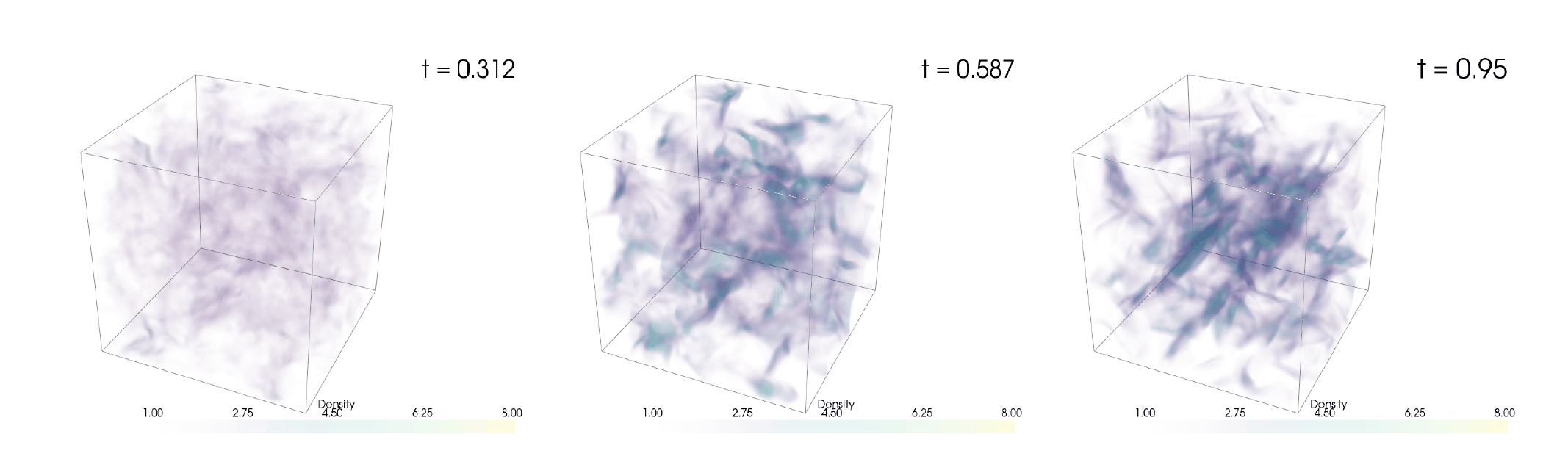}
    \includegraphics[width=1.0\linewidth]{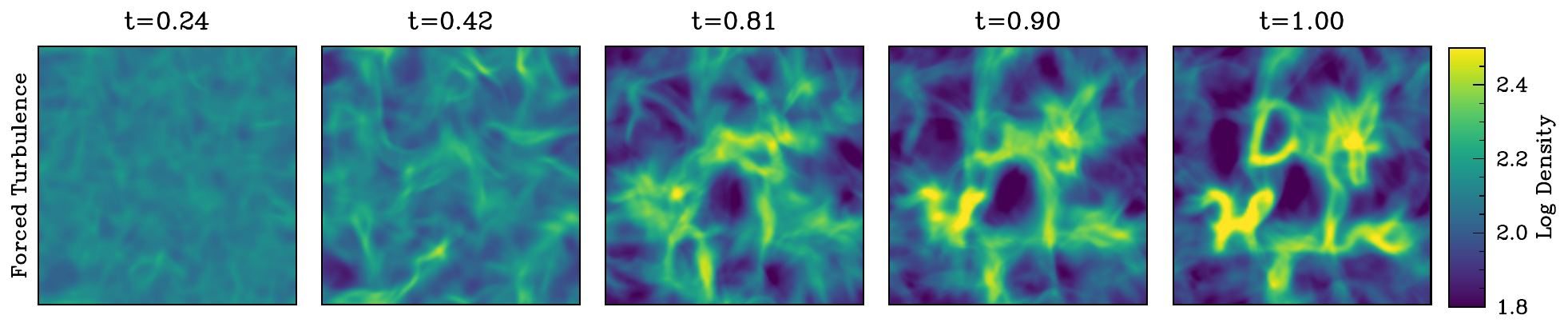}
    \caption{Various optimizations in $256^3$ boxes including non-gravitational source terms, shown at selected times (code units) chosen to highlight the dominant stages of the dynamics. \textit{Top row (Turbulence + Cooling):} mid-plane slice of the instantaneous, normalized radiative cooling rate in a run initialized with turbulent ICs and evolved with driven turbulence plus heating/cooling; cooling is initially diffuse but rapidly localizes into a filamentary network coincident with compressive structures. \textit{Middle/Bottom row (Forced Turbulence):} Volume Rendering and projection of  density from the driven-turbulence run, illustrating the growth of a high-contrast density web maintained by the interplay of stochastic forcing and thermal instability. Three dimensional volume renderings and animations are available online at \url{https://github.com/bhorowitz/DiffHydro_public/}.}
    \label{fig:icresults_forcing}
\end{figure*}

\subsection{Turbulent field with Cooling}

We next consider a radiatively cooling reconstruction that uses the same initial--condition parametrization as Sec.~\ref{subsec:icturb}: the control variables are the low--$k$ Fourier phases that define the initial turbulent velocity field. The flow is then evolved with the differentiable heating/cooling model of Sec.~\ref{sec:cooling}. Unlike the density--matching cases, however, the likelihood is not applied to $\rho$ (or its projection). Instead, after the forward run we \emph{compute the local cooling rate} from the final thermodynamic state (using the same cooling operator as in the source update) and compare this derived scalar field to a target cooling--rate map. Gradients of this mismatch are backpropagated through the cooling operator and the full hydrodynamical evolution to the initial turbulent phases.

The top row of Fig.~\ref{fig:icresults_forcing} shows a \emph{central slice} through the 3D domain of the resulting cooling--rate reconstruction. The slice highlights that the optimizer is steering the turbulent realization so that, by the final time, compressions, shocks, and shear--driven structures occur in precisely those locations and amplitudes that yield the desired pattern of radiative losses. Regions of enhanced cooling correspond to gas that the optimized turbulence has driven into the appropriate high--density/low--temperature part of phase space, while low--cooling regions trace volumes kept warmer or more rarefied by the turbulent dynamics. Because the cooling rate depends nonlinearly on the local thermodynamic history, this inversion is more indirect than matching density itself: the optimizer must adjust initial phases to control \emph{where} strong compressions form, \emph{how long} they persist against turbulent disruption, and \emph{what temperatures} they reach under radiative losses, so that the final cooling--rate field matches the target at slice level.

Overall, this example demonstrates that \texttt{diffhydro} supports gradient--based inference on \emph{derived physical observables} that are not state variables of the Euler system. By differentiating through both the stiff cooling source terms and the turbulent cascade that seeds multiphase structure, the method can reconstruct initial turbulent phases consistent with a specified cooling--rate morphology. 

Beyond being a convenient derived diagnostic, the cooling rate is also closely tied to directly observable emission. In the optically thin limit relevant for much of the CGM/ICM and hot ISM, the volumetric radiative loss rate scales as $\mathcal{L}\propto n_e n_i \Lambda(T,Z)$, i.e. the bolometric emissivity of the plasma; band--limited observables such as soft X--ray surface brightness correspond to the high--energy portion of the same cooling function \citep{2012MNRAS.420.3545B,2016ascl.soft08002Z}. Thus, reconstructing a target cooling--rate morphology can be viewed as a key step toward inference on physically motivated emission proxies, and provides a natural bridge between differentiable simulations and multi--wavelength data products.

\subsection{Forced Turbulence}
\label{subsec:IC_FT}
In the forced--turbulence reconstruction, we again treat the low--$k$ turbulent phases $\phi_i(\mathbf{k})$ as control variables and initialize the box using the Fourier--space procedure of Sec.~\ref{subsubsec:turb}. The subsequent evolution, however, is not freely decaying: we maintain a statistically steady cascade using the Ornstein--Uhlenbeck (OU) driving described in Sec. \ref{subsec:OUprocess}, in which the Fourier--space acceleration field evolves as a correlated stochastic process with forcing restricted to a different $k$--band and solenoidal/compressive mixture than that used for initialization. 

The Markov--chain character of this update means that each forcing realization depends explicitly on the previous one, so the flow retains a memory of the driving history rather than being a sequence of independent kicks. For the inverse problem, we therefore \emph{fix the forcing seed and OU history} throughout optimization, and backpropagate gradients only through the hydrodynamical response to that specific stochastic trajectory. Operationally, the optimizer searches over initial phases that, when evolved under a single reproducible OU forcing path, yield a final density map matching the target. 

The bottom row of Fig.~\ref{fig:icresults_forcing} (Forced Turbulence) illustrates the resulting dynamics in projected log--density. At early times ($t\simeq 0.24$) the fixed OU acceleration quickly imprints large--scale compressions and shear, producing a familiar driven--turbulence texture but already biased by the optimized phases toward the target morphology. As the run proceeds ($t\simeq0.42$--$0.81$), the continuous energy injection sustains strong shocks and vorticity, so overdense sheets and filaments are repeatedly created, disrupted, and re--formed. Because the forcing is temporally correlated, small changes to the initial phase configuration alter not only the immediate placement of early compressions but also the subsequent interaction between those structures and the evolving driving field. In practice this makes the mapping $\boldsymbol\phi \mapsto \rho(t_f)$ more sensitive than in the decaying case: trajectories that are initially close can decorrelate rapidly once the OU chain amplifies differences through sustained stirring. The intermediate panels show this sensitivity as a rapid reorganization of filaments into the target’s strokes, with the optimizer leveraging the persistent forcing to keep selected shock complexes coherent while allowing others to wash out. By the final time ($t=1$), the projection reproduces the target pattern despite the continually driven, chaotic background. This example highlights a complementary capability to Sec. \ref{subsec:icturb}: rather than relying on free decay to amplify a carefully chosen initial realization, the optimizer must coordinate the initial low--$k$ phases \emph{with a history--dependent forcing path}. 

\textbf{Note on Stochasticity and Backpropagation:} The successful reconstruction in this case demonstrates that \texttt{diffhydro} can propagate stable gradients through OU--driven turbulence and can solve IC inference problems even when the forward model contains temporally correlated stochastic terms. Crucially, the OU forcing enters as a \emph{Markovian} stochastic process: the acceleration field at step $t+\Delta t$ is generated from the state at $t$ plus an independent Gaussian increment. This has two practical consequences for inverse problems. First, the stochastic drive can be treated as part of the differentiable computation graph without requiring any precomputed noise series. Rather than storing a full forcing history (which would be memory prohibitive at high resolution), we only need to re-generate the local increment at each step, so the inverse solve remains tractable even for long integrations (see also \citet{li2020scalable}). Second, because the forcing is Markovian, gradients with respect to the ICs do not require backpropagating through an explicitly stored random trajectory; instead they flow through the deterministic update map conditioned on the sampled increments. In this sense, OU forcing behaves like a temporally correlated but still reparameterizable noise source, enabling end-to-end differentiation through driven turbulence.

In the context of broader gradient-based sampling techniques, the random seed controlling the forcing can be treated as an auxiliary latent variable. Fixing the seed yields a well-defined conditional posterior over ICs, while varying it produces an ensemble that marginalizes over stochastic realizations of the drive. This provides a natural route to expectation values under the full stochastic forward model.\footnote{This is analogous to averaging over dropout masks or minibatch noise in machine learning.} From a sampling perspective (e.g., in HMC or related methods), changing the seed corresponds to drawing a new noise path and therefore exploring a different conditional energy landscape. Practically, this allows us to assess posterior robustness to stochasticity and to reduce variance by considering multiple seeds, in direct analogy with the ``reparametrization trick'' commonly used for training neural networks, where gradients are taken through a deterministic transformation of an explicit noise variable. See also \citet{2024MNRAS.529.2473H} for related cases with discrete stochastic terms.

Finally, one may contemplate a fully joint reconstruction of both the ICs and the driving noise field (e.g. \citet{2025arXiv251105484R}). In principle, this would correspond to optimizing (or sampling) over the entire sequence of OU increments in addition to the ICs, yielding a maximum a posteriori estimate of the latent forcing realization. However, for Markov forcing this is substantially more challenging: the latent noise must be specified at \emph{every} timestep, leading to an optimization problem whose dimensionality scales with the total number of steps and which is therefore both memory intensive and highly degenerate (multiple noise histories can yield similar macroscopic states). It is possible this approach may become feasible if a more physical prescription turbulent forcing formalism is developed that doesn't require so many latents. %For this reason, we focus on conditioning on a seed and/or marginalizing over seeds, which captures the stochastic uncertainty while keeping the inverse problem computationally feasible.

%\subsection{Sampling Initial Conditions and Parameters}

\section{Example: Solver-in-the-loop}
\label{sec:cil}

\begin{figure}
    \centering
    \includegraphics[width=0.99\linewidth]{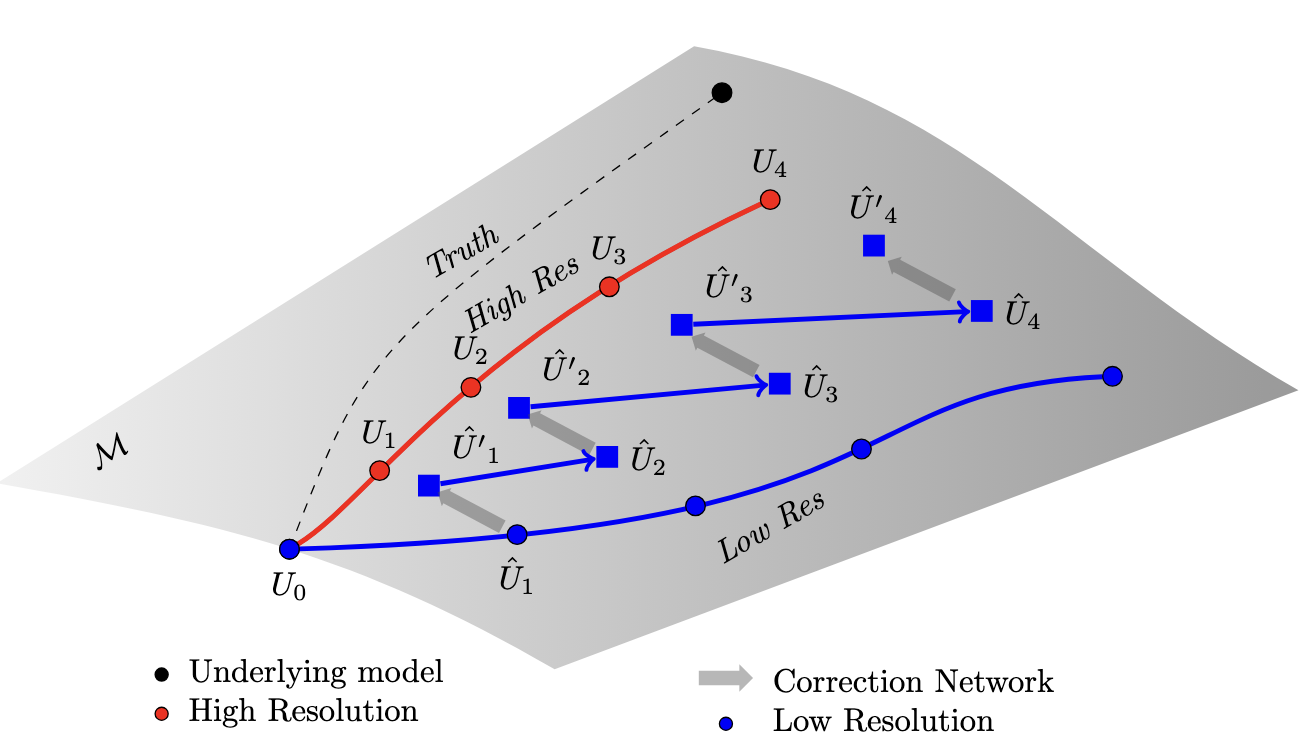}
    \caption{\textbf{Solver-in-the-loop:} Schematic diagram showing how an integrated machine learning model can be used to correct a ``low resolution" simulation to recover the properties of a ``high resolution" simulation at significantly reduced computational cost. We visualize the evolution of the system as a path on the solution manifold $\mathcal{M}$, with each point representing a system state. The corrector network, $\mathcal{B}_\theta$, can be viewed as a geodesic flow on the manifold that pushes the low-resolution solution closer to either the reference high-resolution solution or the underlying truth (if suitable data are available). The augmentation network, $\mathcal{B}_\theta$ can be trained dynamically by running the low-resolution simulation to the end state, $\hat{U}'_f$ and then snapshots along the simulation can be compared with the reference simulation $U_i$. The derivatives of the loss can be used to update the parameters of the correction network, $\theta$. }
    \label{fig:sil_diagram}
\end{figure}

Machine-learning (ML) models are increasingly employed to accelerate or augment hydrodynamical simulations, but high fidelity typically requires that the learned model interacts consistently with the numerical solver. \emph{Solver-in-the-loop} (SiL) methods achieve this by training ML components while they are embedded directly within the hydrodynamics time-integration cycle. This ensures that gradients, prediction errors, and stability constraints are evaluated with respect to the \emph{true} numerical update operator rather than an offline surrogate \citep{2020arXiv200700016U}. If the hydrodynamic solver advances the conserved state vector $\mathbf{U}$ according to
\begin{equation}
    \mathbf{U}^{n+1} = \mathcal{H}\!\left( \mathbf{U}^n,\, \mathcal{B}_{\theta} \right),
\end{equation}
where $\mathcal{B}_{\theta}$ is an ML component with parameters $\theta$, then SiL training optimizes $\theta$ with respect to the composite operator $\mathcal{H}$, not $\mathcal{B}_{\theta}$ in isolation. We show this in terms of a flow on a solution space $\mathcal{M}$ in Figure \ref{fig:sil_diagram}, where the correction network $\mathcal{B}_\theta$ pushes the low resolution trajectory, $\mathbf{\hat{U}}$, towards the reference high resolution trajectory $\mathbf{U}$.

In practice, SiL approaches can be applied to a range of hydrodynamic operators, including sub-grid closures, flux corrections, and stiff source terms. For a a generic finite-volume discretization with volume elements $\Delta V_i$ and area elements $A_{i,j}$, the standard update equation
\begin{equation}
    \mathbf{U}^{n+1}_i = 
    \mathbf{U}^{n}_i 
    - \frac{\Delta t}{\Delta V_i}
    \sum_{f} 
    \mathbf{F}_{i,f}\, A_{i,f}
    + \Delta t\,\mathbf{S}_i,
    \label{fig:sil}
\end{equation}
may be modified by modifying either the fluxes $\mathbf{F}_{i,f}$ or source terms $\mathbf{S}_i$ with predictions from $\mathcal{B}_\theta$. During SiL training, the solver is unrolled for several time steps, and gradients are obtained via automatic differentiation or adjoint methods. This allows physically meaningful losses—e.g.\ enforcing conservation, accurate shock profiles, or preservation of turbulent spectra—to be evaluated on the evolved fields rather than on pointwise training targets, yielding models that remain stable when coupled to the solver's reconstruction scheme, Riemann solver, and CFL timestep.

A major advantage of SiL optimization is that it mitigates distribution shift between training data and the dynamical states encountered during long integrations. Because $\mathcal{B}_{\theta}$ is trained \emph{in situ}, it learns to correct errors arising from numerical diffusion, coarse resolution, or operator splitting under the exact operating conditions of the simulation. SiL frameworks also enable the incorporation of physics-based regularization directly within the training loop—for example enforcing positivity of density and pressure, symmetry constraints, or divergence-free conditions at each time step. This leads to hybrid solvers that retain the interpretability and guarantees of classical hydrodynamics while exploiting the flexibility and efficiency of data-driven models.

\begin{figure*}
    \centering
    \includegraphics[width=0.99\linewidth]{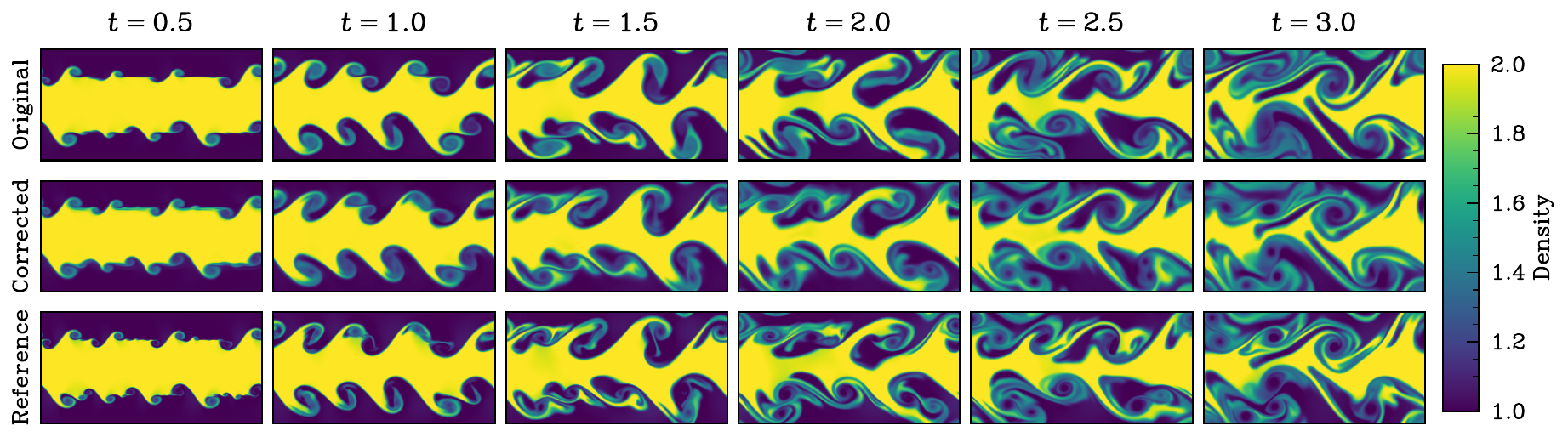}
    \includegraphics[width=0.99\linewidth]{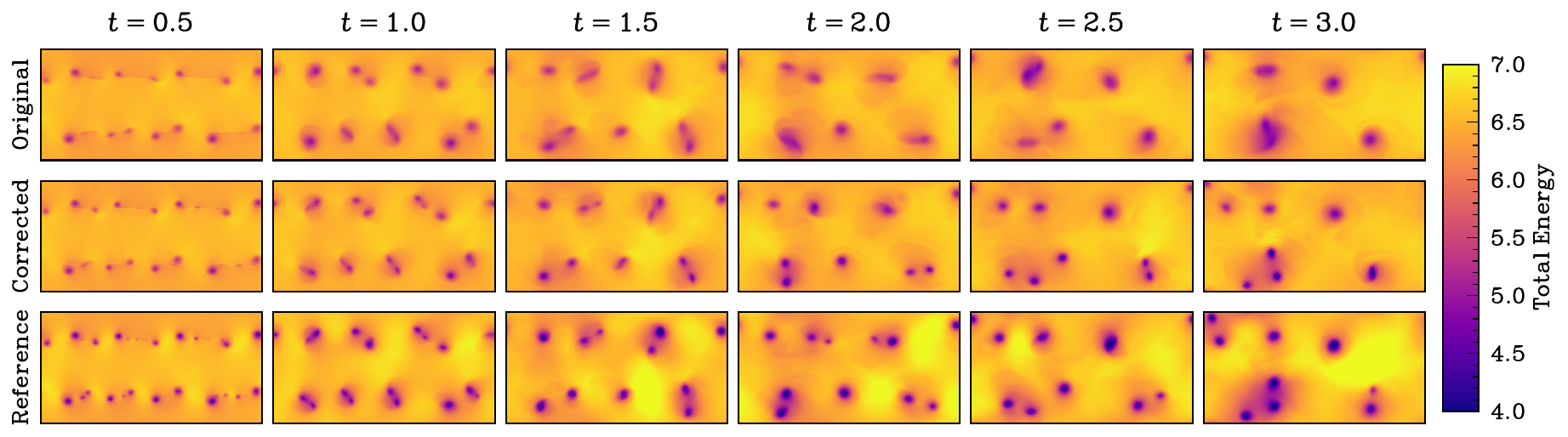}
    \caption{A visualization of the Kelvin Helmholtz flux corrector network on a test initial conditions set. Despite only training our model up until $t=1.0$, we are able to maintain numerical stability well past $t=3.0$. This is seen both in terms of the recovered density structures as well as the position/number of the vortexes seen in the energy field.}
    \label{fig:kh_bigplot}
\end{figure*}

\begin{figure}
    \centering
    \includegraphics[width=1.0\linewidth]{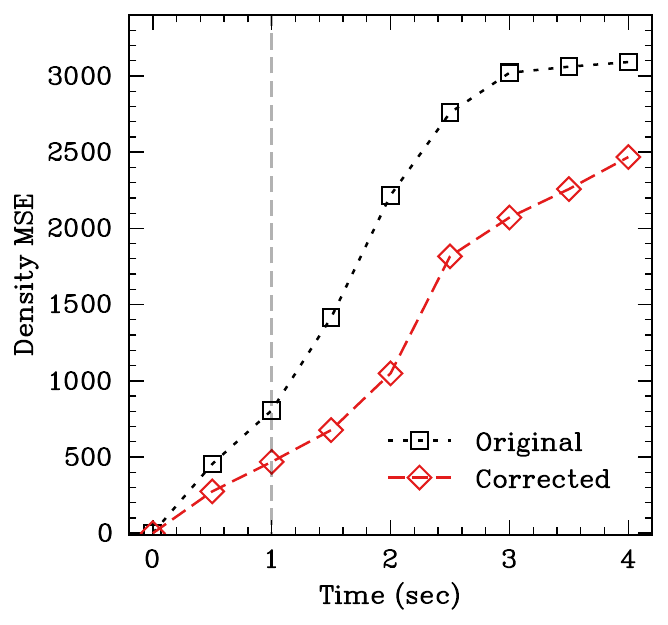}
    \caption{Performance of the Kelvin Helmholtz flux corrector network is compared over time compared to the uncorrected network on a test set. While the network was only trained until $t=1.0$ ($\sim 1000$ timesteps), we find significant improvement over the baseline until $t\sim3.0$ which is far into the nonlinear regime.}
    \label{fig:kh_perf}
\end{figure}
\subsection{Kelvin Helmholtz Corrector}

To assess the ability of our solver-in-the-loop (SiL) framework to recover unresolved vortical structure, we apply it to a two-dimensional Kelvin--Helmholtz instability benchmark. In this setup, an ML-based flux correction term $\mathcal{B}_{\theta}$ augments the base finite-volume fluxes in order to reproduce the small-scale shear-layer morphology present in a high-resolution adaptive mesh refinement (AMR) reference solution. During training, the reference AMR simulation is downsampled to the fixed coarse grid used by the baseline solver, and the SiL model is optimized by unrolling the full hydrodynamic update from Eq. \ref{fig:sil} without additional source terms in 2d,

\begin{equation}
\begin{aligned}
&\mathbf{U}^{n+1}_{i,j}
= \mathbf{U}^{n}_{i,j}\\
&- \Delta t \Bigg[
        \frac{
            \mathbf{F}^{(\mathrm{base})}_{i+\frac12,j}
            + \mathcal{B}_{\theta}(\mathbf{U}^n)_{i+\frac12,j}
            -
            \mathbf{F}^{(\mathrm{base})}_{i-\frac12,j}
            - \mathcal{B}_{\theta}(\mathbf{U}^n)_{i-\frac12,j}
        }{\Delta x} \\
&
    + \frac{
            \mathbf{G}^{(\mathrm{base})}_{i,j+\frac12}
            + \mathcal{B}_{\theta}(\mathbf{U}^n)_{i,j+\frac12}
            -
            \mathbf{G}^{(\mathrm{base})}_{i,j-\frac12}
            - \mathcal{B}_{\theta}(\mathbf{U}^n)_{i,j-\frac12}
        }{\Delta y}
\Bigg],
\end{aligned}
\end{equation}
and backpropagating losses defined on the evolved, downsampled AMR fields. In practice, we add an additional flux term to our hydro-solver which takes in the state vector and the solved convective flux. This a~posteriori, trajectory-level training forces $\mathcal{B}_{\theta}$ to reconstruct the secondary billows, rolled-up vortices, and sharp shear features that the coarse solver would otherwise diffuse away. Because the optimization occurs with the ML term embedded within the full CFL update operator, the learned flux augmentation preserves stability while selectively injecting the fine-scale vorticity required to match the AMR reference. As a result, the SiL-enhanced solver aims to reproduce the correct vortex merger sequence, power-spectrum slope, and small-scale intermittency far more faithfully than either the coarse baseline simulation or an offline-trained model.

We generate training sample of five 2-dimensional Kelvin-Helmholtz simulations with \textsc{Athena++} on a $128 \times 256$ grid with AMR run up until $t=5$. In order to control for the high solver-specific numerical sensitivity of the initialization points of the instabilities, we only choose our effective initial conditions to correspond to $t=0.5$. This corresponds to a point where the instability has not yet initiated but the initial white noise has started to propagate to larger scales.

For our corrector flux, we employ a convolutional neural network, shown in Table \ref{table:network}, consisting of four initial convolutional blocks with $3\times3$, $5\times5$, $5\times5$, and $3\times3$ kernels and $(12,24,32,32)$ output channels, using SeLU, $\tanh$, SeLU and SeLU non-linearities, respectively, all with periodic padding and unit stride. The feature map after the first convolutional block (12 channels) is concatenated along the channel dimension with the output of the fourth block (32 channels) as a skip-connection, yielding a 44-channel representation. This is followed by two additional $3\times3$ convolutions with 16 and 5 output channels, using SeLU and a final linear activation, respectively. The specific architecture was chosen to be generic and relatively small, containing only 43333 parameters. It is also fully convolutional, allowing training and application on different sized boxes.

We use a mean-squared likelihood over all 4 channels ($\delta$,$v_x$,$v_y$, $E_{tot}$) upweighting $\delta$ to have a comparable dynamic range as $E_{tot}$. We train our model using the ADAM optimizer described in Sec. \ref{sec:opt_ic} over our 5 training samples without gradient accumulation. Due to the numerical sensitivity of the coupled system, we find fairly small learning rates are needed to achieve stable training, so we use a decay schedule starting from $10^{-3}$ and decaying to $10^{-4}$. 
\begin{table}
\centering
\caption{Architecture of the convolutional neural network used in this work.
All convolutions use periodic padding and unit stride; no pooling layers are
employed. The skip connection concatenates the feature maps after the first
and fourth convolutional blocks along the channel dimension.}
\begin{tabular}{lllll}
\hline
Layer & Operation & Kernel & Channels (in,out) & Activation \\
\hline
1 & Conv & $3\times3$ & $C_{\rm in},\,12$ & SeLU \\
2 & Conv & $5\times5$ & $12,\,24$          & $\tanh$ \\
3 & Conv & $5\times5$ & $24,\,32$          & SeLU \\
4 & Conv & $3\times3$ & $32,\,32$          & SeLU \\
5 & Concat & --       & $12+32,\,44$       & -- \\
6 & Conv & $3\times3$ & $44,\,16$          & SeLU \\
7 & Conv & $3\times3$ & $16,\,5$           & linear \\
\hline
\label{table:network}
\end{tabular}
\label{tab:cnn-arch}
\end{table}
Figure~\ref{fig:kh_bigplot} compares the evolution of the Kelvin--Helmholtz instability under the baseline solver, the solver augmented with the learned flux-correction network, and the downsampled high-resolution AMR reference on a test initial condition realization. Across all times shown, the corrector substantially improves the reconstruction of small-scale vortical structure that is otherwise lost in the coarse simulation. By $t \approx 1$, the last step used for training, the corrected solution already exhibits sharper shear-layer roll-up and more distinct secondary billows, closely matching the morphology and phase of the reference. At later stages ($t \gtrsim 2$), when the uncorrected solver becomes visibly over-diffusive and suppresses fine-scale filamentation, the solver-in-the-loop (SiL) enhanced model preserves coherent vortex cores, tighter spiral arms, and more faithful small-scale enstrophy. The corrected density field follows the reference merger sequence and large-scale mode structure more accurately than the baseline, with significantly reduced phase drift.

A similar trend is observed in the total-energy field: the coarse simulation rapidly smooths out localized kinetic-energy concentrations, whereas the corrected evolution maintains the compact high-energy spots and their advection trajectories seen in the AMR reference. While some small deviations remain in the precise location and strength of energy peaks, the corrected model preserves their number, spacing, and overall spatial distribution far more accurately than the baseline. Overall, the solver-in-the-loop corrector enables the coarse simulation to reproduce both the qualitative morphology and the temporal progression of the Kelvin--Helmholtz instability with markedly improved fidelity, recovering structure that would otherwise require substantially higher spatial resolution.

These trends can also be seen quantitatively in terms of the pixel mean squared error in Fig. \ref{fig:kh_perf}. We see significantly improved error up until time $t\sim 3.0.$ At this point small errors accumulate enough for small offsets to develop which the MSE severely penalizes. 
%\subsection{Physics Discovery: Supernova}

\section{Conclusions}
\label{sec:conc}
We have presented an expanded \texttt{diffhydro}, a fully differentiable, accelerator-native hydrodynamics framework designed for data-driven astrophysical fluid modeling. Implemented in JAX, \texttt{diffhydro} provides a modular finite-volume solver for the compressible Euler equations on uniform Cartesian meshes, with interchangeable reconstruction methods, approximate Riemann solvers, and explicit Runge–Kutta integrators compiled by XLA into fused GPU/TPU kernels. The solver is organized around composable flux and force “tasks,” making it straightforward to incorporate additional physical modules or learned components while preserving end-to-end differentiability and performance.

A central contribution is the demonstration of practical reverse-mode differentiation for astrophysically relevant flows. To keep adjoint calculations tractable for long 3D integrations, we combined checkpointing with custom adjoints for memory-intensive operators, including FFT- and multigrid-based self-gravity solvers and stochastic Ornstein–Uhlenbeck turbulence forcing. This yields gradients consistent with the discrete numerical update at manageable memory and wall-time cost, enabling PDE-constrained optimization and simulation-based inference with realistic physics.

We validated \texttt{diffhydro} on standard benchmark problems. Sedov–Taylor blast waves, Kelvin–Helmholtz instabilities, and driven/decaying turbulence reproduce expected solutions and show close agreement with Athena++ under matched setups, with discrepancies limited to solver-sensitive fine-scale structure. We further demonstrated stable operator-split evolution with a differentiable two-branch ISM heating/cooling model, confirming that \texttt{diffhydro} can evolve multiphase radiatively cooling flows without sacrificing differentiability.

These capabilities enable entirely new routes to connect simulations directly to observations. We illustrated this with two examples. First, we performed gradient-based inverse problems to reconstruct complex initial conditions in turbulent, self-gravitating, and radiatively cooling systems, highlighting the suitability of \texttt{diffhydro} for likelihood-free inference, optimal control, and differentiable forward modeling. Second, we trained a solver-in-the-loop neural corrector that compensates for coarse-grid diffusion during time integration. In a Kelvin–Helmholtz test, the corrector improves small-scale morphology and delays phase drift while preserving stability, demonstrating how differentiable solvers can support robust hybrid closures trained under the same numerical conditions in which they are applied.

Several limitations remain in the current release. \texttt{diffhydro} presently targets uniform Cartesian grids and explicit time integration; curvilinear geometries, AMR or moving meshes, and more general implicit, relativistic, or radiation-hydrodynamic operators are not yet supported. While the framework includes gravity, turbulence driving, conduction, heating/cooling, and early MHD components, additional microphysics (e.g., larger chemistry networks or radiative transfer) will require further differentiable implementations and validation.

Looking ahead, extending the elliptic infrastructure to variable-coefficient and coupled systems will unlock differentiable diffusion, cosmic-ray transport, and radiation-like operators. More sophisticated thermal models and more complete MHD/anisotropic conduction are natural next steps. On the ML side, \texttt{diffhydro} provides a controlled setting for developing learned closures and correctors that are trained, tested, and deployed within the same discretized PDE update, allowing systematic study of stability and physical regularization in data-constrained regimes.

The broader JAX ecosystem in astrophysics has rapidly matured into a genuinely end-to-end differentiable stack: on the analysis side, packages such as \texttt{jax-cosmo} \citep{2023OJAp....6E..15C} for cosmological theory and likelihoods, \texttt{jfof} \citep{2025jfof} for clustering, \texttt{jaxspec} \citep{2024A&A...690A.317D} and related tools for fully Bayesian spectral inference, and JAX-native GP/time-series frameworks like \texttt{gallifrey} \citep{2025A&A...699A..42B} have made gradient-based inference on large datasets possible; in parallel, differentiable forward models such as \texttt{PMWD} \citep{li2022pmwd}, \texttt{JaxPM}, and \texttt{Disco-DJ} \citep{2024JCAP...06..063H,2025arXiv251005206L} have modeled large scale structure formation. Recently, there has been a marked increase of activity with various differentiable simulation components created in \texttt{jax}, including ray tracing \citep{2025rayrace}, synchrotron emission \citep{2025sync}, astrochemical interactions \citep{2025astrochemical}, and stellar winds/cosmic rays \citep{2024stellarwinds}.
In this landscape, \texttt{diffhydro} fills a key gap by providing an astrophysics-oriented, shock-capturing, multiphysics hydrodynamics solver that is differentiable by construction and scalable.

In summary, \texttt{diffhydro} makes differentiable, multiphysics astrophysical gas dynamics feasible on modern accelerators, and is intended as a practical engine for simulation-based inference and data-constrained modeling in turbulent, self-gravitating, radiatively cooling flows. We hope it will accelerate efforts to integrate high-fidelity numerics with gradient-based learning and observational data in a unified, end-to-end differentiable workflow.

\section*{Acknowledgments}
We would like to thank Francois Lanusse, Philip Mocz, Tilman Troster, Leonard Storcks, and Tobias Buck for helpful discussions on differentiable hydrodynamics. This work is supported by the MEXT/JSPS KAKENHI Grant Numbers JP22K21349, 24H00002, 24H00241, and 25K01032 (K.N.). This research used resources of the National Energy Research Scientific Computing Center (NERSC), a Department of Energy User Facility. This work made use of AI tools, including ChatGPT and Claude-AI. 

\section*{Data Availability}

No new data were generated or analyzed in support of this research. All code related to this work is available at \url{https://github.com/bhorowitz/DiffHydro_public}.

\appendix

\section{Hydrodynamical Numerical Methods and Implementation}
\label{ap:hydro}

Here we provide an overview of some of the hydrodynamical numerical methods used in this paper. The hydrodynamic update consists of MUSCL-type spatial reconstruction, numerical flux
evaluation, and either a dimensionally split or unsplit time integrator. Two evolution
paths exist in the solver:
(i) an unsplit method-of-lines (MOL) scheme using explicit Runge--Kutta (RK) integrators, and
(ii) a dimensionally split two-stage RK2 sweep (matching that in HL25).
Both approaches are implemented in \texttt{hydro\_core.py}.

\subsection{Reconstruction}

By default, we use the second-order MUSCL reconstruction performed on the primitive variables
$\mathbf{W} = (\rho,\,\mathbf{v},\,P)$.
Slopes are limited using
\begin{equation}
\Delta W_i = \mathrm{limiter}(W_{i+1}-W_i,\; W_i-W_{i-1}),
\end{equation}
from which left/right interface states follow as
\begin{equation}
W^{L}_{i+1/2} = W_i + \tfrac12 \Delta W_i,
\qquad
W^{R}_{i+1/2} = W_{i+1} - \tfrac12 \Delta W_{i+1}.
\end{equation}
Several slope limiters are available, including \textsc{minmod}, MC, Van~Leer, and
Superbee, as well as a piecewise-constant reconstruction corresponding to a
first-order, highly diffusive scheme.

In addition to MUSCL-type reconstruction, the code provides several other
higher-order reconstruction stencils implemented in \texttt{diffhydro/solver/recon.py}.
These include a piecewise-linear PLM formulation, the Colella--Woodward
piecewise-parabolic method (PPM), a first-order WENO1 scheme, and a fifth-order
targeted ENO (TENO5) reconstruction. All reconstruction methods are implemented
as vectorized, differentiable \textsc{jax} kernels and may be selected
interchangeably at runtime, allowing the user to trade off sharpness of
discontinuities, robustness, performance, and accuracy.

\subsection{Riemann solver}

Interface fluxes are obtained by solving approximate Riemann problems at the
cell faces. By default, the code employs the HLLC solver for hydrodynamics.
Wave speeds are estimated using
\begin{equation}
S_L = \min(v_L - c_L,\; v_R - c_R), \qquad
S_R = \max(v_L + c_L,\; v_R + c_R),
\end{equation}
with the adiabatic sound speed $c_s^2 = \gamma P/\rho$ or, optionally, a more
general effective sound-speed model when additional physics is active.
Flux contributions from conduction or other operators are included by
summation in the flux stack.

Additional approximate Riemann solvers are available in
\texttt{diffhydro/solver/stencil.py}, including the Lax--Friedrichs
(Rusanov) flux and the two-wave HLL solver. All solvers share a unified interface and are
fully vectorized, enabling efficient fusion under \textsc{jit} compilation.
\subsection{Explicit Runge--Kutta time integration}
The primary integrators available are explicit method-of-lines RK schemes provided through
\texttt{INTEGRATOR\_DICT}. These include
\begin{itemize}
    \item second-order RK2 (Heun's method),
    \item third-order SSPRK3,
    \item optional first-order Euler for testing.
\end{itemize}
The unsplit MOL update applies the chosen integrator to the right-hand side
\texttt{rhs\_unsplit}:
\begin{equation}
\frac{\mathrm{d}U}{\mathrm{d}t}
=
-\sum_{a=1}^{3}
\frac{1}{\Delta x_a}
\left(F_a - \mathrm{roll}_{a}(F_a)\right),
\end{equation}
where the roll includes halo exchange for multi-gpu/host via \texttt{roll\_with\_halo}.
The resulting semi-discrete ODE is integrated by RK2 or SSPRK3 through calls such as
\begin{equation}
U^{n+1} = \texttt{integrator}(\texttt{rhs\_unsplit},\, U^n,\, \Delta t,\, \text{params}),
\end{equation}
as implemented in \texttt{mol\_solve\_step}.

\begin{figure}
    \centering
    \includegraphics[width=0.50\linewidth]{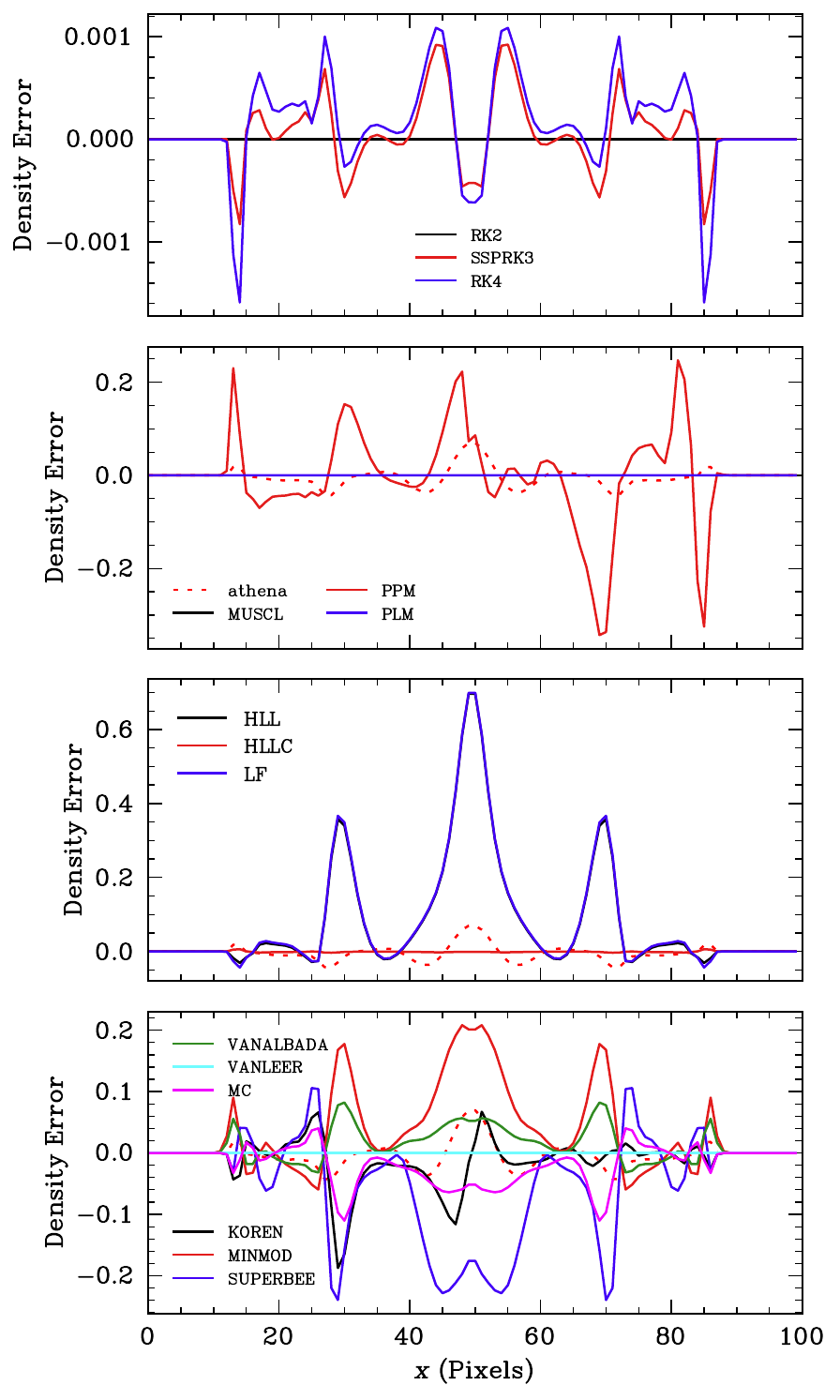}
    \caption{Comparison of various numerical choices implemented in \texttt{diffhydro} in terms of fractional error in density through the center of a Sedov blast wave. We choose for our fiducial model RK2 integrator, HLLC Riemann solver, MUSCL reconstruction with a Vanleer flux limiter. We also show a reference \textsc{Athena++} run with the same initial conditions for comparison. \emph{(Top)}: Choice of integrator used. \emph{(Upper middle)}: choice of flux reconstruction method used. \emph{(Lower middle)}: choice of Reimann solver used. \emph{(Bottom)}: flux limiter used.}
    \label{fig:sedov_compare}
\end{figure}

\begin{figure}
    \centering
    \includegraphics[width=0.50\linewidth]{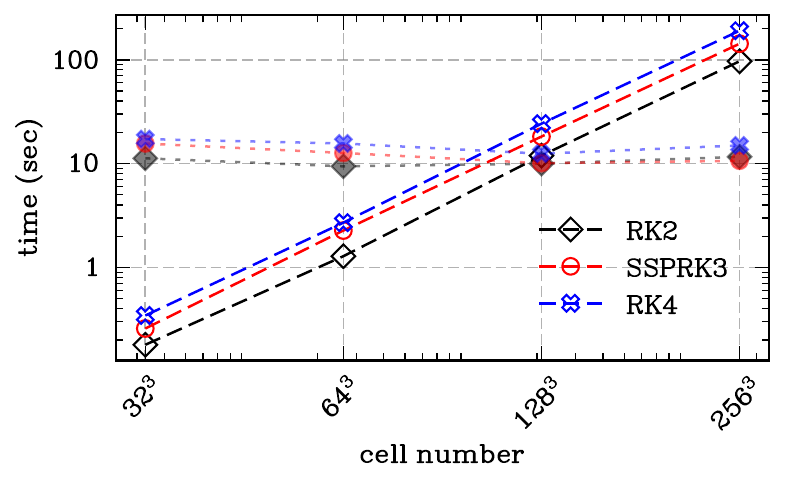}
    \caption{Blast timings as a function of grid size on a single GPU for 1000 timesteps. We show compilation time as filled symbols, and forward evaluations as open symbols. Notably, we find near constant compilation time for this particular problem setup due to the scale invariant nature of the Sedov blast wave.}
    \label{fig:sedov_timing}
\end{figure}
\subsubsection{Dimensional splitting (optional RK2 sweeps)}
If the user selects the split-sweep mode (\texttt{use\_mol=False}), the solver performs
a classical directional splitting across the cyclic axis-ordering patterns defined in
\texttt{splitting\_schemes}.
Each directional stage uses a two-stage RK2 update,
\begin{equation}
U^{(1)} = U^n - \frac{\Delta t}{2\Delta x}\left(F - \mathrm{roll}(F)\right),
\end{equation}
followed by a corrector step,
\begin{equation}
U^{n+1} = U^n - \frac{\Delta t}{\Delta x}\left(F^{(1)} - \mathrm{roll}(F^{(1)})\right),
\end{equation}
implemented in \texttt{split\_solve\_step}. This integration scheme matches that used in HL25, but it is not used for any example presented in this work.

\subsection{CFL condition}
The timestep is computed from all active fluxes and forces using
\begin{equation}
\Delta t = C_{\mathrm{CFL}}\,
\min\frac{\Delta x}{|v_a| + c_s},
\end{equation}
obtained through the per-flux timestep routines inside \texttt{flux.timestep()} and
\texttt{force.timestep()}.
Gradients through the timestep are can be optionally stopped to improve the stability of
differentiable training loops.

\subsection{JAX vectorization and differentiability}
All reconstruction, Riemann, and update operations are implemented as pure functions suitable
for \textsc{jit} compilation. Multi-device parallelism is handled by explicit sharding and
halo exchange (via \texttt{shard\_map}), while memory-efficient differentiation uses
\texttt{jax.checkpoint}.
This structure ensures that both forward evolution and reverse-mode differentiation remain
scalable for large 3D problems.

\section{Examples and Comparisons}
\label{ap:examples}
In this section we provide some examples of pure hydrodynamical \texttt{diffhydro} and comparisons to \texttt{Athena++}, our primary benchmark code. For all our benchmarks (unless specifically metioned), we will use an HLLC Riemann solver with a MUSCL stencil and a Van Leer flux limiter. We will use a second order Runga-Kutta integrator without dimensional splitting.

\subsection{Sedov Taylor Blast Wave}
The Sedov--Taylor blast wave provides a stringent verification problem for hydrodynamical solvers owing to its strongly nonlinear character and the availability of an analytic, self-similar solution. In this configuration, a fixed amount of energy is deposited impulsively into an initially uniform, cold medium. The subsequent evolution is governed by a spherical shock that expands as a power-law in time,
$R \propto (E t^{2}/\rho_{0})^{1/5}$, with the interior thermodynamic and kinematic variables following unique similarity profiles determined solely by the adiabatic index. This test probes a code's ability to capture strong shocks, conserve energy, and maintain the symmetry and scaling structure of the solution even at low resolution.

The fidelity with which the Sedov solution is recovered depends sensitively on the underlying Riemann solver, which we show in Figure \ref{fig:sedov_compare} and show time scaling in Figure \ref{fig:sedov_timing}. Approximate solvers with relatively low dissipation, such as Roe or HLLC, typically reproduce a sharper shock front and more accurate post-shock density peak than more diffusive schemes like HLLE. Exact solvers can, in principle, deliver the most accurate reconstruction of the similarity solution, but in practice may struggle with robustness when confronted with the extreme pressure ratios inherent to the problem. The balance between robustness and sharp-shock capturing therefore places the choice of Riemann solver at the forefront of code-verification considerations.

Spatial reconstruction and slope limiting also play a central role in the accuracy of the test. First-order or highly diffusive second-order schemes tend to broaden the shock and smear the central high-pressure region, which delays the shock propagation and suppresses the peak values of density and pressure. More sophisticated reconstruction strategies---such as PLM or higher-order MUSCL methods---are able to maintain a sharper discontinuity and more faithfully reproduce the internal similarity structure, particularly when paired with moderate limiters (e.g.\ MC or van Leer) that avoid excessive numerical diffusion while preventing Gibbs-type oscillations. The Sedov blast thus serves not only as a shock test, but also as a valuable assessment of how reconstruction and limiting manage steep gradients and strong rarefactions.

%Temporal integration schemes exert a subtler but still important influence. First-order methods introduce significant phase errors and exacerbate shock broadening, whereas second-order Runge--Kutta schemes generally provide sufficient accuracy for this problem. When higher-order reconstruction is employed, strong-stability-preserving integrators (such as SSP-RK3) help maintain the overall order of accuracy and reduce temporal dissipation. 

\subsection{Kelvin Helmholtz Instability}

\begin{figure}
    \centering
    \includegraphics[width=0.5\linewidth]{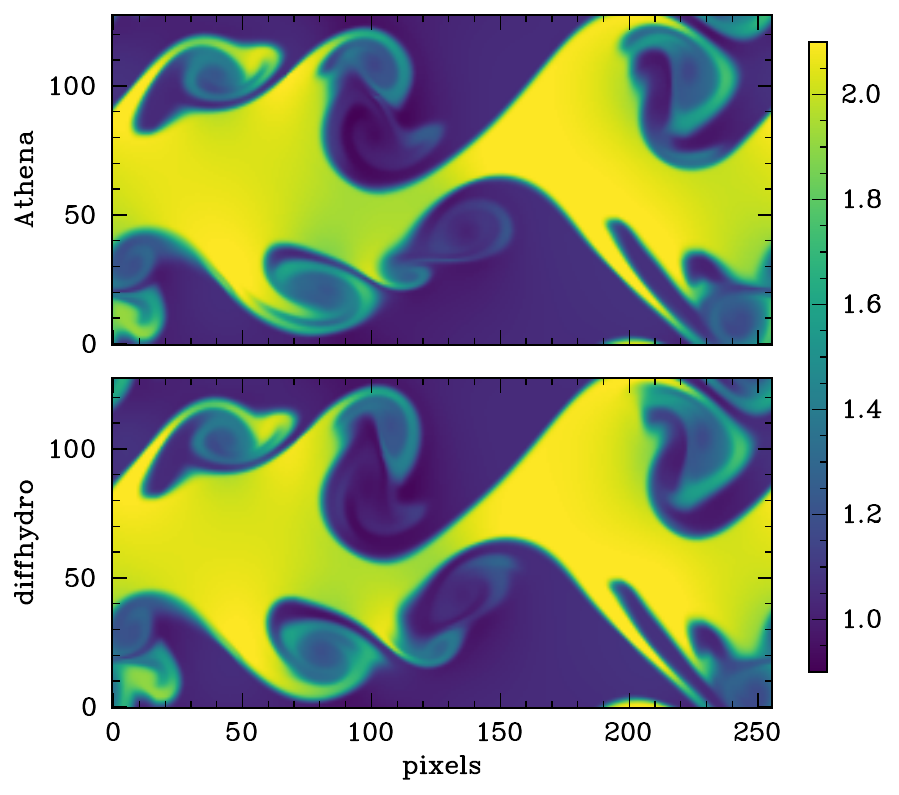}
    \caption{Two-dimensional Kelvin–Helmholtz instability at the same simulation time, comparing the reference \textsc{Athena++} solution (top) with the \textsc{diffhydro} result (bottom). Both runs use identical initial conditions obtained from \textsc{Athena++}. The large-scale shear layer evolution and vortex structures are in close qualitative agreement, while small differences arise from the distinct flux-reconstruction schemes employed by the two codes.}
    \label{fig:kh}
\end{figure}

The Kelvin--Helmholtz (KH) instability develops at interfaces between fluids of differing velocities, and is therefore ubiquitous in astrophysical environments containing winds, jets, cloud--flow interactions, or shear layers in the ISM and CGM. 
In the linear regime, perturbations of wavelength $\lambda$ grow at a characteristic rate 
\begin{equation}
    t_{\mathrm{KH}}^{-1} \sim k\,|\Delta v|,
\end{equation}
where $k=2\pi/\lambda$ is the perturbation wavenumber and $\Delta v$ is the velocity jump across the interface. 
In numerical simulations, the KH instability serves as a stringent test of a scheme's ability to advect sharp contact discontinuities without excessive diffusion, to capture vorticity, and to resolve the roll-up of vortices as the system enters the nonlinear regime. 
Accurate modelling requires both sufficient spatial resolution and reconstruction schemes that minimise numerical dissipation, as overly diffusive methods artificially suppress or delay the onset of KH growth. In astrophysical applications, the nonlinear evolution of the instability governs entrainment and mixing between phases, promotes the formation of multiphase gas in winds and jets, and mediates momentum and energy exchange in cloud--wind and filament–flow interactions. As such, faithful reproduction of KH dynamics is essential for capturing the correct mixing properties and thermodynamic evolution of shear-driven astrophysical flows.

We show an example of a Kelvin–Helmholtz (KH) instability evolved with \texttt{diffhydro} initialized with initial conditions generated by \textsc{Athena++} in Figure~\ref{fig:kh}, and compare the resulting evolution to the corresponding \textsc{Athena++} simulation. The two panels exhibit a high degree of qualitative similarity: the large-scale billows form at comparable locations, the secondary vortices appear with similar morphology, and the overall shear-layer evolution proceeds in a nearly identical fashion. This agreement is notable given that the two codes employ different flux–reconstruction schemes and distinct underlying discretizations.

Despite this close correspondence, we do not expect the solutions to match \textsc{Athena++} exactly. Subtle details of the KH instability—such as the timing and amplitude of secondary roll-ups, the sharpness of contact discontinuities, and the development of small-scale turbulent structure—are well known to depend sensitively on the numerical method, including reconstruction order, limiter choice, and Riemann solver treatment. Differences of this kind have been documented extensively in the literature \citep[e.g.,][]{san2015evaluation}. Our comparison therefore demonstrates that, while the two methods inevitably diverge at the finest scales, the overall dynamical evolution is robust and the qualitative fidelity of the KH flow is well preserved.

\section{V-cycles and W-cycles in multigrid gravity solvers}
\label{ap:vwcycle}

Multigrid algorithms differ in how aggressively they target coarse-grid error. Two standard variants are
the V-cycle and the W-cycle, which behave similarly on fine grids but differ substantially in the treatment
of long-wavelength errors.

\paragraph*{V-cycles:}

A V-cycle consists of a single descent through the grid hierarchy,
\begin{equation}
\ell = 0 \rightarrow 1 \rightarrow 2 \rightarrow \cdots \rightarrow L,
\end{equation}
followed by a single ascent back to the fine grid,
\begin{equation}
L \rightarrow L-1 \rightarrow \cdots \rightarrow 0.
\end{equation}
Only one recursive call is made at each coarse level. This strategy is computationally inexpensive
and rapidly damps high-frequency errors, but convergence can be limited by slowly varying components
of the residual, which are characteristic of periodic Poisson problems.

\paragraph*{W-cycles:}

A W-cycle performs two recursive traversals of the coarse levels:
\begin{equation}
0 \rightarrow 1 \rightarrow 2
\rightarrow (3 \rightarrow 4 \rightarrow 3)
\rightarrow 2
\rightarrow (3 \rightarrow 4 \rightarrow 3)
\rightarrow 1 \rightarrow 0.
\end{equation}
The result is a much stronger coarse-grid correction. If a V-cycle corresponds to an operator
$\mathcal{V}$, then a W-cycle behaves approximately like
\begin{equation}
\mathcal{W} \approx \mathcal{V} + \mathcal{V}^2,
\end{equation}
providing a more accurate approximation to $A^{-1}$ at the cost of additional work.

\textbf{Choice of W-cycles in this work:}  For the periodic Poisson equation encountered in self-gravitating hydrodynamics, the residual error after smoothing is dominated by long-wavelength modes. These modes benefit disproportionately from the enhanced coarse-grid treatment of W-cycles. In practice, W-cycles reduce residual norms by an order of magnitude per cycle even on large grids, whereas V-cycles may require substantially more cycles or smoothing steps to achieve comparable accuracy.

Furthermore, in differentiable simulations the stronger coarse-grid correction improves the conditioning of both the forward and adjoint Poisson solves. For these reasons, we adopt W-cycles as the default strategy in our multigrid gravity implementation in this work.

\bibliography{example_b}{}

@book{pontryagin2018mathematical,
  title={Mathematical theory of optimal processes},
  author={Pontryagin, Lev Semenovich},
  year={1962},
  publisher={Routledge}
}

@ARTICLE{2024A&C....4800858D,
       author = {{Ding}, J. and {Horowitz}, B. and {Luki{\'c}}, Z.},
        title = "{TensorFlow Hydrodynamics Analysis for Ly-{\ensuremath{\alpha}} Simulations}",
      journal = {Astronomy and Computing},
         year = 2024,
        month = jul,
       volume = {48},
          eid = {100858},
        pages = {100858},
          doi = {10.1016/j.ascom.2024.100858},
archivePrefix = {arXiv},
       eprint = {2407.16009},
 primaryClass = {astro-ph.CO},
       adsurl = {https://ui.adsabs.harvard.edu/abs/2024A&C....4800858D}
}

@ARTICLE{2020arXiv200304962X,
       author = {{XRISM Science Team}},
        title = "{Science with the X-ray Imaging and Spectroscopy Mission (XRISM)}",
      journal = {arXiv e-prints},
         year = 2020,
        month = mar,
          eid = {arXiv:2003.04962},
        pages = {arXiv:2003.04962},
          doi = {10.48550/arXiv.2003.04962},
archivePrefix = {arXiv},
       eprint = {2003.04962},
 primaryClass = {astro-ph.HE},
       adsurl = {https://ui.adsabs.harvard.edu/abs/2020arXiv200304962X}
}

@ARTICLE{2019ApJ...873..111I,
       author = {{Ivezi{\'c}}, {\v{Z}}eljko and {Kahn}, Steven M. and {Tyson}, J. Anthony and {Abel}, Bob and {Acosta}, Emily and {Allsman}, Robyn and {Alonso}, David and {AlSayyad}, Yusra and {Anderson}, Scott F. and {Andrew}, John and {Angel}, James Roger P. and {Angeli}, George Z. and {Ansari}, Reza and {Antilogus}, Pierre and {Araujo}, Constanza and {Armstrong}, Robert and {Arndt}, Kirk T. and {Astier}, Pierre and {Aubourg}, {\'E}ric and {Auza}, Nicole and {Axelrod}, Tim S. and {Bard}, Deborah J. and {Barr}, Jeff D. and {Barrau}, Aurelian and {Bartlett}, James G. and {Bauer}, Amanda E. and {Bauman}, Brian J. and {Baumont}, Sylvain and {Bechtol}, Ellen and {Bechtol}, Keith and {Becker}, Andrew C. and {Becla}, Jacek and {Beldica}, Cristina and {Bellavia}, Steve and {Bianco}, Federica B. and {Biswas}, Rahul and {Blanc}, Guillaume and {Blazek}, Jonathan and {Blandford}, Roger D. and {Bloom}, Josh S. and {Bogart}, Joanne and {Bond}, Tim W. and {Booth}, Michael T. and {Borgland}, Anders W. and {Borne}, Kirk and {Bosch}, James F. and {Boutigny}, Dominique and {Brackett}, Craig A. and {Bradshaw}, Andrew and {Brandt}, William Nielsen and {Brown}, Michael E. and {Bullock}, James S. and {Burchat}, Patricia and {Burke}, David L. and {Cagnoli}, Gianpietro and {Calabrese}, Daniel and {Callahan}, Shawn and {Callen}, Alice L. and {Carlin}, Jeffrey L. and {Carlson}, Erin L. and {Chandrasekharan}, Srinivasan and {Charles-Emerson}, Glenaver and {Chesley}, Steve and {Cheu}, Elliott C. and {Chiang}, Hsin-Fang and {Chiang}, James and {Chirino}, Carol and {Chow}, Derek and {Ciardi}, David R. and {Claver}, Charles F. and {Cohen-Tanugi}, Johann and {Cockrum}, Joseph J. and {Coles}, Rebecca and {Connolly}, Andrew J. and {Cook}, Kem H. and {Cooray}, Asantha and {Covey}, Kevin R. and {Cribbs}, Chris and {Cui}, Wei and {Cutri}, Roc and {Daly}, Philip N. and {Daniel}, Scott F. and {Daruich}, Felipe and {Daubard}, Guillaume and {Daues}, Greg and {Dawson}, William and {Delgado}, Francisco and {Dellapenna}, Alfred and {de Peyster}, Robert and {de Val-Borro}, Miguel and {Digel}, Seth W. and {Doherty}, Peter and {Dubois}, Richard and {Dubois-Felsmann}, Gregory P. and {Durech}, Josef and {Economou}, Frossie and {Eifler}, Tim and {Eracleous}, Michael and {Emmons}, Benjamin L. and {Fausti Neto}, Angelo and {Ferguson}, Henry and {Figueroa}, Enrique and {Fisher-Levine}, Merlin and {Focke}, Warren and {Foss}, Michael D. and {Frank}, James and {Freemon}, Michael D. and {Gangler}, Emmanuel and {Gawiser}, Eric and {Geary}, John C. and {Gee}, Perry and {Geha}, Marla and {Gessner}, Charles J.~B. and {Gibson}, Robert R. and {Gilmore}, D. Kirk and {Glanzman}, Thomas and {Glick}, William and {Goldina}, Tatiana and {Goldstein}, Daniel A. and {Goodenow}, Iain and {Graham}, Melissa L. and {Gressler}, William J. and {Gris}, Philippe and {Guy}, Leanne P. and {Guyonnet}, Augustin and {Haller}, Gunther and {Harris}, Ron and {Hascall}, Patrick A. and {Haupt}, Justine and {Hernandez}, Fabio and {Herrmann}, Sven and {Hileman}, Edward and {Hoblitt}, Joshua and {Hodgson}, John A. and {Hogan}, Craig and {Howard}, James D. and {Huang}, Dajun and {Huffer}, Michael E. and {Ingraham}, Patrick and {Innes}, Walter R. and {Jacoby}, Suzanne H. and {Jain}, Bhuvnesh and {Jammes}, Fabrice and {Jee}, M. James and {Jenness}, Tim and {Jernigan}, Garrett and {Jevremovi{\'c}}, Darko and {Johns}, Kenneth and {Johnson}, Anthony S. and {Johnson}, Margaret W.~G. and {Jones}, R. Lynne and {Juramy-Gilles}, Claire and {Juri{\'c}}, Mario and {Kalirai}, Jason S. and {Kallivayalil}, Nitya J. and {Kalmbach}, Bryce and {Kantor}, Jeffrey P. and {Karst}, Pierre and {Kasliwal}, Mansi M. and {Kelly}, Heather and {Kessler}, Richard and {Kinnison}, Veronica and {Kirkby}, David and {Knox}, Lloyd and {Kotov}, Ivan V. and {Krabbendam}, Victor L. and {Krughoff}, K. Simon and {Kub{\'a}nek}, Petr and {Kuczewski}, John and {Kulkarni}, Shri and {Ku}, John and {Kurita}, Nadine R. and {Lage}, Craig S. and {Lambert}, Ron and {Lange}, Travis and {Langton}, J. Brian and {Le Guillou}, Laurent and {Levine}, Deborah and {Liang}, Ming and {Lim}, Kian-Tat and {Lintott}, Chris J. and {Long}, Kevin E. and {Lopez}, Margaux and {Lotz}, Paul J. and {Lupton}, Robert H. and {Lust}, Nate B. and {MacArthur}, Lauren A. and {Mahabal}, Ashish and {Mandelbaum}, Rachel and {Markiewicz}, Thomas W. and {Marsh}, Darren S. and {Marshall}, Philip J. and {Marshall}, Stuart and {May}, Morgan and {McKercher}, Robert and {McQueen}, Michelle and {Meyers}, Joshua and {Migliore}, Myriam and {Miller}, Michelle and {Mills}, David J.},
        title = "{LSST: From Science Drivers to Reference Design and Anticipated Data Products}",
      journal = {\apj},
         year = 2019,
        month = mar,
       volume = {873},
       number = {2},
          eid = {111},
        pages = {111},
          doi = {10.3847/1538-4357/ab042c},
archivePrefix = {arXiv},
       eprint = {0805.2366},
 primaryClass = {astro-ph},
       adsurl = {https://ui.adsabs.harvard.edu/abs/2019ApJ...873..111I}
}

@ARTICLE{2006SSRv..123..485G,
       author = {{Gardner}, Jonathan P. and {Mather}, John C. and {Clampin}, Mark and {Doyon}, Rene and {Greenhouse}, Matthew A. and {Hammel}, Heidi B. and {Hutchings}, John B. and {Jakobsen}, Peter and {Lilly}, Simon J. and {Long}, Knox S. and {Lunine}, Jonathan I. and {McCaughrean}, Mark J. and {Mountain}, Matt and {Nella}, John and {Rieke}, George H. and {Rieke}, Marcia J. and {Rix}, Hans-Walter and {Smith}, Eric P. and {Sonneborn}, George and {Stiavelli}, Massimo and {Stockman}, H.~S. and {Windhorst}, Rogier A. and {Wright}, Gillian S.},
        title = "{The James Webb Space Telescope}",
      journal = {\ssr},
         year = 2006,
        month = apr,
       volume = {123},
       number = {4},
        pages = {485-606},
          doi = {10.1007/s11214-006-8315-7},
archivePrefix = {arXiv},
       eprint = {astro-ph/0606175},
 primaryClass = {astro-ph},
       adsurl = {https://ui.adsabs.harvard.edu/abs/2006SSRv..123..485G}
}

@article{mcnamara2004fluid,
  title={Fluid control using the adjoint method},
  author={McNamara, Antoine and Treuille, Adrien and Popovi{\'c}, Zoran and Stam, Jos},
  journal={ACM Transactions On Graphics (TOG)},
  volume={23},
  number={3},
  pages={449--456},
  year={2004},
  publisher={ACM New York, NY, USA}
}

@ARTICLE{2024MNRAS.529.2473H,
       author = {{Horowitz}, Benjamin and {Hahn}, ChangHoon and {Lanusse}, Francois and {Modi}, Chirag and {Ferraro}, Simone},
        title = "{Differentiable stochastic halo occupation distribution}",
      journal = {\mnras},
         year = 2024,
        month = apr,
       volume = {529},
       number = {3},
        pages = {2473-2482},
          doi = {10.1093/mnras/stae350},
archivePrefix = {arXiv},
       eprint = {2211.03852},
 primaryClass = {astro-ph.CO},
       adsurl = {https://ui.adsabs.harvard.edu/abs/2024MNRAS.529.2473H}
}

@ARTICLE{2025arXiv250202294H,
       author = {{Horowitz}, Benjamin and {Luki\'c}, Zarija},
        title = "{Differentiable Cosmological Hydrodynamics for Field-Level Inference and High Dimensional Parameter Constraints}",
      journal = {arXiv e-prints},
         year = 2025,
        month = feb,
          eid = {arXiv:2502.02294},
        pages = {arXiv:2502.02294},
          doi = {10.48550/arXiv.2502.02294},
archivePrefix = {arXiv},
       eprint = {2502.02294},
 primaryClass = {astro-ph.CO},
       adsurl = {https://ui.adsabs.harvard.edu/abs/2025arXiv250202294H}
}

@ARTICLE{2021A&C....3700505M,
       author = {{Modi}, C. and {Lanusse}, F. and {Seljak}, U.},
        title = "{FlowPM: Distributed TensorFlow implementation of the FastPM cosmological N-body solver}",
      journal = {Astronomy and Computing},
         year = 2021,
        month = oct,
       volume = {37},
          eid = {100505},
        pages = {100505},
          doi = {10.1016/j.ascom.2021.100505},
archivePrefix = {arXiv},
       eprint = {2010.11847},
 primaryClass = {astro-ph.CO},
       adsurl = {https://ui.adsabs.harvard.edu/abs/2021A&C....3700505M}
}

@ARTICLE{2024arXiv240916053S,
       author = {{Stone}, James M. and {Mullen}, Patrick D. and {Fielding}, Drummond and {Grete}, Philipp and {Guo}, Minghao and {Kempski}, Philipp and {Most}, Elias R. and {White}, Christopher J. and {Wong}, George N.},
        title = "{AthenaK: A Performance-Portable Version of the Athena++ AMR Framework}",
      journal = {arXiv e-prints},
         year = 2024,
        month = sep,
          eid = {arXiv:2409.16053},
        pages = {arXiv:2409.16053},
          doi = {10.48550/arXiv.2409.16053},
archivePrefix = {arXiv},
       eprint = {2409.16053},
 primaryClass = {astro-ph.IM},
       adsurl = {https://ui.adsabs.harvard.edu/abs/2024arXiv240916053S}
}

@ARTICLE{2025ApJ...990...49G,
       author = {{Guo}, Minghao and {Kim}, Chang-Goo and {Stone}, James M.},
        title = "{Evolution of Supernova Remnants in a Cloudy Multiphase Interstellar Medium}",
      journal = {\apj},
         year = 2025,
        month = sep,
       volume = {990},
       number = {1},
          eid = {49},
        pages = {49},
          doi = {10.3847/1538-4357/adeb85},
archivePrefix = {arXiv},
       eprint = {2411.12809},
 primaryClass = {astro-ph.GA},
       adsurl = {https://ui.adsabs.harvard.edu/abs/2025ApJ...990...49G}
}

@ARTICLE{2022ApJS..262....9O,
       author = {{Oku}, Yuri and {Tomida}, Kengo and {Nagamine}, Kentaro and {Shimizu}, Ikkoh and {Cen}, Renyue},
        title = "{Osaka Feedback Model. II. Modeling Supernova Feedback Based on High-resolution Simulations}",
      journal = {\apjs},
         year = 2022,
        month = sep,
       volume = {262},
       number = {1},
          eid = {9},
        pages = {9},
          doi = {10.3847/1538-4365/ac77ff},
archivePrefix = {arXiv},
       eprint = {2201.00970},
 primaryClass = {astro-ph.GA},
       adsurl = {https://ui.adsabs.harvard.edu/abs/2022ApJS..262....9O}
}

@article{giles2000introduction,
  title={An introduction to the adjoint approach to design},
  author={Giles, Michael B and Pierce, Niles A},
  journal={Flow, turbulence and combustion},
  volume={65},
  number={3},
  pages={393--415},
  year={2000},
  publisher={Springer}
}

@article{Lukic2015,
       author = {{Luki{\'c}}, Zarija and {Stark}, Casey W. and {Nugent}, Peter and {White}, Martin and {Meiksin}, Avery A. and {Almgren}, Ann},
        title = "{The Lyman {\ensuremath{\alpha}} forest in optically thin hydrodynamical simulations}",
      journal = {\mnras},
         year = 2015,
        month = feb,
       volume = {446},
       number = {4},
        pages = {3697-3724},
          doi = {10.1093/mnras/stu2377},
archivePrefix = {arXiv},
       eprint = {1406.6361},
 primaryClass = {astro-ph.CO},
       adsurl = {https://ui.adsabs.harvard.edu/abs/2015MNRAS.446.3697L}
}

@ARTICLE{2019TARDIS,
   author = {{Horowitz}, B. and {Lee}, K.-G. and {White}, M. and {Krolewski}, A. and 
	{Ata}, M.},
    title = "{TARDIS Paper I: A Constrained Reconstruction Approach to Modeling the z\~{}2.5 Cosmic Web Probed by Lyman-alpha Forest Tomography}",
  journal = {arXiv e-prints},
archivePrefix = "arXiv",
   eprint = {1903.09049},
     year = 2019,
    month = mar,
   adsurl = {http://adsabs.harvard.edu/abs/2019arXiv190309049H}
}

@article{KH,
	Adsurl = {http://adsabs.harvard.edu/abs/2013MNRAS.435L..78K},
	Archiveprefix = {arXiv},
	Author = {{Kitaura}, F.-S. and {He{\ss}}, S.},
	Doi = {10.1093/mnrasl/slt101},
	Eprint = {1212.3514},
	Journal = {\mnras},
	Month = aug,
	Pages = {L78-L82},
	Title = {{Cosmological structure formation with augmented Lagrangian perturbation theory}},
	Volume = 435,
	Year = 2013}

@book{NumRec,
	Adsurl = {http://adsabs.harvard.edu/abs/2002nrca.book.....P},
	Author = {{Press}, W.~H. and {Teukolsky}, S.~A. and {Vetterling}, W.~T. and {Flannery}, B.~P.},
	Booktitle = {Numerical recipes in C++ : the art of scientific computing by William H.~Press.~xxviii, 1,002 p.~: ill.~; 26 cm.~ Includes bibliographical references and index.~ISBN : 0521750334},
	Title = {{Numerical recipes in C++ : the art of scientific computing}},
	Year = 2002}

@misc{li2022pmwd,
      title={pmwd: A Differentiable Cosmological Particle-Mesh $N$-body Library}, 
      author={Yin Li and Libin Lu and Chirag Modi and Drew Jamieson and Yucheng Zhang and Yu Feng and Wenda Zhou and Ngai Pok Kwan and François Lanusse and Leslie Greengard},
      year={2022},
      eprint={2211.09958},
      archivePrefix={arXiv},
      primaryClass={astro-ph.IM},
      url={https://arxiv.org/abs/2211.09958}
}

@ARTICLE{Gizmo,
   author = {{Hopkins}, P.~F.},
    title = "{A new class of accurate, mesh-free hydrodynamic simulation methods}",
  journal = {\mnras},
archivePrefix = "arXiv",
   eprint = {1409.7395},
     year = 2015,
    month = jun,
   volume = 450,
    pages = {53-110},
      doi = {10.1093/mnras/stv195},
   adsurl = {http://adsabs.harvard.edu/abs/2015MNRAS.450...53H}
}

@ARTICLE{Arepo,
   author = {{Springel}, V.},
    title = "{E pur si muove: Galilean-invariant cosmological hydrodynamical simulations on a moving mesh}",
  journal = {\mnras},
archivePrefix = "arXiv",
   eprint = {0901.4107},
     year = 2010,
    month = jan,
   volume = 401,
    pages = {791-851},
      doi = {10.1111/j.1365-2966.2009.15715.x},
   adsurl = {http://adsabs.harvard.edu/abs/2010MNRAS.401..791S}
}

@article{Arepo_public,
    author = {Weinberger, Rainer and Springel, Volker and Pakmor, R\"udiger},
    title = "{The Arepo public code release}",
    eprint = "1909.04667",
    archivePrefix = "arXiv",
    primaryClass = "astro-ph.IM",
    doi = "10.3847/1538-4365/ab908c",
    journal = "Astrophys. J. Suppl.",
    volume = "248",
    number = "2",
    pages = "32",
    year = "2020"
}

@ARTICLE{Ramses,
       author = {{Teyssier}, R.},
        title = "{Cosmological hydrodynamics with adaptive mesh refinement. A new high resolution code called RAMSES}",
      journal = {\aap},
         year = 2002,
        month = apr,
       volume = {385},
        pages = {337-364},
          doi = {10.1051/0004-6361:20011817},
archivePrefix = {arXiv},
       eprint = {astro-ph/0111367},
 primaryClass = {astro-ph},
       adsurl = {https://ui.adsabs.harvard.edu/abs/2002A&A...385..337T}
}

@ARTICLE{2015ApJ...814....4L,
       author = {{Li}, Miao and {Ostriker}, Jeremiah P. and {Cen}, Renyue and {Bryan}, Greg L. and {Naab}, Thorsten},
        title = "{Supernova Feedback and the Hot Gas Filling Fraction of the Interstellar Medium}",
      journal = {\apj},
         year = 2015,
        month = nov,
       volume = {814},
       number = {1},
          eid = {4},
        pages = {4},
          doi = {10.1088/0004-637X/814/1/4},
archivePrefix = {arXiv},
       eprint = {1506.07180},
 primaryClass = {astro-ph.GA},
       adsurl = {https://ui.adsabs.harvard.edu/abs/2015ApJ...814....4L}
}

@ARTICLE{2017ApJ...834...25K,
       author = {{Kim}, Chang-Goo and {Ostriker}, Eve C. and {Raileanu}, Roberta},
        title = "{Superbubbles in the Multiphase ISM and the Loading of Galactic Winds}",
      journal = {\apj},
         year = 2017,
        month = jan,
       volume = {834},
       number = {1},
          eid = {25},
        pages = {25},
          doi = {10.3847/1538-4357/834/1/25},
archivePrefix = {arXiv},
       eprint = {1610.03092},
 primaryClass = {astro-ph.GA},
       adsurl = {https://ui.adsabs.harvard.edu/abs/2017ApJ...834...25K}
}

@ARTICLE{Kim2015,
       author = {{Kim}, Chang-Goo and {Ostriker}, Eve C.},
        title = "{Momentum Injection by Supernovae in the Interstellar Medium}",
      journal = {\apj},
         year = 2015,
        month = apr,
       volume = {802},
       number = {2},
          eid = {99},
        pages = {99},
          doi = {10.1088/0004-637X/802/2/99},
archivePrefix = {arXiv},
       eprint = {1410.1537},
 primaryClass = {astro-ph.GA},
       adsurl = {https://ui.adsabs.harvard.edu/abs/2015ApJ...802...99K}
}

@ARTICLE{2018A&A...617A.133M,
       author = {{Mini{\`e}re}, J. and {Bouquet}, S.~E. and {Michaut}, C. and {Sanz}, J. and {Mancini}, M.},
        title = "{Numerical study of the Vishniac instability in cooled supernova remnants}",
      journal = {\aap},
         year = 2018,
        month = sep,
       volume = {617},
          eid = {A133},
        pages = {A133},
          doi = {10.1051/0004-6361/201832663},
       adsurl = {https://ui.adsabs.harvard.edu/abs/2018A&A...617A.133M}
}

@ARTICLE{1974ApJ...188..501C,
       author = {{Chevalier}, Roger A.},
        title = "{The Evolution of Supernova Remnants. Spherically Symmetric Models}",
      journal = {\apj},
         year = 1974,
        month = mar,
       volume = {188},
        pages = {501-516},
          doi = {10.1086/152740},
       adsurl = {https://ui.adsabs.harvard.edu/abs/1974ApJ...188..501C}
}

@ARTICLE{1983ApJ...274..152V,
       author = {{Vishniac}, E.~T.},
        title = "{The dynamic and gravitational instabilities of spherical shocks}",
      journal = {\apj},
         year = 1983,
        month = nov,
       volume = {274},
        pages = {152-167},
          doi = {10.1086/161433},
       adsurl = {https://ui.adsabs.harvard.edu/abs/1983ApJ...274..152V}
}

@ARTICLE{2024ApJ...965..168R,
       author = {{Romano}, Leonard E.~C. and {Behrendt}, Manuel and {Burkert}, Andreas},
        title = "{Cloud Formation by Supernova Implosion}",
      journal = {\apj},
         year = 2024,
        month = apr,
       volume = {965},
       number = {2},
          eid = {168},
        pages = {168},
          doi = {10.3847/1538-4357/ad2c05},
archivePrefix = {arXiv},
       eprint = {2402.05796},
 primaryClass = {astro-ph.GA},
       adsurl = {https://ui.adsabs.harvard.edu/abs/2024ApJ...965..168R}
}

@ARTICLE{Koyama2002,
       author = {{Koyama}, Hiroshi and {Inutsuka}, Shu-ichiro},
        title = "{An Origin of Supersonic Motions in Interstellar Clouds}",
      journal = {\apjl},
         year = 2002,
        month = jan,
       volume = {564},
       number = {2},
        pages = {L97-L100},
          doi = {10.1086/338978},
archivePrefix = {arXiv},
       eprint = {astro-ph/0112420},
 primaryClass = {astro-ph},
       adsurl = {https://ui.adsabs.harvard.edu/abs/2002ApJ...564L..97K}
}

@ARTICLE{2017MNRAS.466.2217S,
       author = {{Smith}, Britton D. and {Bryan}, Greg L. and {Glover}, Simon C.~O. and {Goldbaum}, Nathan J. and {Turk}, Matthew J. and {Regan}, John and {Wise}, John H. and {Schive}, Hsi-Yu and {Abel}, Tom and {Emerick}, Andrew and {O'Shea}, Brian W. and {Anninos}, Peter and {Hummels}, Cameron B. and {Khochfar}, Sadegh},
        title = "{GRACKLE: a chemistry and cooling library for astrophysics}",
      journal = {\mnras},
         year = 2017,
        month = apr,
       volume = {466},
       number = {2},
        pages = {2217-2234},
          doi = {10.1093/mnras/stw3291},
archivePrefix = {arXiv},
       eprint = {1610.09591},
 primaryClass = {astro-ph.CO},
       adsurl = {https://ui.adsabs.harvard.edu/abs/2017MNRAS.466.2217S}
}

@ARTICLE{1998PASP..110..761F,
       author = {{Ferland}, G.~J. and {Korista}, K.~T. and {Verner}, D.~A. and {Ferguson}, J.~W. and {Kingdon}, J.~B. and {Verner}, E.~M.},
        title = "{CLOUDY 90: Numerical Simulation of Plasmas and Their Spectra}",
      journal = {\pasp},
         year = 1998,
        month = jul,
       volume = {110},
       number = {749},
        pages = {761-778},
          doi = {10.1086/316190},
       adsurl = {https://ui.adsabs.harvard.edu/abs/1998PASP..110..761F}
}

@ARTICLE{2023arXiv230113062S,
       author = {{Snider}, Daniel and {Liang}, Ruofan},
        title = "{Operator Fusion in XLA: Analysis and Evaluation}",
      journal = {arXiv e-prints},
         year = 2023,
        month = jan,
          eid = {arXiv:2301.13062},
        pages = {arXiv:2301.13062},
          doi = {10.48550/arXiv.2301.13062},
archivePrefix = {arXiv},
       eprint = {2301.13062},
 primaryClass = {cs.LG},
       adsurl = {https://ui.adsabs.harvard.edu/abs/2023arXiv230113062S}
}

@article{ashari2015optimizing,
  title={On optimizing machine learning workloads via kernel fusion},
  author={Ashari, Arash and Tatikonda, Shirish and Boehm, Matthias and Reinwald, Berthold and Campbell, Keith and Keenleyside, John and Sadayappan, P},
  journal={ACM SIGPLAN Notices},
  volume={50},
  number={8},
  pages={173--182},
  year={2015},
  publisher={ACM New York, NY, USA}
}

@inproceedings{frostig2019compiling,
  title={Compiling machine learning programs via high-level tracing},
  author={Frostig, Roy and Johnson, Matthew James and Leary, Chris},
  booktitle={SysML conference 2018},
  year={2019}
}

@article{san2015evaluation,
  title={Evaluation of Riemann flux solvers for WENO reconstruction schemes: Kelvin--Helmholtz instability},
  author={San, Omer and Kara, Kursat},
  journal={Computers \& Fluids},
  volume={117},
  pages={24--41},
  year={2015},
  publisher={Elsevier}
}

@ARTICLE{Sutherland1993,
       author = {{Sutherland}, Ralph S. and {Dopita}, M.~A.},
        title = "{Cooling Functions for Low-Density Astrophysical Plasmas}",
      journal = {\apjs},
         year = 1993,
        month = sep,
       volume = {88},
        pages = {253},
          doi = {10.1086/191823},
       adsurl = {https://ui.adsabs.harvard.edu/abs/1993ApJS...88..253S}
}

@ARTICLE{1930PhRv...36..823U,
       author = {{Uhlenbeck}, G.~E. and {Ornstein}, L.~S.},
        title = "{On the Theory of the Brownian Motion}",
      journal = {Physical Review},
         year = 1930,
        month = sep,
       volume = {36},
       number = {5},
        pages = {823-841},
          doi = {10.1103/PhysRev.36.823},
       adsurl = {https://ui.adsabs.harvard.edu/abs/1930PhRv...36..823U}
}

@ARTICLE{Eswaran1988,
       author = {{Eswaran}, V. and {Pope}, S.~B.},
        title = "{An examination of forcing in direct numerical simulations of turbulence}",
      journal = {Computers and Fluids},
         year = 1988,
        month = jan,
       volume = {16},
       number = {3},
        pages = {257-278},
          doi = {10.1016/0045-7930(88)90013-8},
       adsurl = {https://ui.adsabs.harvard.edu/abs/1988CF.....16..257E}
}

@ARTICLE{Federrath2010,
       author = {{Federrath}, C. and {Roman-Duval}, J. and {Klessen}, R.~S. and {Schmidt}, W. and {Mac Low}, M.-M.},
        title = "{Comparing the statistics of interstellar turbulence in simulations and observations. Solenoidal versus compressive turbulence forcing}",
      journal = {\aap},
         year = 2010,
        month = mar,
       volume = {512},
          eid = {A81},
        pages = {A81},
          doi = {10.1051/0004-6361/200912437},
archivePrefix = {arXiv},
       eprint = {0905.1060},
 primaryClass = {astro-ph.SR},
       adsurl = {https://ui.adsabs.harvard.edu/abs/2010A&A...512A..81F}
}

@ARTICLE{1998ApJ...508L..99S,
       author = {{Stone}, James M. and {Ostriker}, Eve C. and {Gammie}, Charles F.},
        title = "{Dissipation in Compressible Magnetohydrodynamic Turbulence}",
      journal = {\apjl},
         year = 1998,
        month = nov,
       volume = {508},
       number = {1},
        pages = {L99-L102},
          doi = {10.1086/311718},
archivePrefix = {arXiv},
       eprint = {astro-ph/9809357},
 primaryClass = {astro-ph},
       adsurl = {https://ui.adsabs.harvard.edu/abs/1998ApJ...508L..99S}
}

@ARTICLE{2004ARA&A..42..211E,
       author = {{Elmegreen}, Bruce G. and {Scalo}, John},
        title = "{Interstellar Turbulence I: Observations and Processes}",
      journal = {\araa},
         year = 2004,
        month = sep,
       volume = {42},
       number = {1},
        pages = {211-273},
          doi = {10.1146/annurev.astro.41.011802.094859},
archivePrefix = {arXiv},
       eprint = {astro-ph/0404451},
 primaryClass = {astro-ph},
       adsurl = {https://ui.adsabs.harvard.edu/abs/2004ARA&A..42..211E}
}

@ARTICLE{2025arXiv251105484R,
       author = {{Rubira}, Henrique and {Schmidt}, Fabian},
        title = "{Non-Gaussian Galaxy Stochasticity and the Noise-Field Formulation}",
      journal = {arXiv e-prints},
         year = 2025,
        month = nov,
          eid = {arXiv:2511.05484},
        pages = {arXiv:2511.05484},
          doi = {10.48550/arXiv.2511.05484},
archivePrefix = {arXiv},
       eprint = {2511.05484},
 primaryClass = {astro-ph.CO},
       adsurl = {https://ui.adsabs.harvard.edu/abs/2025arXiv251105484R}
}

@inproceedings{li2020scalable,
  title={Scalable gradients and variational inference for stochastic differential equations},
  author={Li, Xuechen and Wong, Ting-Kam Leonard and Chen, Ricky TQ and Duvenaud, David K},
  booktitle={Symposium on Advances in Approximate Bayesian Inference},
  pages={1--28},
  year={2020},
  organization={PMLR}
}

@ARTICLE{1995ApJS..100..269P,
       author = {{Pen}, Ue-Li},
        title = "{A Linear Moving Adaptive Particle-Mesh N-Body Algorithm}",
      journal = {\apjs},
         year = 1995,
        month = sep,
       volume = {100},
        pages = {269},
          doi = {10.1086/192219},
       adsurl = {https://ui.adsabs.harvard.edu/abs/1995ApJS..100..269P}
}

@article{ibeid2020fft,
  title={FFT, FMM, and multigrid on the road to exascale: Performance challenges and opportunities},
  author={Ibeid, Huda and Olson, Luke and Gropp, William},
  journal={Journal of Parallel and Distributed Computing},
  volume={136},
  pages={63--74},
  year={2020},
  publisher={Elsevier}
}

@ARTICLE{2008ApJ...688L..79F,
       author = {{Federrath}, Christoph and {Klessen}, Ralf S. and {Schmidt}, Wolfram},
        title = "{The Density Probability Distribution in Compressible Isothermal Turbulence: Solenoidal versus Compressive Forcing}",
      journal = {\apjl},
         year = 2008,
        month = dec,
       volume = {688},
       number = {2},
        pages = {L79},
          doi = {10.1086/595280},
archivePrefix = {arXiv},
       eprint = {0808.0605},
 primaryClass = {astro-ph},
       adsurl = {https://ui.adsabs.harvard.edu/abs/2008ApJ...688L..79F}
}

@ARTICLE{2007ApJ...665..416K,
       author = {{Kritsuk}, Alexei G. and {Norman}, Michael L. and {Padoan}, Paolo and {Wagner}, Rick},
        title = "{The Statistics of Supersonic Isothermal Turbulence}",
      journal = {\apj},
         year = 2007,
        month = aug,
       volume = {665},
       number = {1},
        pages = {416-431},
          doi = {10.1086/519443},
archivePrefix = {arXiv},
       eprint = {0704.3851},
 primaryClass = {astro-ph},
       adsurl = {https://ui.adsabs.harvard.edu/abs/2007ApJ...665..416K}
}

@ARTICLE{2013MNRAS.436.1245F,
       author = {{Federrath}, Christoph},
        title = "{On the universality of supersonic turbulence}",
      journal = {\mnras},
         year = 2013,
        month = dec,
       volume = {436},
       number = {2},
        pages = {1245-1257},
          doi = {10.1093/mnras/stt1644},
archivePrefix = {arXiv},
       eprint = {1306.3989},
 primaryClass = {astro-ph.SR},
       adsurl = {https://ui.adsabs.harvard.edu/abs/2013MNRAS.436.1245F}
}

@ARTICLE{2020arXiv200700016U,
       author = {{Um}, Kiwon and {Brand}, Robert and {Yun} and {Fei} and {Holl}, Philipp and {Thuerey}, Nils},
        title = "{Solver-in-the-Loop: Learning from Differentiable Physics to Interact with Iterative PDE-Solvers}",
      journal = {arXiv e-prints},
         year = 2020,
        month = jun,
          eid = {arXiv:2007.00016},
        pages = {arXiv:2007.00016},
          doi = {10.48550/arXiv.2007.00016},
archivePrefix = {arXiv},
       eprint = {2007.00016},
 primaryClass = {physics.comp-ph},
       adsurl = {https://ui.adsabs.harvard.edu/abs/2020arXiv200700016U}
}

@article{Bezgin2025,
   author = {Deniz A. Bezgin and Aaron B. Buhendwa and Nikolaus A. Adams},
   doi = {10.1016/j.cpc.2024.109433},
   issn = {00104655},
   journal = {Computer Physics Communications},
   month = {3},
   pages = {109433},
   title = {JAX-Fluids 2.0: Towards HPC for differentiable CFD of compressible two-phase flows},
   volume = {308},
   url = {https://linkinghub.elsevier.com/retrieve/pii/S0010465524003564},
   year = {2025},
}

@ARTICLE{2025arXiv250523940F,
       author = {{Fan}, Xiantao and {Liu}, Xinyang and {Wang}, Meng and {Wang}, Jian-Xun},
        title = "{Diff-FlowFSI: A GPU-Optimized Differentiable CFD Platform for High-Fidelity Turbulence and FSI Simulations}",
      journal = {arXiv e-prints},
         year = 2025,
        month = may,
          eid = {arXiv:2505.23940},
        pages = {arXiv:2505.23940},
          doi = {10.48550/arXiv.2505.23940},
archivePrefix = {arXiv},
       eprint = {2505.23940},
 primaryClass = {physics.flu-dyn},
       adsurl = {https://ui.adsabs.harvard.edu/abs/2025arXiv250523940F}
}

@ARTICLE{2024A&A...690A.317D,
       author = {{Dupourqu{\'e}}, S. and {Barret}, D. and {Diez}, C.~M. and {Guillot}, S. and {Quintin}, E.},
        title = "{jaxspec: A fast and robust Python library for X-ray spectral fitting}",
      journal = {\aap},
         year = 2024,
        month = oct,
       volume = {690},
          eid = {A317},
        pages = {A317},
          doi = {10.1051/0004-6361/202451736},
       adsurl = {https://ui.adsabs.harvard.edu/abs/2024A&A...690A.317D}
}

@ARTICLE{2012MNRAS.420.3545B,
       author = {{Biffi}, V. and {Dolag}, K. and {B{\"o}hringer}, H. and {Lemson}, G.},
        title = "{Observing simulated galaxy clusters with PHOX: a novel X-ray photon simulator}",
      journal = {\mnras},
         year = 2012,
        month = mar,
       volume = {420},
       number = {4},
        pages = {3545-3556},
          doi = {10.1111/j.1365-2966.2011.20278.x},
archivePrefix = {arXiv},
       eprint = {1112.0314},
 primaryClass = {astro-ph.CO},
       adsurl = {https://ui.adsabs.harvard.edu/abs/2012MNRAS.420.3545B}
}

@software{2016ascl.soft08002Z,
       author = {{ZuHone}, John A. and {Hallman}, Eric. J.},
        title = "{pyXSIM: Synthetic X-ray observations generator}",
 howpublished = {Astrophysics Source Code Library, record ascl:1608.002},
         year = 2016,
        month = aug,
          eid = {ascl:1608.002},
archivePrefix = {ascl},
       eprint = {1608.002},
       adsurl = {https://ui.adsabs.harvard.edu/abs/2016ascl.soft08002Z}
}

@ARTICLE{2020ApJS..249....4S,
       author = {{Stone}, James M. and {Tomida}, Kengo and {White}, Christopher J. and {Felker}, Kyle G.},
        title = "{The Athena++ Adaptive Mesh Refinement Framework: Design and Magnetohydrodynamic Solvers}",
      journal = {\apjs},
         year = 2020,
        month = jul,
       volume = {249},
       number = {1},
          eid = {4},
        pages = {4},
          doi = {10.3847/1538-4365/ab929b},
archivePrefix = {arXiv},
       eprint = {2005.06651},
 primaryClass = {astro-ph.IM},
       adsurl = {https://ui.adsabs.harvard.edu/abs/2020ApJS..249....4S}
}

@ARTICLE{2021arXiv210905237T,
       author = {{Thuerey}, N. and {Holzschuh}, B. and {Holl}, P. and {Kohl}, G. and {Lino}, M. and {Liu}, Q. and {Schnell}, P. and {Trost}, F.},
        title = "{Physics-based Deep Learning}",
      journal = {arXiv e-prints},
         year = 2021,
        month = sep,
          eid = {arXiv:2109.05237},
        pages = {arXiv:2109.05237},
          doi = {10.48550/arXiv.2109.05237},
archivePrefix = {arXiv},
       eprint = {2109.05237},
 primaryClass = {stat.ML},
       adsurl = {https://ui.adsabs.harvard.edu/abs/2021arXiv210905237T}
}

@ARTICLE{2023CoPhC.28208527B,
       author = {{Bezgin}, Deniz A. and {Buhendwa}, Aaron B. and {Adams}, Nikolaus A.},
        title = "{JAX-Fluids: A fully-differentiable high-order computational fluid dynamics solver for compressible two-phase flows}",
      journal = {Computer Physics Communications},
         year = 2023,
        month = jan,
       volume = {282},
          eid = {108527},
        pages = {108527},
          doi = {10.1016/j.cpc.2022.108527},
archivePrefix = {arXiv},
       eprint = {2203.13760},
 primaryClass = {physics.flu-dyn},
       adsurl = {https://ui.adsabs.harvard.edu/abs/2023CoPhC.28208527B}
}

@ARTICLE{2023arXiv230713533K,
       author = {{Kim}, Hojin and {Shankar}, Varun and {Viswanathan}, Venkatasubramanian and {Maulik}, Romit},
        title = "{Generalizable data-driven turbulence closure modeling on unstructured grids with differentiable physics}",
      journal = {arXiv e-prints},
         year = 2023,
        month = jul,
          eid = {arXiv:2307.13533},
        pages = {arXiv:2307.13533},
          doi = {10.48550/arXiv.2307.13533},
archivePrefix = {arXiv},
       eprint = {2307.13533},
 primaryClass = {physics.flu-dyn},
       adsurl = {https://ui.adsabs.harvard.edu/abs/2023arXiv230713533K}
}

@ARTICLE{2025arXiv250904067C,
       author = {{Chaikin}, Evgenii and {Schaye}, Joop and {Schaller}, Matthieu and {Ploeckinger}, Sylvia and {Bah{\'e}}, Yannick M. and {Ben{\'\i}tez-Llambay}, Alejandro and {Correa}, Camila and {Forouhar Moreno}, Victor J. and {Frenk}, Carlos S. and {Hu{\v{s}}ko}, Filip and {Kugel}, Roi and {McGibbon}, Robert and {Richings}, Alexander J. and {Trayford}, James W. and {Borrow}, Josh and {Crain}, Robert A. and {Helly}, John C. and {Lacey}, Cedric G. and {Ludlow}, Aaron and {Nobels}, Folkert S.~J.},
        title = "{COLIBRE: calibrating subgrid feedback in cosmological simulations that include a cold gas phase}",
      journal = {arXiv e-prints},
         year = 2025,
        month = sep,
          eid = {arXiv:2509.04067},
        pages = {arXiv:2509.04067},
          doi = {10.48550/arXiv.2509.04067},
archivePrefix = {arXiv},
       eprint = {2509.04067},
 primaryClass = {astro-ph.GA},
       adsurl = {https://ui.adsabs.harvard.edu/abs/2025arXiv250904067C}
}

@ARTICLE{2023MNRAS.526.6103K,
       author = {{Kugel}, Roi and {Schaye}, Joop and {Schaller}, Matthieu and {Helly}, John C. and {Braspenning}, Joey and {Elbers}, Willem and {Frenk}, Carlos S. and {McCarthy}, Ian G. and {Kwan}, Juliana and {Salcido}, Jaime and {van Daalen}, Marcel P. and {Vandenbroucke}, Bert and {Bah{\'e}}, Yannick M. and {Borrow}, Josh and {Chaikin}, Evgenii and {Hu{\v{s}}ko}, Filip and {Jenkins}, Adrian and {Lacey}, Cedric G. and {Nobels}, Folkert S.~J. and {Vernon}, Ian},
        title = "{FLAMINGO: calibrating large cosmological hydrodynamical simulations with machine learning}",
      journal = {\mnras},
         year = 2023,
        month = dec,
       volume = {526},
       number = {4},
        pages = {6103-6127},
          doi = {10.1093/mnras/stad2540},
archivePrefix = {arXiv},
       eprint = {2306.05492},
 primaryClass = {astro-ph.CO},
       adsurl = {https://ui.adsabs.harvard.edu/abs/2023MNRAS.526.6103K}
}

@ARTICLE{2021MNRAS.506.4011E,
       author = {{Elliott}, Edward J. and {Baugh}, Carlton M. and {Lacey}, Cedric G.},
        title = "{Efficient exploration and calibration of a semi-analytical model of galaxy formation with deep learning}",
      journal = {\mnras},
         year = 2021,
        month = sep,
       volume = {506},
       number = {3},
        pages = {4011-4030},
          doi = {10.1093/mnras/stab1837},
archivePrefix = {arXiv},
       eprint = {2103.01072},
 primaryClass = {astro-ph.GA},
       adsurl = {https://ui.adsabs.harvard.edu/abs/2021MNRAS.506.4011E}
}

@ARTICLE{2025MNRAS.536..254M,
       author = {{Modi}, Chirag and {Pandey}, Shivam and {Ho}, Matthew and {Hahn}, ChangHoon and {R{\'e}galdo-Saint Blancard}, Bruno and {Wandelt}, Benjamin},
        title = "{Sensitivity analysis of simulation-based inference for galaxy clustering}",
      journal = {\mnras},
         year = 2025,
        month = jan,
       volume = {536},
       number = {1},
        pages = {254-265},
          doi = {10.1093/mnras/stae2473},
archivePrefix = {arXiv},
       eprint = {2309.15071},
 primaryClass = {astro-ph.CO},
       adsurl = {https://ui.adsabs.harvard.edu/abs/2025MNRAS.536..254M}
}

@ARTICLE{2022ApJ...941...42H,
       author = {{Horowitz}, Benjamin and {Dornfest}, Max and {Luki{\'c}}, Zarija and {Harrington}, Peter},
        title = "{HYPHY: Deep Generative Conditional Posterior Mapping of Hydrodynamical Physics}",
      journal = {\apj},
         year = 2022,
        month = dec,
       volume = {941},
       number = {1},
          eid = {42},
        pages = {42},
          doi = {10.3847/1538-4357/ac9ea7},
archivePrefix = {arXiv},
       eprint = {2106.12675},
 primaryClass = {astro-ph.CO},
       adsurl = {https://ui.adsabs.harvard.edu/abs/2022ApJ...941...42H}
}

@ARTICLE{2004RvMP...76..125M,
       author = {{Mac Low}, Mordecai-Mark and {Klessen}, Ralf S.},
        title = "{Control of star formation by supersonic turbulence}",
      journal = {Reviews of Modern Physics},
         year = 2004,
        month = jan,
       volume = {76},
       number = {1},
        pages = {125-194},
          doi = {10.1103/RevModPhys.76.125},
archivePrefix = {arXiv},
       eprint = {astro-ph/0301093},
 primaryClass = {astro-ph},
       adsurl = {https://ui.adsabs.harvard.edu/abs/2004RvMP...76..125M}
}

@ARTICLE{2025arXiv251005206L,
       author = {{List}, Florian and {Hahn}, Oliver and {Fl{\"o}ss}, Thomas and {Winkler}, Lukas},
        title = "{DISCO-DJ II: a differentiable particle-mesh code for cosmology}",
      journal = {arXiv e-prints},
         year = 2025,
        month = oct,
          eid = {arXiv:2510.05206},
        pages = {arXiv:2510.05206},
          doi = {10.48550/arXiv.2510.05206},
archivePrefix = {arXiv},
       eprint = {2510.05206},
 primaryClass = {astro-ph.CO},
       adsurl = {https://ui.adsabs.harvard.edu/abs/2025arXiv251005206L}
}

@ARTICLE{2024JCAP...06..063H,
       author = {{Hahn}, Oliver and {List}, Florian and {Porqueres}, Natalia},
        title = "{DISCO-DJ I: a differentiable Einstein-Boltzmann solver for cosmology}",
      journal = {\jcap},
         year = 2024,
        month = jun,
       volume = {2024},
       number = {6},
          eid = {063},
        pages = {063},
          doi = {10.1088/1475-7516/2024/06/063},
archivePrefix = {arXiv},
       eprint = {2311.03291},
 primaryClass = {astro-ph.CO},
       adsurl = {https://ui.adsabs.harvard.edu/abs/2024JCAP...06..063H}
}

@ARTICLE{2023OJAp....6E..15C,
       author = {{Campagne}, Jean-Eric and {Lanusse}, Fran{\c{c}}ois and {Zuntz}, Joe and {Boucaud}, Alexandre and {Casas}, Santiago and {Karamanis}, Minas and {Kirkby}, David and {Lanzieri}, Denise and {Peel}, Austin and {Li}, Yin},
        title = "{JAX-COSMO: An End-to-End Differentiable and GPU Accelerated Cosmology Library}",
      journal = {The Open Journal of Astrophysics},
         year = 2023,
        month = apr,
       volume = {6},
          eid = {15},
        pages = {15},
          doi = {10.21105/astro.2302.05163},
archivePrefix = {arXiv},
       eprint = {2302.05163},
 primaryClass = {astro-ph.CO},
       adsurl = {https://ui.adsabs.harvard.edu/abs/2023OJAp....6E..15C}
}

@ARTICLE{2025astrochemical,
       author = {{Vermari{\"e}n}, Gijs and {Grassi}, Tommaso and {Van de Sande}, Marie and {Viti}, Serena and {Bovino}, Stefano and {Lupi}, Alessandro and {Ruf}, Alexander and {Branca}, Lorenzo and {Walsh}, Catherine},
        title = "{Carbox: an end-to-end differentiable astrochemical simulation framework}",
      journal = {arXiv e-prints},
         year = 2025,
        month = nov,
          eid = {arXiv:2511.10558},
        pages = {arXiv:2511.10558},
          doi = {10.48550/arXiv.2511.10558},
archivePrefix = {arXiv},
       eprint = {2511.10558},
 primaryClass = {astro-ph.GA},
       adsurl = {https://ui.adsabs.harvard.edu/abs/2025arXiv251110558V}
}

@ARTICLE{2025rayrace,
       author = {{Branca}, Lorenzo and {Rost}, Rune and {Buck}, Tobias},
        title = "{Ray-trax: Fast, Time-Dependent, and Differentiable Ray Tracing for On-the-fly Radiative Transfer in Turbulent Astrophysical Flows}",
      journal = {arXiv e-prints},
         year = 2025,
        month = nov,
          eid = {arXiv:2511.09389},
        pages = {arXiv:2511.09389},
          doi = {10.48550/arXiv.2511.09389},
archivePrefix = {arXiv},
       eprint = {2511.09389},
 primaryClass = {astro-ph.IM},
       adsurl = {https://ui.adsabs.harvard.edu/abs/2025arXiv251109389B}
}

@ARTICLE{2025A&A...699A..42B,
       author = {{Boettner}, C.},
        title = "{gallifrey: JAX-based Gaussian process structure learning for astronomical time series}",
      journal = {\aap},
         year = 2025,
        month = jul,
       volume = {699},
          eid = {A42},
        pages = {A42},
          doi = {10.1051/0004-6361/202554518},
archivePrefix = {arXiv},
       eprint = {2505.20394},
 primaryClass = {astro-ph.IM},
       adsurl = {https://ui.adsabs.harvard.edu/abs/2025A&A...699A..42B}
}

@ARTICLE{2025jfof,
       author = {{Horowitz}, Benjamin and {Bayer}, Adrian E.},
        title = "{jFoF: GPU Cluster Finding with Gradient Propagation}",
      journal = {arXiv e-prints},
         year = 2025,
        month = oct,
          eid = {arXiv:2510.26851},
        pages = {arXiv:2510.26851},
          doi = {10.48550/arXiv.2510.26851},
archivePrefix = {arXiv},
       eprint = {2510.26851},
 primaryClass = {astro-ph.IM},
       adsurl = {https://ui.adsabs.harvard.edu/abs/2025arXiv251026851H}
}

@ARTICLE{2025sync,
       author = {{Diao}, Kangning and {Li}, Zack and {Grumitt}, Richard D.~P. and {Mao}, Yi},
        title = "{synax: A Differentiable and GPU-accelerated Synchrotron Simulation Package}",
      journal = {\apjs},
         year = 2025,
        month = may,
       volume = {278},
       number = {1},
          eid = {25},
        pages = {25},
          doi = {10.3847/1538-4365/adc5ff},
archivePrefix = {arXiv},
       eprint = {2410.01136},
 primaryClass = {astro-ph.IM},
       adsurl = {https://ui.adsabs.harvard.edu/abs/2025ApJS..278...25D}
}

@ARTICLE{2024stellarwinds,
       author = {{Storcks}, Leonard and {Buck}, Tobias},
        title = "{Differentiable Conservative Radially Symmetric Fluid Simulations and Stellar Winds -- jf1uids}",
      journal = {arXiv e-prints},
         year = 2024,
        month = oct,
          eid = {arXiv:2410.23093},
        pages = {arXiv:2410.23093},
          doi = {10.48550/arXiv.2410.23093},
archivePrefix = {arXiv},
       eprint = {2410.23093},
 primaryClass = {physics.flu-dyn},
       adsurl = {https://ui.adsabs.harvard.edu/abs/2024arXiv241023093S}
}
\bibliographystyle{aasjournalv7}



\end{document}